\documentclass[structabstract]{aa} 
\usepackage{longtable}
\usepackage{latexsym}
\usepackage{amssymb}
\usepackage{lscape}
\usepackage[]{natbib}

\usepackage{color}

\usepackage{graphicx}
\usepackage{txfonts}
\def\lsim{\mathrel{\rlap{\lower 3pt \hbox{$\sim$}} \raise 2.0pt \hbox{$<$}}}
\def\gsim{\mathrel{\rlap{\lower 3pt \hbox{$\sim$}} \raise 2.0pt \hbox{$>$}}}

\title{H$\alpha$ imaging observations of early-type galaxies from the $\rm ATLAS^{3D}$ survey \thanks{based on observations taken at the
Mexican Observatorio Astronomico Nacional and at the Loiano telescope belonging to the Bologna Observatory.}
}

\subtitle{}

\author{G. Gavazzi  \inst{1}                               
\and G. Consolandi \inst{1}                               
\and S. Pedraglio \inst{1}
\and M. Fossati \inst{2,3}                                  
\and M. Fumagalli \inst{4}                                
\and A. Boselli \inst{5}                                    
}
\authorrunning{G. Gavazzi et al.}
\titlerunning{H$\alpha$ imaging of $\rm ATLAS^{3D}$ galaxies}
\institute{Universit\`a degli Studi di Milano-Bicocca, Piazza della Scienza 3, 20126 Milano, Italy\\
\email {giuseppe.gavazzi@mib.infn.it}
\and
Max-Planck-Institut f{\"u}r Extraterrestrische Physik, Giessenbachstrasse, D-85748 Garching, Germany\\
\email {mfossati@mpe.mpg.de}
\and
Universit{\"a}ts-Sternwarte M{\"u}nchen, Schenierstrasse 1, D-81679 M{\"u}nchen, Germany. 
\and
Institute for Computational Cosmology, and Centre for Extragalactic Astronomy, Durham University, South Road, Durham, DH1 3LE, UK\\
\email {michele.fumagalli@durham.ac.uk}
\and
Aix Marseille Universit\'e, CNRS, LAM (Laboratoire d'Astrophysique de Marseille) UMR 7326, F-13388, Marseille, France\\
\email {alessandro.boselli@lam.fr}
}
\usepackage{amstext}

\begin{document}

\date{Received; accepted}

\abstract
{The traditional knowledge of the mechanisms that caused the formation and evolution of early-type galaxies (ETG) in a hierarchical universe was 
challenged by the unexpected finding by $\rm ATLAS^{3D}$ that 86\%  of the ETGs show signs of a fast-rotating disk at their interior. This implies a common origin of most spiral galaxies, followed by a quenching phase, 
while only a minority of the most massive systems are slow rotators
and were likely to be the products of merger events. }
{Our aim is to improve our knowledge on the content and distribution of ionized hydrogen and their usage to form stars
in a representative sample of ETGs
for which the kinematics and detailed morphological classification were known from $\rm ATLAS^{3D}$.}
{Using narrow-band filters centered on the redshifted H$\alpha$ line along with a broad-band ($r$-Gunn) filter to recover the stellar continuum, 
we observed or collected existing imaging observations 
for 147 ETGs (including members of the Virgo cluster) that are representative of the whole $\rm ATLAS^{3D}$ survey.}
{Fifty-five ETGs (37\%) were detected in the H$\alpha$ line above our detection threshold, (H$\alpha E.W. \leq -1 \AA$), and 21 harbor a strong source
(H$\alpha E.W. \leq -5 \AA$) .}
{The strong H$\alpha$ emitters appear associated with mostly low-mass ($M_*\sim 10^{10} M_\odot$) S0 galaxies that contain conspicuous 
stellar and gaseous discs. These harbor 
significant star formation at their interior, including their nuclei. The weak H$\alpha$ emitters are almost one order of magnitude more massive, 
contain gas-poor discs and harbor an AGN at their centers. Their emissivity is dominated by [NII] and does not imply star formation. 
The 92 undetected ETGs constitute the majority in our sample and are gas-free systems that lack a disc and exhibit 
passive spectra even in their nuclei.
These pieces of evidence reinforce the conclusion made previously that the evolution of ETGs followed the secular channel for the less
massive systems and the dry merging channel for the most massive galaxies at the center of clusters of galaxies.}
 

\maketitle
%

\section{Introduction}
Our understanding of the processes that caused the formation of galaxies and their subsequent evolution 
must cope with the observational evidence that today, galaxies are distributed in a bimodal population (Kauffmann et al. 2003, Balogh et al. 2004, 
Baldry et al. 2004):
the blue cloud, composed of star-forming gas-rich disky systems, and the red sequence made of 
quiescent gas-poor "red and dead" galaxies.
Galaxies are assumed to migrate from the star-forming blue cloud to the red sequence owing to a variety of quenching mechanisms.
It is common belief that in a hierarchical universe the main process that brought early-type galaxies (ETGs) across the 
green valley was merging of disky systems (late-type galaxies;
LTGs). These catastrophic events should have dissipated 
angular momentum of the pre-merging LTGs, producing non- or slowly rotating elliptical dispersion-dominated galaxies. 

 However, until the advent of integral field spectrographs, the fraction of true ellipticals (i.e., slow rotators, the outcome of mergers) 
could only be deduced by the optical morphology of galaxies without any knowledge of 
the stellar kinematics. 
Owing to the SAURON IFU spectrograph, the recent $\rm ATLAS^{3D}$ survey (Cappellari et al. 2011) derived 
resolved stellar kinematic maps of 260 ETGs and for the first time studied the kinematic morphology-density relation  using fast and slow rotators instead of ellipticals (E) and lenticulars (S0) 
(Cappellari et al. 2011b).
The survey showed that ETGs are dominated (86\%) by fast rotators up to intermediate stellar masses (Emsellem et al. 2011), and from the lowest density environments up to 
the dense core of the Virgo cluster where only a small increase in the fraction of slow rotators is found.
True ellipticals (i.e., slow rotators) are only found among the most massive ETGs and always have an ellipticity lower than 0.4, while lenticulars (i.e., fast rotators)
are less massive galaxies and span all possible different ellipticities (Cappellari et al. 2016). 
This new evidence challenged the paradigm stating that most ETGs are mainly created by merging events, and it opened  a revision of the classic tuning-fork scheme that includes the
kinematic information. Here, the fast rotators (namely S0s) follow a sequence parallel to spirals. 
The final outcome of the $\rm ATLAS^{3D}$ survey can be summarized as in Cappellari (2016):  
Fast-rotator ETGs were originally star-forming disks and evolved owing to 
secular phenomena, e.g. gas accretion, bulge growth, and quenching. 
On the opposite slow rotators assembled around massive halos at high redshift via gas poor merging. 
The  dichotomy in stellar mass between the fast rotators and the more massive slow rotators 
has also been reproduced in recent simulations (Bois et al. 2011; Penoyre et al. 2017) and
was reinforced by the findings of even more recent surveys such as the MASSIVE
survey (Veale et al. 2017), which demonstrates that only the most massive ETGs ($M_*>10^{11.5} M_\odot$) are true slow rotators and are perhaps the only genuine outcomes 
of equal-mass merger events.
These ETGs represent the majority of central massive haloes, such as cD galaxies in clusters of galaxies. 
On the other hand, less massive fast rotators, which represent the vast majority of ETGs, likely emerge from less dramatic
evolutionary paths that leave them 
enough angular momentum to be supported by rotation.\\
Hence, in addition to mergers (Kauffmann et al.1993), other secular (e.g., bar instability; Gavazzi et al. 2015) and environmental processes (Boselli \& Gavazzi 2006, 2014) 
need to be invoked to explain the population of the color-luminosity plane.
Which processes caused the migration of galaxies across the green valley (perhaps in both directions; Y{\i}ld{\i}z et al. 2017) is still a matter of debate.
This controversy has been tried to be resolved through many lines of research, which were focused on understanding the processes that govern the transformation efficiency of gas (HI and $\rm H_2$) into stars as a function of stellar mass, morphology, and environment.
Ongoing surveys that exploit integral field spectroscopy, such as CALIFA (Sanchez et al. 2012), 
SAMI (Bryant et al. 2015), and MaNGA (Belfiore et al. 2016), are about to provide us with spatially resolved diagnostic diagrams that will
eventually contribute to solving the current controversy.\\ 
As the SAURON IFU does not include the H$\alpha$ line in its bandpass,
we decided to investigate in this work the ionized hydrogen content of ETGs 
that were selected by $\rm ATLAS^{3D}$  by means of imaging observations taken through  narrow-band (80 \AA) filters, combined with 
a spectroscopic investigation of their nuclear activity. To this end, we combined existing spectroscopy (for the most part from Ho et al, 1995 and from SDSS
DR13 Albareti et al. 2016)
with observations obtained using the Loiano telescope at the Bologna Observatory.
The imaging observations, carried out at the 2.1m telescope at San Pedro Martir (SPM), are described in Section 3, 
and their calibration is described in Section 4, while our nuclear spectroscopy
is given in Section 5. The criterion to separate strong from weak H$\alpha$ detections and from undetected targets is described in Section 6.
The results are presented and discussed in Sections 7. We adopt a flat $\Lambda CDM$  cosmology with $\Omega_M=0.3$,
 $\Omega_\Lambda=0.7$ and $H_o$=73 $\rm km~s^{-1}Mpc^{-1}$. Magnitudes are given in the AB system.

\section{Sample}

\begin{figure*}
\centering
\includegraphics[angle=0, scale=0.8]{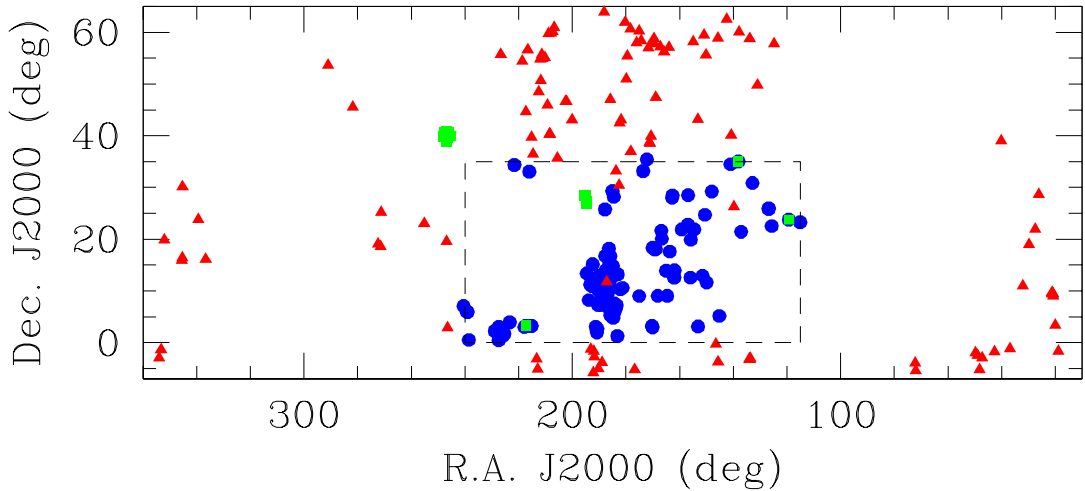}
\caption{Full $\rm ATLAS^{3D}$ survey (red triangles) and the subsample of 147 targets observed in H$\alpha$  
as part of this work (blue circles). Green squares represent the 20 filler targets (not in  
$\rm ATLAS^{3D}$).}
\label{atlas}  
\end{figure*}

The sample of ETGs analyzed in this work was extracted from the $\rm ATLAS^{3D}$ whole-sky catalog of 260 ETGs (see Figure
\ref{atlas}). The selection criteria adopted by Cappellari et al (2011) for including objects in $\rm ATLAS^{3D}$ are that they belong to
a volume-limited sample of 42 Mpc radius, are brighter than $M_k$ = -21.5 mag, and have $-6^o<\delta<64^o$.
The selection criterion for including them in our investigation is purely positional, according to target visibility in March-April from the
site of SPM (latitude = +31 deg). ETGs were selected in the range $10^h<R.A.<16^h$; $0^o<Dec.<35^o$; this contains the Virgo cluster.
Of the 260 galaxies in the $\rm ATLAS^{3D}$ survey,  151 fall within these boundaries:  29 of them were observed in previous H$\alpha$ 
imaging campaigns, 4 were not observed because of  bright stars in their vicinity, and the remaining 118 are the subject of current 
observational campaigns, as illustrated in Figure \ref{atlas} (blue symbols). 
To fill in the morning  hours, we selected 20 filler targets in the Coma and Hercules superclusters in 2016 (green symbols).
They mainly contribute to the setup of the H$\alpha$ reduction procedure (see Section \ref{Calibrations}.) 
The selection of 147 of 260 $\rm ATLAS^{3D}$ targets does not bias the distribution
of the selected galaxies for the morphological and the kinematical type. The morphological type mix in $\rm ATLAS^{3D}$
is 74\%  S0 and 26\% E; in our sample the two percentages become 68\% and 32\%. 
In $\rm ATLAS^{3D}$ , 14\%  of the ETGs are slow rotators (SR), and  the remaining 86\% are fast rotators (FR); in our sample the two percentages become 17\% and 83\%.
However, the observed sample contains the Virgo cluster and is more biased in favor of cluster galaxies (52 \%) than the 
entire $\rm ATLAS^{3D}$ survey (29 \%). 

\section{Imaging observations}

\begin{figure}
\centering
\includegraphics[angle=0, scale=0.39]{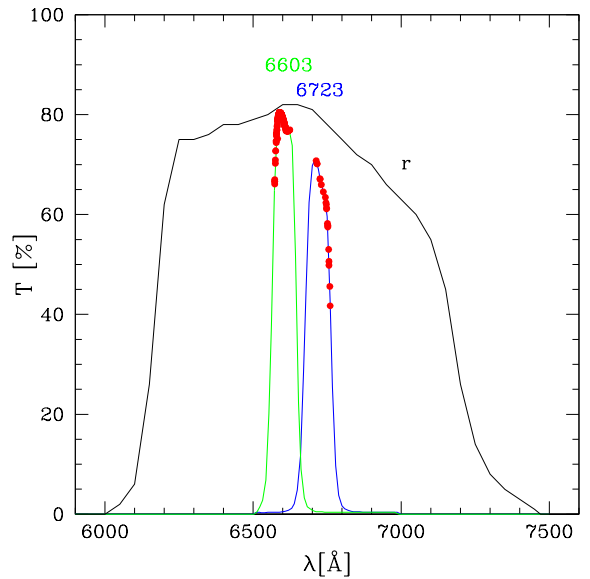}
\caption{Transmission profiles of the narrow-band filters (6603 \AA : green line; 6723 \AA : blue line) and  of the $r$-band filter (black line). 
The observed galaxies (red dots) are overlaid on the narrow-band filter profiles at the wavelength given by their redshift.}
\label{filters}  
\end{figure}

\begin{figure}
\centering
\includegraphics[angle=0, scale=0.39]{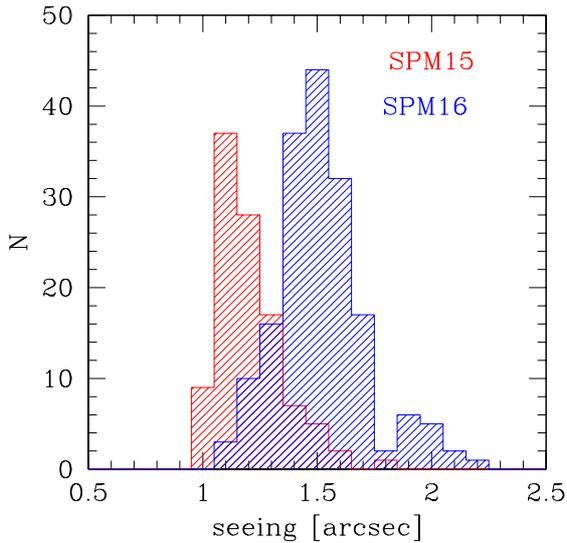}
\caption{Seeing conditions in 2015 (red) and 2016 (blue) measured on the final images. On average, the seeing in 2015 was 1.2 arcsec, 
while in 2016 it was 1.5 arcsec.}
\label{seeing}  
\end{figure}

Narrow-band imaging of the H$\alpha$ line emission (rest frame $\lambda$ = 6562.8 \AA) of 118 galaxies 
in the main program (+ 20 filler targets) was obtained using the 2.1m telescope
at the SPM, belonging to the Mexican Observatorio Astron\'omico Nacional (OAN).
The H$\alpha$  measurements of the remaining 29 targets included in the present investigation 
are taken from the literature. 
The observations were scheduled in two observing runs of eight nights each in 2015 (March 17-24) and 2016 (April 7-14), 
both including new-moon periods. The weather phenomenon el Ni\~no on the Pacific caused both runs to be severely affected by bad whether.
In 2015, in particular, four nights were clear, and only 1.5 of them were photometric. 
In 2016 we worked in nearly photometric conditions during five nights out of eight.
In both years, the Marconi CCD type e2vm2 was used with 2048x2048 pixel, binned twice with a pixel scale of 0.35 arcsec in a field of view
of 5.5x5.5 arcmin.
Each galaxy was observed using a narrow-band interference filter, whose bandpass included the redshifted wavelength of 
 the H$\alpha$ line and also the [NII] lines (ON-band frame). These filters maximize the throughput at the galaxy redshift 
 (see Fig. \ref{filters}) \footnote{The transmission profiles of the interferometric filters 
 plotted in Figure \ref{filters} refer to the their nominal values, as measured in 2000 at 20 degrees Celsius. The transmission, however, is known to drift toward
 the blue by approximately 1 $\AA$ per 3 degrees Celsius. The filter transmission is also known to change slightly 
 with time and to migrate toward the blue with beam convergence. These last two effects were not taken into account when we computed the
 transmissivity at the galaxy redshift. They should be negligible, however, 
 given the F8.5 focal ratio of the 2.1m telescope.}.
For each galaxy, we acquired three ON-band exposures with an integration time ranging 
 from 5 to 10 min, according to the seeing conditions and  source brightness.
 The stellar continuum subtraction was secured by means of shorter (typically three times 1 min) 
 exposures taken through a broad-band ($\lambda$ 6231 \AA, $\Delta\lambda\sim 1200$ \AA) $r$-Gunn filter (OFF-band frames).\\
While the median seeing of the SPM site is $\sim 0''.6$, 
 the final FWHM for point sources in the images is affected by the poor telescope guiding 
 and dome seeing. The final distribution ranges from $\sim 1''$ to $\sim 2''$, 
 with a mean seeing of $1''.20 \pm 0''.02$ in 2015 and  $1''.50 \pm 0''.02$  in 2016, as shown  in Fig. \ref{seeing} 
\footnote{The sky-subtracted H$\alpha$ NET and OFF-band (normalized $r$) images are made available via the http://goldmine.mib.infn.it/ site. 
The ON-band image can be obtained by adding the NET to the OFF image. The adopted photometric zero-point of the H$\alpha$  images
and the seeing are stored in the headers.}.

\section{Data reduction}

 We reduced the CCD frames following the procedure  described in Gavazzi et al. (2012), 
 based on the STSDAS and GALPHOT IRAF packages. We refer to that paper for details and give here only a brief 
 summary of the data reduction procedures.
 The methods for extracting the photometry of the H$\alpha$+[NII] line (flux and equivalent width)
 and for estimating its error budget can also be found in Section 4.2 of Gavazzi et al. (2012). 
 
 In short, each image was bias-subtracted and flat-field corrected using sky exposures obtained during twilight 
 in sky regions devoid of stars. 
 When three exposures on the same object were available, we adopted a median combination of the realigned images 
 to help reject the cosmic-ray hits in the final stack. Otherwise, we removed cosmic rays by direct inspection of the frames.
 We subtracted a mean sky background, 
 computed around the galaxy, using the GALPHOT tasks MARKSKY and SKYFIT.
 The flat-fielded ON-frames were aligned with the OFF-frames using field stars. At this stage, the seeing was determined
 independently on the two sets of images. After normalization of the OFF-band frames (see Section \ref{Calibrations}),
 NET images were produced by subtracting the OFF from the ON-frames.

 Given the high Galactic latitude of the observed sample, no flux correction for Galactic extinction was applied. 
 We did not attempt deblending of H$\alpha$ from [NII] lines
either, and no correction for internal extinction was applied.
 The corrections computed with the scaling relations given in Section 4.3 of Gavazzi et al. (2012) refer to spiral galaxies,
 therefore they do not necessarily apply to our sample of ETGs.
 When we refer to  H$\alpha$ measurements in the following, we mean H$\alpha$+[NII], uncorrected for internal dust extinction.
 
 \subsection{Calibrations}
\label{Calibrations}

\begin{figure*}
\centering
\includegraphics[angle=0, scale=0.42]{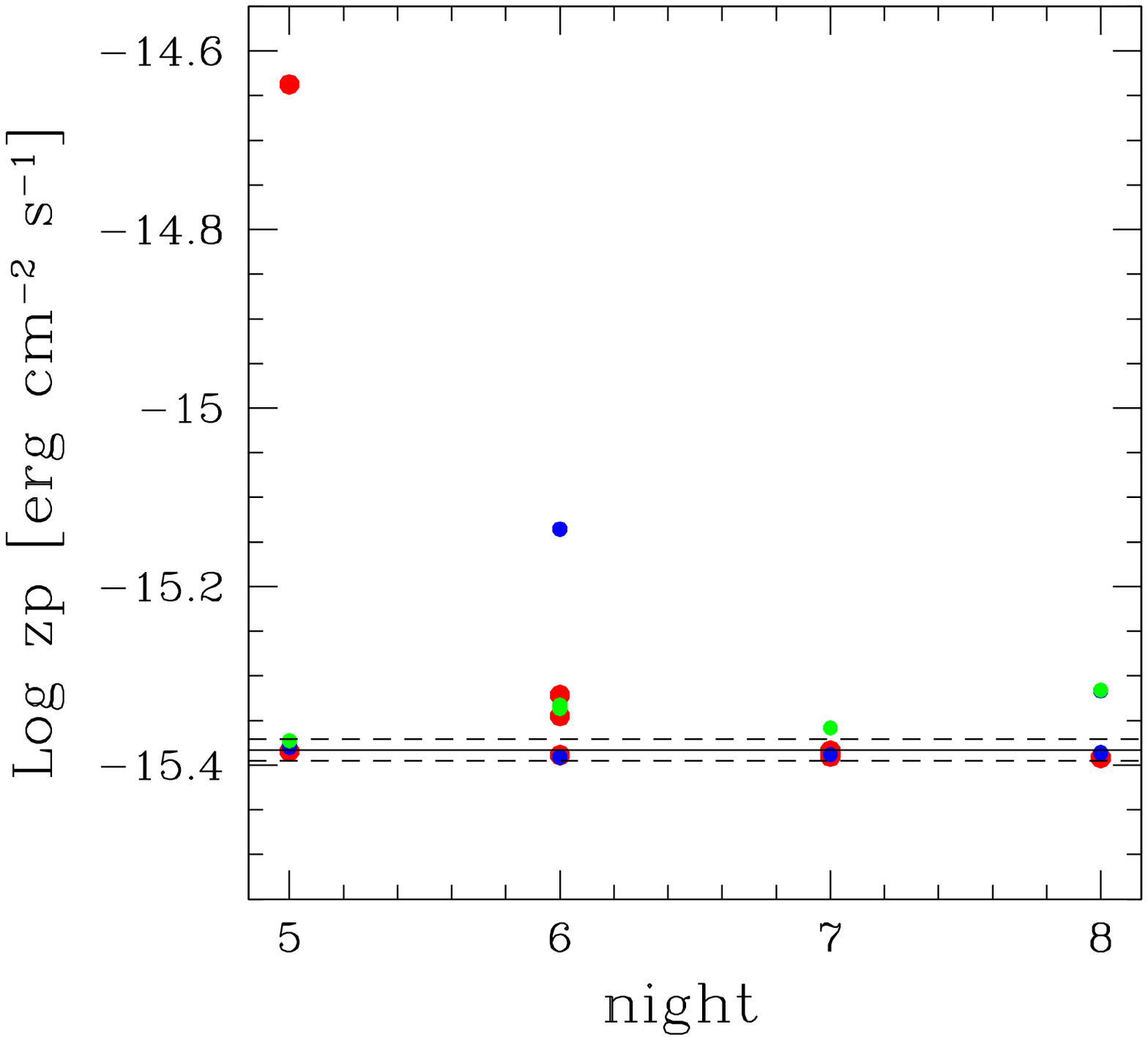}\includegraphics[angle=0, scale=0.42]{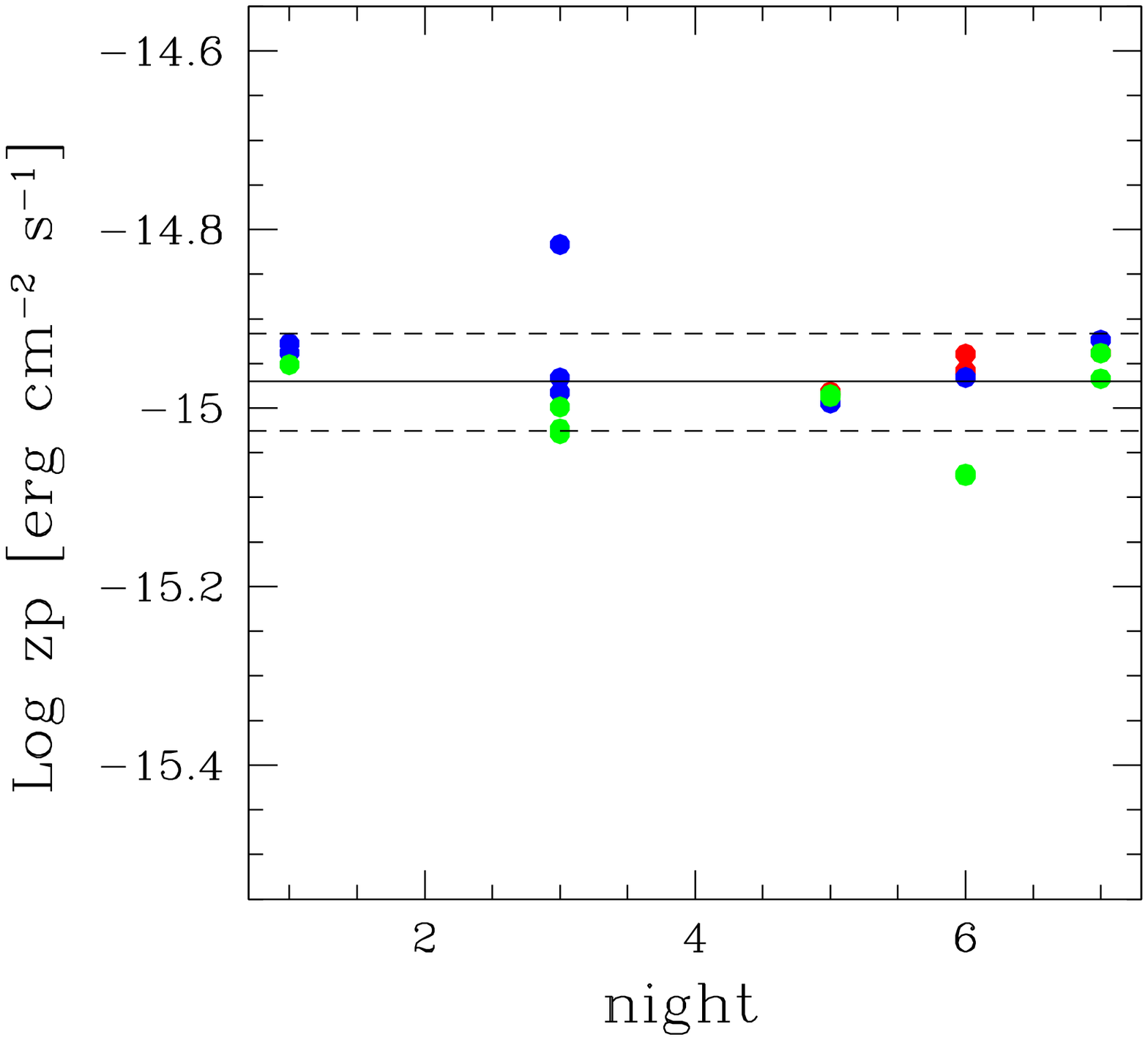}
\caption{Log of the photometric ZP (in $\rm erg ~cm^{-2} sec^{-1}$) in 2015 (left) and 2016 (right), separately for the stars BD33 (blue), 
HZ44 (green), and FG34 (red). The dashed lines show the  $1\sigma$ error bars computed using only the photometric periods in 2015 (night 7 and part of night 8)
and all measurements taken in 2016.}
\label{ZP}  
\end{figure*}

We calibrated the absolute flux scale using the standard stars Feige34, HZ44,  and BD33 from the catalog of 
Massey et al. (1988), observed $\text{approximately
every two}$  hours.
As shown in Figure \ref{ZP}, where the log of Zero Point (ZP) (in $\rm erg ~cm^{-2} sec^{-1}$) is displayed,  
only night 7 and the first part of night 8 in 2015 can be considered photometric. For this reason, except for these two periods,
most targets observed in 2015 were reobserved in 2016.
Conversely, the 2016 run was clear in all five observable nights, with an acceptable uncertainty of 5 \% on the zero-point.

\subsection{Second-order calibration}
\label{seccal}
\begin{figure*}
\centering
\includegraphics[angle=0, scale=0.39]{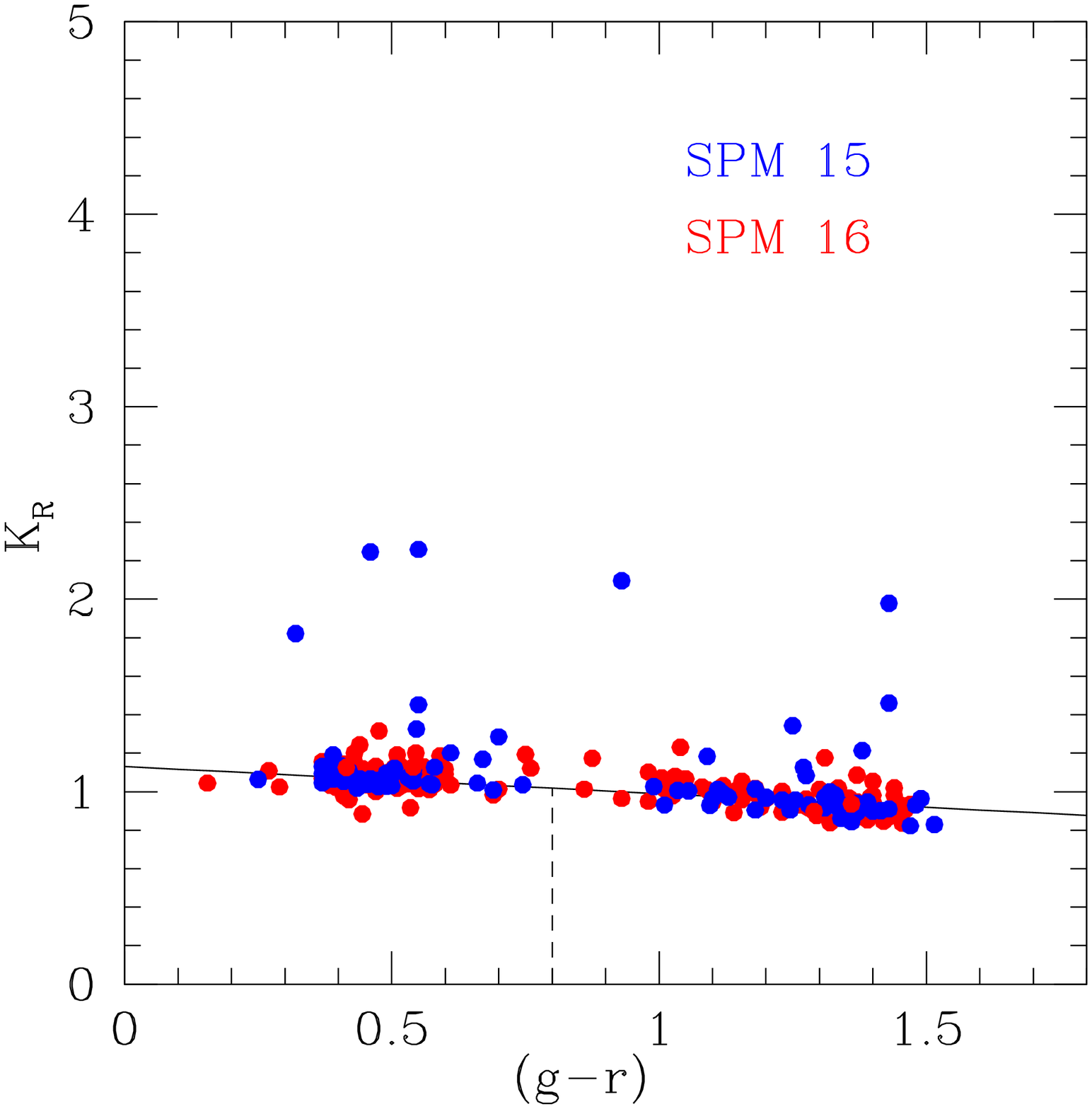}\includegraphics[angle=0, scale=0.39]{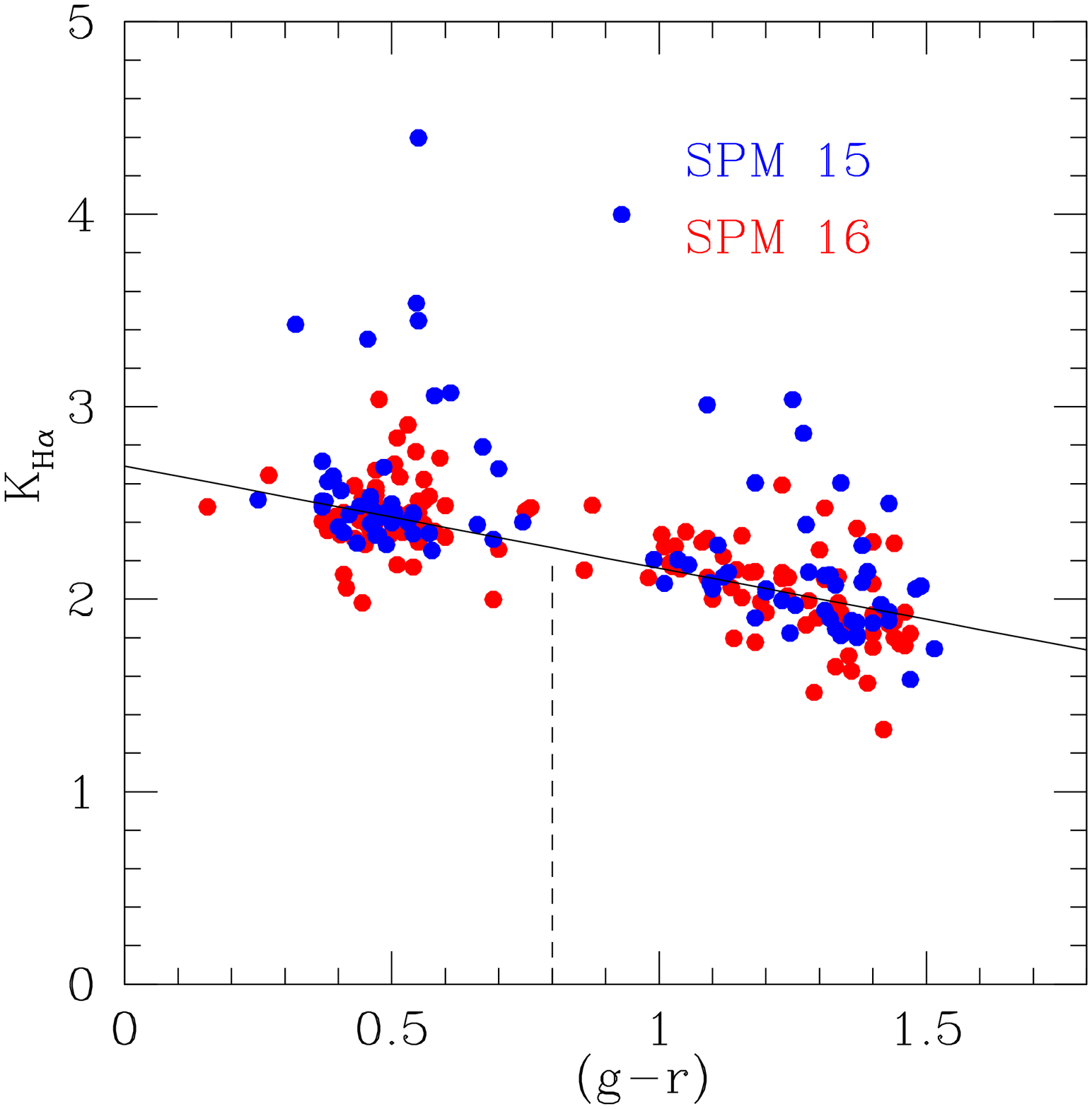}
\caption{(Left panel) Flux ratio $K_R$ in $r$ band between SDSS and this work plotted as a function of $g-r$ color of stars in the field.
The 2016 data are plotted in red and the 2015 data in blue, including some non-photometric measurements. These can
be corrected for using the 2016 fit by adjusting their $K_R$ coefficient to the value derived from the 2016 fit computed at $g-r$=0.8 
using the vertical dashed line drawn at $g-r$=0.8, which  intersects the best-fit relation of the 2016 data.
(Right panel): same for the $K_{H\alpha}$ coefficient.}
\label{Krstars}  
\end{figure*}

Owing to the photometric instability in 2015 and because of the full-sky coverage of the SDSS survey (York et al. 2000), 
we checked and corrected the calibration of our $r$ -band images by comparing the photometry of stars in each
field with their SDSS magnitudes. Using the SDSS navigator tool, we inspected each target field and identified at least ten stars, five
bluer and five redder than approximately $g-r$=0.8 mag. For these stars we compared the SDSS $r$ mag with the one measured in our $r$ frames
assuming the ZP derived for that night from the calibration stars. For each field we separately computed the median $g-r$ 
of the five blue and five red stars so that for each field, only two median measurements were considered.
The median difference between the two sets of magnitudes ($K_R$) is plotted in Figure
\ref{Krstars} as a function of the star color ($g-r$). While most measurements taken in 2016 appear to be accurate, we confirm the presence of few non-photometric
measurements taken in 2015. We fit the 2016 data and used the ratio of the individual data taken in 2015 to the 2016 fit to determine the correct $K_R$ coefficient.
A similar method was used to check and correct the calibration of the ON-band data taken through the narrow-band filters.

In ideal conditions, the flux ratio of field stars should reflect the 
ratio in filter width combined with the ratio of integration time. In our case, the width ratio between our $r$ -band filter and the narrow-band filters is
approximately 11.5 (we compensated for this large difference by adopting an approximately five times longer integration time for the ON-band observation).
As remarked by Spector et al. (2012), however, the normalization factor ($K_{H\alpha}$) depends on the color of the stars, and the
 $K_{H\alpha}$ coefficient to be adopted should correspond to the actual color of the target galaxy (in our case, for ETGs $g-r\sim 0.8$),
 as illustrated in Figure \ref{Krstars}. By applying this method we were able to improve the calibration for both
 the $r$ band and the H$\alpha$ filters for the 118 observed galaxies.

\begin{figure}
\centering
\includegraphics[angle=0, scale=0.42]{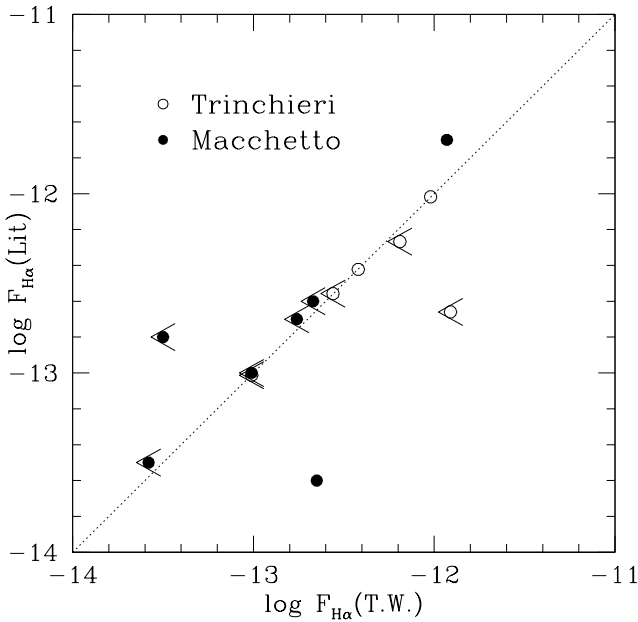}
\caption{Comparison of the H$\alpha$+[NII] flux measured in this work with the flux measured by Macchetto et al. (1996) 
and by Trinchieri \& di Serego Alighieri (1991) for 13 galaxies in common. Only 4 objects were detected by us, the others are upper limits.}
\label{compha}  
\end{figure}

A  check of the quality of our flux calibration is performed in Figure \ref{compha}, where we compare the H$\alpha$+[NII] flux measured in this work
with the value reported in the literature by Trinchieri \& di Serego Alighieri (1991) and  by Macchetto et al. (1996) for 13 galaxies in common (mostly upper limits).
Nine of them are in good agreement, while two show discrepancies by one order of magnitude.

 As a final test, we plot in Figure \ref{flux} a comparison between the flux (left panel)
 and the equivalent width (E.W.) (right panel) derived in our imaging data (separately for strong (blue) and weak (green) detections (see Section \ref{Analysis}), 
 including the 20 filler targets) by integrating the signal in a circular region of 3 arcsec diameter,
 with the values measured in the available nuclear spectra (H$\alpha$+[NII]), showing a satisfactory agreement. 
 The nuclear spectra were taken from SDSS (DR13) when available, or from the NED (for the most part taken from Ho et al. 1995). The remaining 36 targets 
 were observed by us using the Loiano 1.5m telescope, as described next. These are plotted only in the right panel of Figure \ref{flux}, as they were not flux calibrated.
 
Because of the intrinsic shape of ETGs, namely their featureless  H$\alpha$ emission, it is often difficult to assess the robustness 
of the H$\alpha$ emission associated with them
based on the inspection of the NET images. NET frames result from the subtraction of ON-OFF images, both containing bright  
cuspy structures. Small variations in the seeing conditions
combined with slightly (a few percent) imprecise determinations of the normalization coefficient can affect the resulting NET image.
Buson et al. (1993) and Macchetto et al. (1996) adopted a strategy for adjusting the normalization coefficient that was based 
on the absence of negative NET residuals in the external parts of the galaxies themselves. Buson et al. (1993) in particular 
observed a sample of ETGs previously known for
having some H$\alpha$ emission. The authors therefore adjusted the continuum subtraction up to the point that some H$\alpha$ 
residual remained in the NET image, without producing negative residuals in the outer parts.
More quantitative, but based on a similar strategy, was the criterion adopted by Michielsen et al. (2004). 
These authors measured the ON-band and OFF-band flux in a elliptical corona fit to the galaxy
periphery (between two fixed surface brightness levels) and set the normalization coefficient so that the two values were identical.
These two criteria assume that no H$\alpha$ emission is present in the external regions of ETGs. 
We preferred not to adopt such an priori criterion, but to use many field stars to set the normalization coefficient, as described above.

\section{Spectroscopic observations}

Spectroscopic observations of 36 targets were found neither in the SDSS spectroscopic catalog
nor in FITS form within the NED database. These nuclear spectra were obtained by us during several observing runs between 2013 and 2017
using the Bologna Faint Object Spectrograph and Camera (BFOSC, Gualandi \& Merighi 2001) mounted on the 152 cm F/8 Cassini Telescope located
in Loiano, belonging to the Observatory of Bologna. Similarly to previous observations at Loiano (Gavazzi et al. 2011, 2013),
we acquired long-slit spectra taken through a slit of 2 arcsec width and 12.6 arcmin length, combined with an intermediate-resolution 
red-channel grism ($R$ $\sim$ 2200)
covering the 6100 - 8200 \AA ~portion of the spectrum, which includes the $\rm H\alpha$, [NII], and [SII] lines.
BFOSC is equipped with an EEV LN/1300-EB/1 CCD detector of 1300x1340 pixels, reaching 90\% QE near 5500 \AA. 
For the spatial scale of 0.58 arcsec/pixel and a dispersion of 8.8 nm/mm, the resulting spectra have a resolution of 1.6 ~\AA/pix.

\begin{figure*}
\centering
\includegraphics[angle=0, scale=0.39]{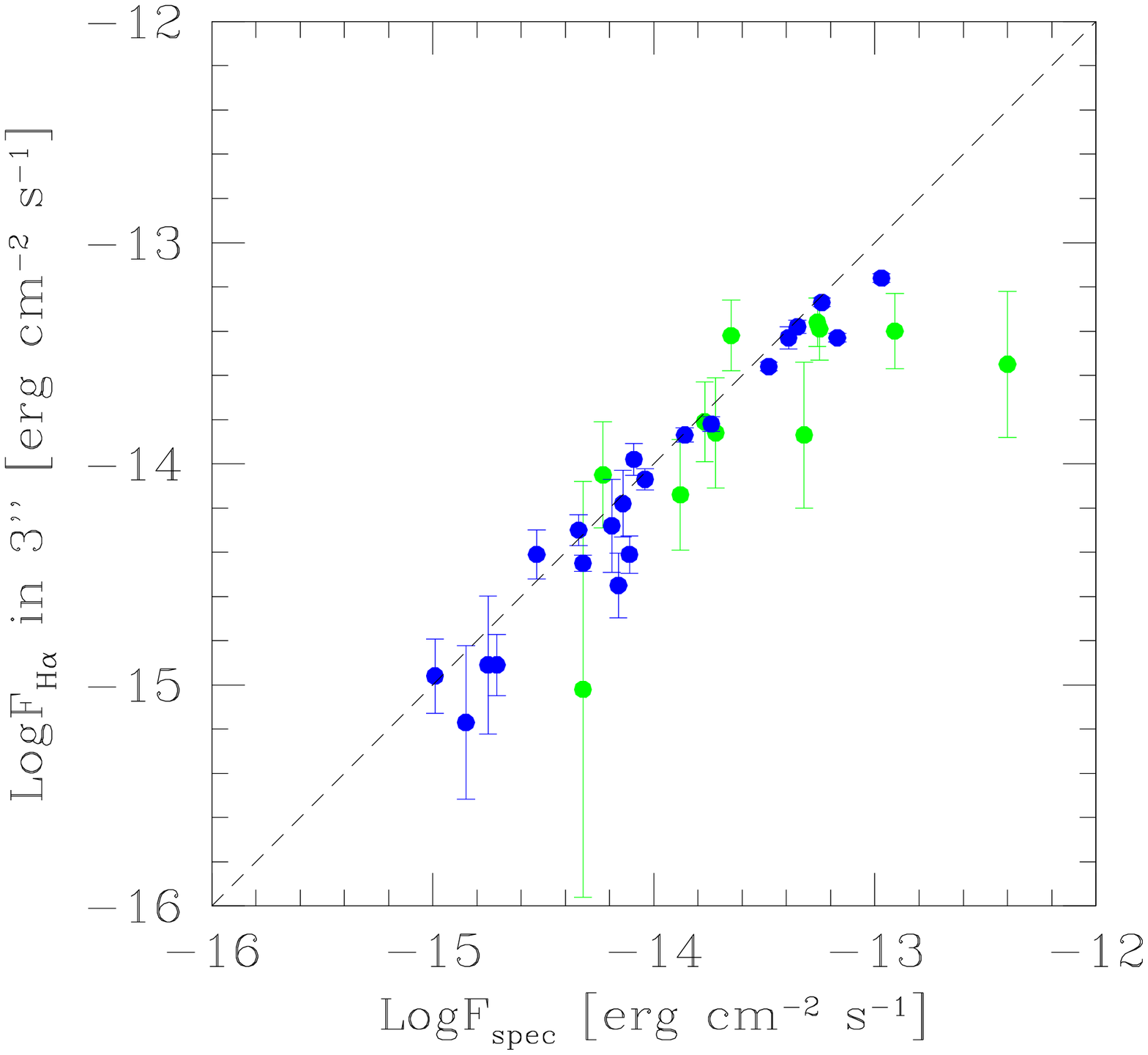}\includegraphics[angle=0, scale=0.39]{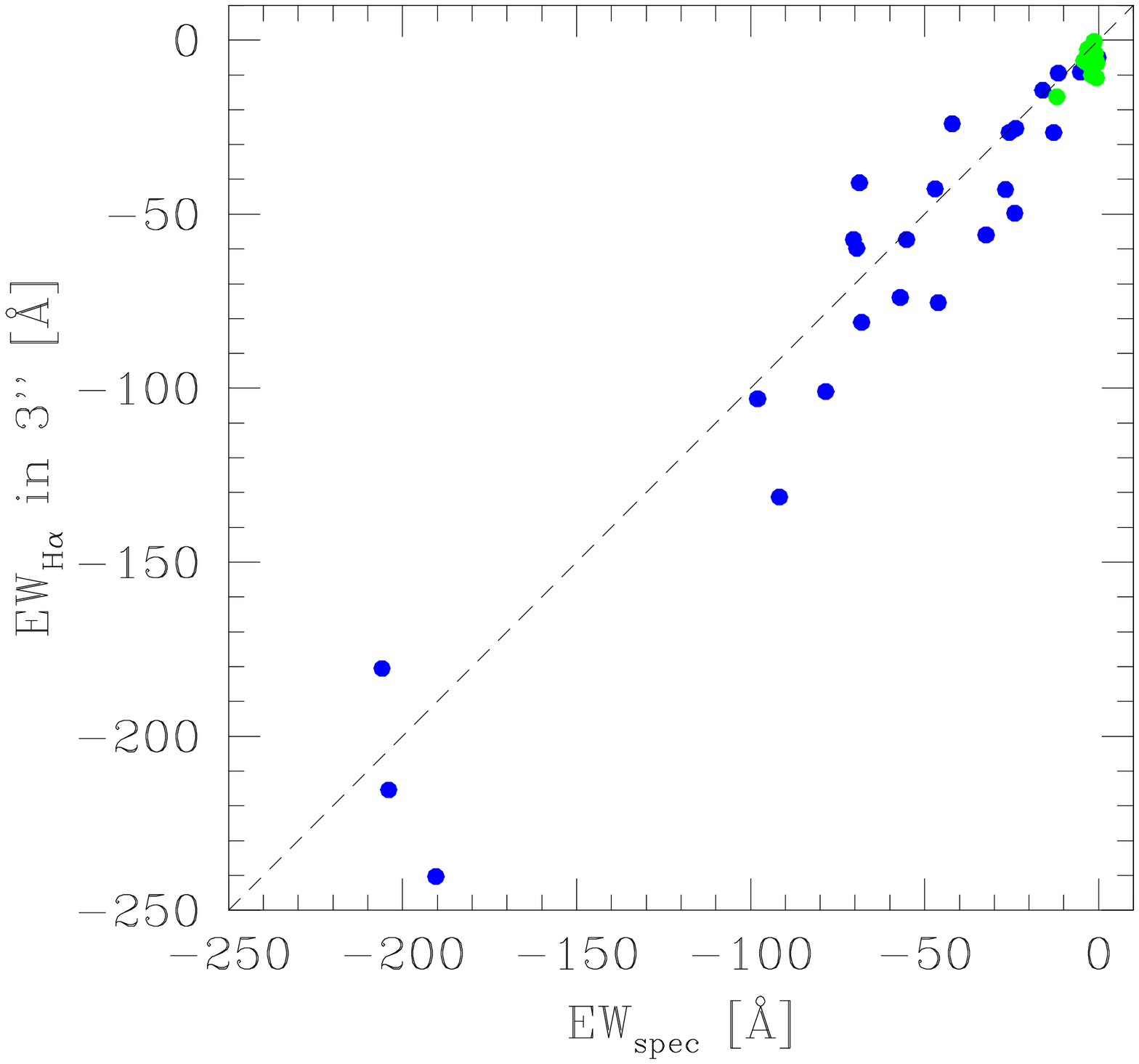}
\caption{Comparison between the H$\alpha$+[NII] flux (left) and E.W. (right) detected in the central 3 arcsec in the 
imaging data and in the nuclear spectra. The right panel includes the 36 spectra taken at Loiano. 
Blue are strong detections, and green are weak detections  (the most discrepant object is N4526).
The sources of observations of nuclear spectra are provided in Table  \ref{photomtab}.
The 45-degree lines serve to guide the eye to the proportionality relations.}
\label{flux}  
\end{figure*}

Exposures of 5-10 minutes were repeated typically three times (to remove cosmic-ray hits). 
The slit was generally set in the E-W direction, except when taken 
along the direction connecting two nearby objects that simultaneously fell in the slit. The wavelength calibration was secured 
by means of frequent exposures of a He-Ar hollow-cathode lamp and further refined using bright OH sky lines.
The spectrograph response was obtained by daily exposures of the star Feige34. 
The typical seeing conditions at Loiano ranged from 1.5" to 2.5". The spectra taken at Loiano were not flux calibrated, and
only measurements of the line EW were derived (right panel of Figure \ref{flux}). \\

The spectra were reduced using standard IRAF procedures. After normalization to the flux in the interval 
6400-6500 $\AA,$ they were shifted to $\lambda_0$ according to their redshift.
Plots of the nuclear spectra obtained at Loiano, covering  approximately from 6200 to 7200 \AA, are given in Figure 11 in the Appendix. 

\section{Results}
\label{results}

The detection threshold that we adopt in this paper comes from a combination of a flux and morphology criterion in the NET images:
we considered detections all sources that have the global H$\alpha$+[NII] EW $<=$ -1 $\AA$ (negative means emission), 
combined with H$\alpha$+[NII] $EW_{3"}$ $<=$ -1 $\AA$ , as determined in the central
3 arcsec aperture. To these sources, we added those (only three objects: 
M86 filamentary, NGC3156, and NGC4435 disky) that, while not meeting the H$\alpha$ 
EW thresholds,  showed clear disc-like or filamentary structures in their NET images.\\
Adopting these criteria, we detect 55 out of 147 galaxies. With a global detection rate of 39\%,
we find 21 strong detections with H$\alpha$+[NII] $EW$ $\leq$ -5 $\AA$, 34 weak detections with $-1 \leq EW<$-5 $\AA$,
and 92 undetected targets with H$\alpha$+[NII] EW$>-1$.

H$\alpha$ continuum subtraction for weak H$\alpha$  emitters is subject to large errors. Colors may vary within galaxies (Spector et al.  2011).
Continuum subtraction errors may therefore be relevant, especially for galaxies that are dominated by central emission (e.g., AGNs).
We note that out of 34 weak candidate detections, a few (NGC2778, NGC4377, and NGC4551) could be spurious, 
as these galaxies show marginal global H$\alpha$ EW, have featureless H$\alpha$ morphology, and do not contain gas.
Conversely, some undetected galaxies (NGC3379, NGC4281, and NGC5813) 
could be missed detections (on the basis of their gas content and/or disk morphology in the HST images). 
Even among the 21 strong detections, two (NGC4429 and NGC4550)  might be partly contaminated by imperfect continuum subtraction (see Figure 12).
Nevertheless, three missing and five spurious detections would not change the conclusions of the present investigation.
We also note that half (50\%) of the weak sources are AGNs (or LIN). This means that their nuclear spectrum is
dominated by [NII] rather than by H$\alpha$, and they therefore
do not contribute to their nuclear star formation (SFR) (e.g., Theios et al. 2016). For these objects, the SFRs given 
in Table \ref{photomtab} must be considered as upper limits.

The NET and OFF images of the 14 galaxies with strong H$\alpha$ emission observed in this work are given in grayscale in Figure 12 in the Appendix.
We note that the ON- and normalized OFF-images were not convolved to the same resolution before subtraction to produce NET images free from seeing effects.
This is made possible by our strategy of observing the ON and OFF frames not only in the same night, but also within minutes from one another.

\begin{figure*}
\centering
\includegraphics[angle=0, scale=0.3]{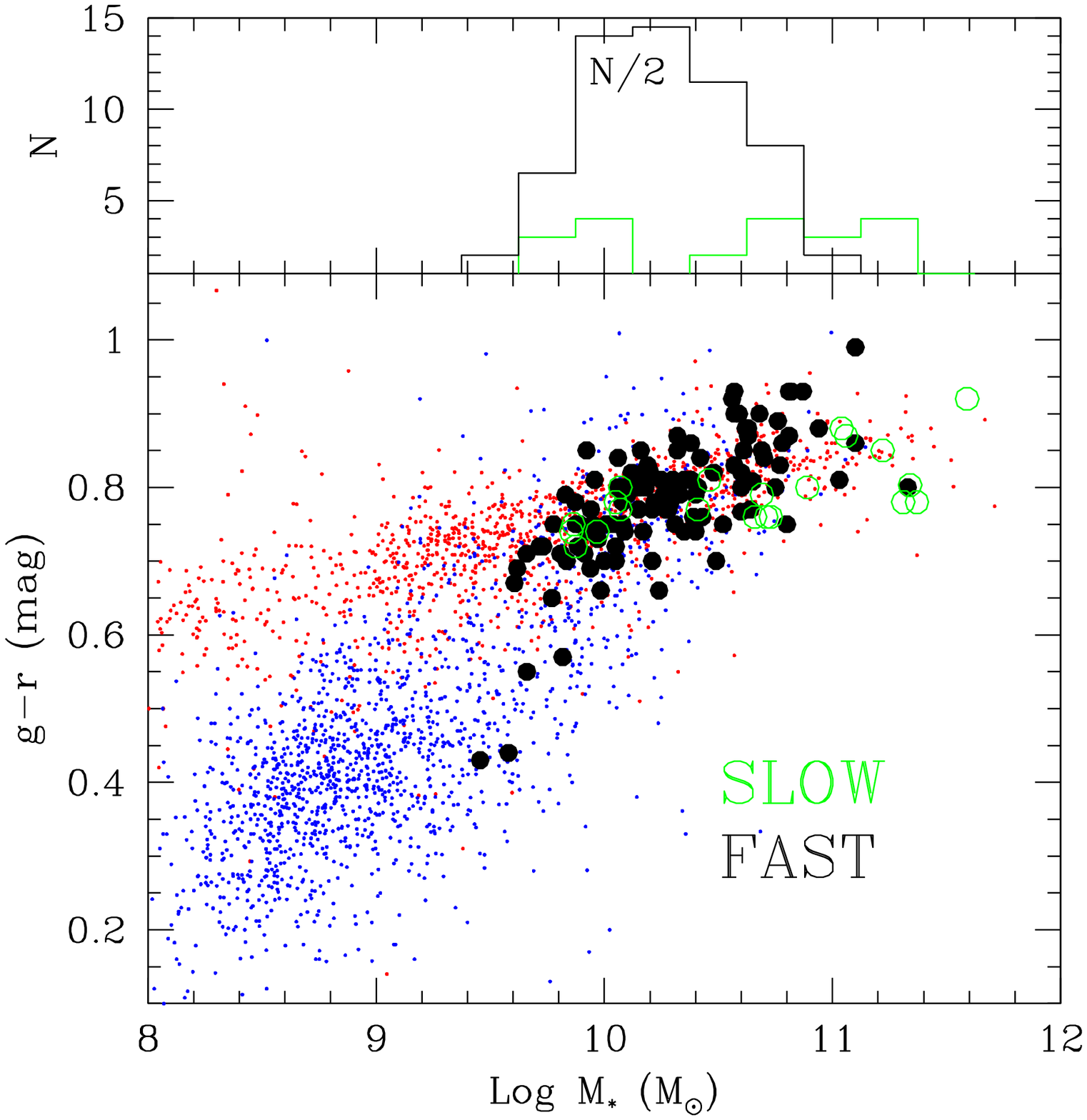} \includegraphics[angle=0, scale=0.3]{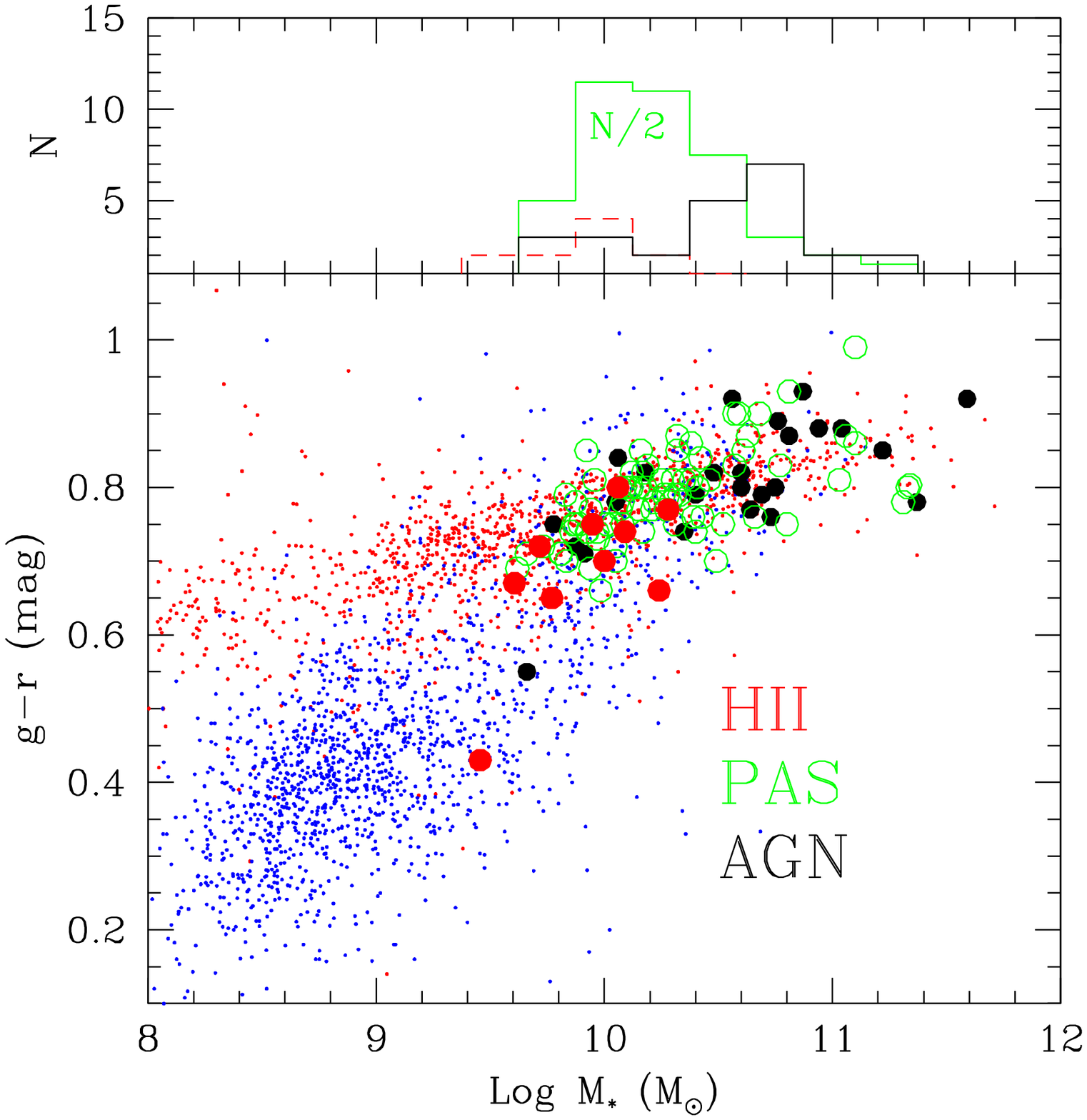} \includegraphics[angle=0, scale=0.3]{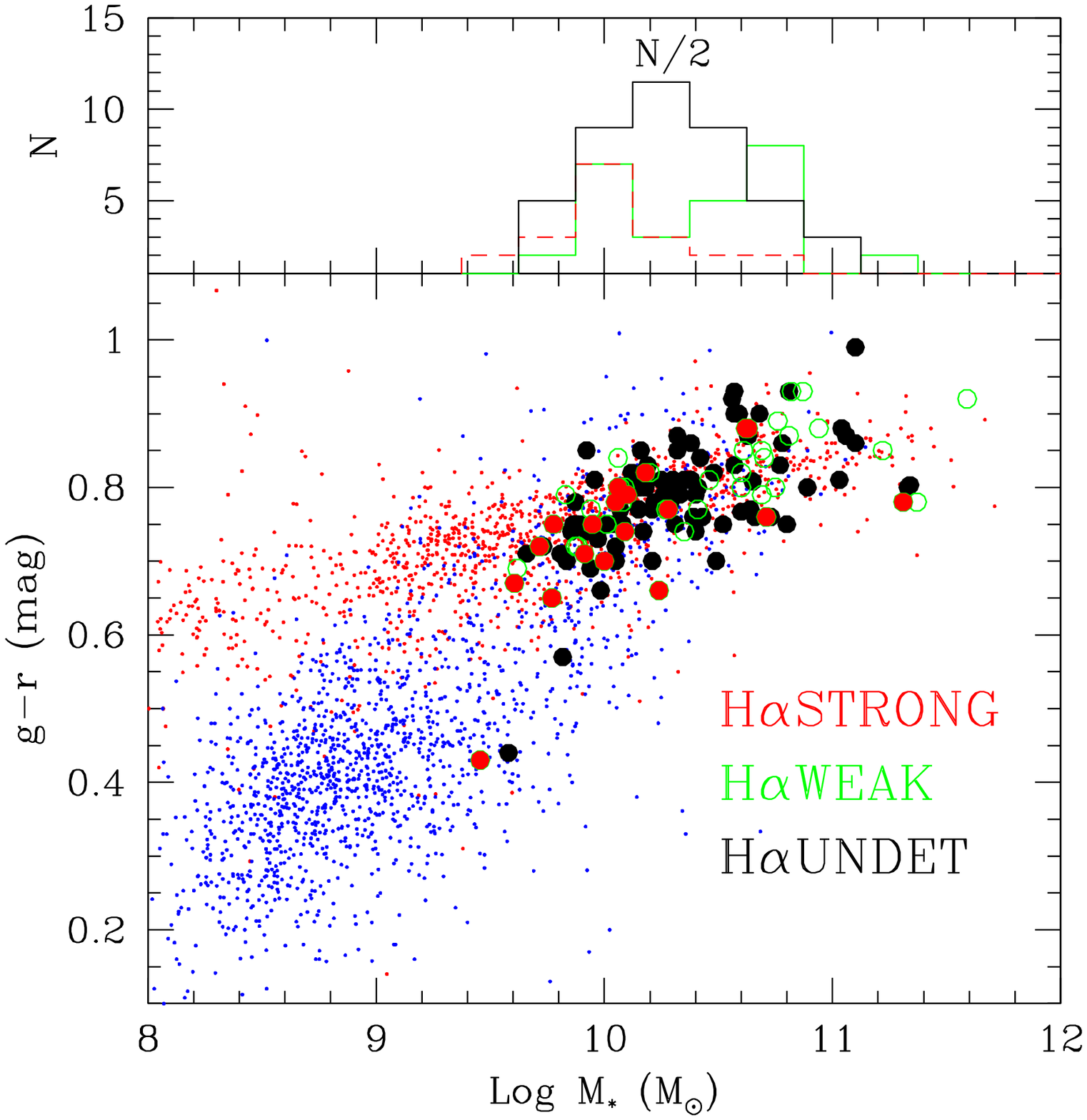} 
\caption{Color-stellar mass relation for  a complete sample of
galaxies in the Coma and Local supercluster from Gavazzi et al. (2010) (small symbols), subdivided into ETGs (red) and LTGs (blue)  to show the separation 
between the red and the blue sequence. 
(Left) The 32 slow-rotator (green open symbols) and the 222 fast-rotator (black filled symbols) ETGs from the whole $\rm ATLAS^{3D}$ survey
(large symbols).
(Center) The 41 AGNs (green open symbols) and the 165 passive (black filled symbols) ETGs from  the whole $\rm ATLAS^{3D}$ survey (large symbols).
(Right) The 22 strong (red), 33 weak (green open symbols), and the 92 undetected (black filled symbols) 
ETGs from the $\rm ATLAS^{3D}$ survey (large symbols).}
\label{colmag}  
\end{figure*}

\begin{table}
\begin{tabular}{l c c c c c c c c}
\hline
            &  S0   & E   &  H$\alpha$d  & HSTd & GAS & AGN     & PAS  &  HII      \\   
            &  \%   & \%  &  \%          &  \%  & \%  & \%      & \%   &  \%       \\    
\hline
\hline
Strong      &  76   &  24 &  47          &  57  &  43  &  24    &  24     &   52   \\
Weak        &  63   &  36 &  27          &  53  &   9  &  50    &  50     &   0    \\
Undetected  &  67   &  32 &  -           &  4   &   0  &   9    &  86     &   0    \\
\hline
\end{tabular}
\caption{Separately for the 21 strong and 34 weak H$\alpha$ detections and for the 92 undetected targets, we give the fraction of S0 and E optical morphology, the percentage of objects with 
an evident disc in their H$\alpha$ morphology, disc or dust in the HST images, the percentage
of HI-H2 detections, the fraction of AGN, and passive or HII region-like nuclei.  }
\label{stat}
\end{table}

Table 1  reports separately for the strong and weak H$\alpha$ detections and for the undetected targets the fraction of objects for each optical morphological class,
the fraction of conspicuous H$\alpha$ 
discs and/or dusty structures in the HST images, the fraction of gas-rich (HI or CO detected) objects and the fraction of AGN, passive and HII region-like nuclei from spectroscopy.
Strong detections appear associated mainly with S0 gas-rich galaxies, with H$\alpha$ and HST discs, and for the most part, they harbor HII region-like nuclei.
AGNs are overabundant among the weak sources. This implies that the emission in this class of sources is not due to star formation associated with the  H$\alpha$ line,
but to relatively strong [NII]. Moreover, undetected targets are gas-free, disk-free systems without star formation, neither extended nor nuclear, without
abundant AGNs, but with passive nuclear spectra.

The strong  H$\alpha$ emitters (H$\alpha EW<-5~\AA$) are significantly (11/21: 52\%) associated with HII region-like spectra. Conversely, 
19 of 34 are weak H$\alpha$ emitters ($-1 \leq EW<$-5 $\AA$): 55\% are AGN, and 15 of 34: 44\% are 
passive (PAS) or retired (RET), none HII.

For the environmental dependence of the ETG properties, we recall that Cappellari (2016) reported that slow rotators constitute a significant 
fraction only in the centers of cluster and groups. 
Our subsample of  $\rm ATLAS^{3D}$ comprises the entire Virgo cluster, therefore we can check whether the fraction of detected and undetected galaxies also correlates with the projected distance from M87. 
We find that 38\% of the strong detected galaxies and 58\% of the undetected galaxies are found inside the Virgo cluster, hinting at a marginal environmental 
anticorrelation between the ionized gas content (and star formation) and the projected galaxy density.

To show that
$\rm ATLAS^{3D}$ targets are genuine red sequence systems, with only little contamination from blue cloud and green valley objects, we plot in
Figure \ref{colmag} the color ($g-r$) versus stellar mass ($M_\odot$) relation for the full $\rm ATLAS^{3D}$ sample (large symbols),
while in the same figure we plot (small symbols)
the color-luminosity relation from a complete sample of SDSS galaxies in the Coma supercluster (selected and morphologically classified by Gavazzi et al. 2010), 
separately for LTGs (blue) and ETGs (red). 

In Figure \ref{colmag} (left panel) we first confirm a result that is known from Emsellem et al. (2011), who showed that slow rotators tend to be massive 
($M_{dyn}>10^{10.5} M_\odot$) 
and dominate the high-mass end of ETGs. This result has been emphasized by Veale et al. (2017), who showed that the fraction of slow rotators, 
that is, a mere 14 \% in the $\rm ATLAS^{3D}$ survey,  reaches 90\% for the massive ($M_{dyn} >10^{10.5}~M_\odot$) ETGs ($M_K<-26$ mag).
For example, the brightest $\rm ATLAS^{3D}$ members of the Virgo cluster that are included in our work
(N4472=M49, N4486=M87, N4374=M84, and N4406=M86) are all slow rotators (see also Boselli et al. 2014).
In agreement with Emsellem et al. (2011), we interpret this result as an indication that massive slow rotators represent the extreme 
instances within the red sequence of galaxies that might have suffered from significant merging 
without being able to rebuild a fast-rotating component. 

Figure \ref{colmag} (central panel) gives the color-mass relation, dividing galaxies with respect to their nuclear spectral classification
according to Gavazzi et al. (2011, 2013)\footnote{Gavazzi et al. (2011) classified the nuclear spectra by dividing them into HII region-like, 
strong AGNs (including SEY), liners (LIN), "retired" nuclei (RET)
that are likely excited by old stars, and passive nuclei (PAS).}. 
AGNs (SEY+AGN) are plotted separately from passive (PAS+LIN+RET) and from HII-like systems. Again there is 
a significant separation between the average luminosity of AGNs, which are brighter than PAS by 0.6 mag on average. 
This result is expected (see, e.g., Gavazzi et al. 2011), and along with the previous finding, it helps explain why slow rotators in $\rm ATLAS^{3D}$ 
 contain a larger fraction of AGNs (31\%) than fast rotators (13\%).
Conversely, HII region-like nuclei are harbored by the least massive galaxies, but the small statistics prevents us from being more quantitative. 
The frequency of AGNs is 30\% among galaxies more massive than $10^{10.5}~M_\odot$, which is significantly higher than the 10\%
found among galaxies less massive than $10^{10.5}~M_\odot$. These values are consistent with those of Kauffmann et al. (2003b), who found
that galaxies with $10^{10.5}~M_\odot$ harbor between 20\%\ and 40\% of AGNs. They also found that the fraction of AGNs among emission-line galaxies 
increases steeply with mass, reaching 100\% when the stellar mass reaches $10^{11.5}~M_\odot$.


In order to understand the difference between ETGs that are detected in H$\alpha$ from the undetected ones, we plot in Figure
\ref{colmag} (right panel) the color-mass relation for the subsample of 147 $\rm ATLAS^{3D}$  observed in H$\alpha$,
subdivided between detected (strong or weak) and undetected objects. Significant mass segregation is evident. 
Weak detections are spread over the entire mass range, and the most massive objects coincide with slow-rotating AGNs. 
The strong detections are associated with low-mass systems, which from the previous analysis were identified with HII region-like nuclei.
These are associated with S0 galaxies and tend to avoid Es. 

\begin{figure}
\centering
\includegraphics[angle=0, scale=0.39]{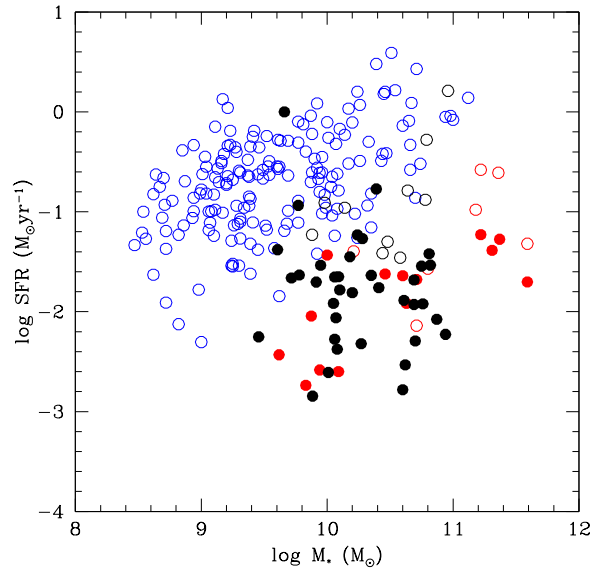}
\caption{Relation between the stellar mass and the SFR separately for a set  of HRS galaxies (empty symbols): (blue: LTG,
black: S0+S0a, and red: E)  and the ETGs from this work (filled symbols): red: E, and black: S0.}
\label{SFR}  
\end{figure}

In Figure \ref{SFR} we compare the SFR derived for a set of galaxies in the HRS catalog of Boselli et al. (2015; mainly composed of LTGs)
with the SFR obtained for ETGs in this work (limited to the 55 galaxies detected in our H$\alpha$ imaging campaign). Its is clear that
LTGs have, for any given stellar mass, an SFR\ higher by about a factor of 20 than ETGs, and that ETGs in our survey agree well with the few ETGs in the HRS survey.\\

\begin{figure}
\centering
\includegraphics[angle=0, scale=0.39]{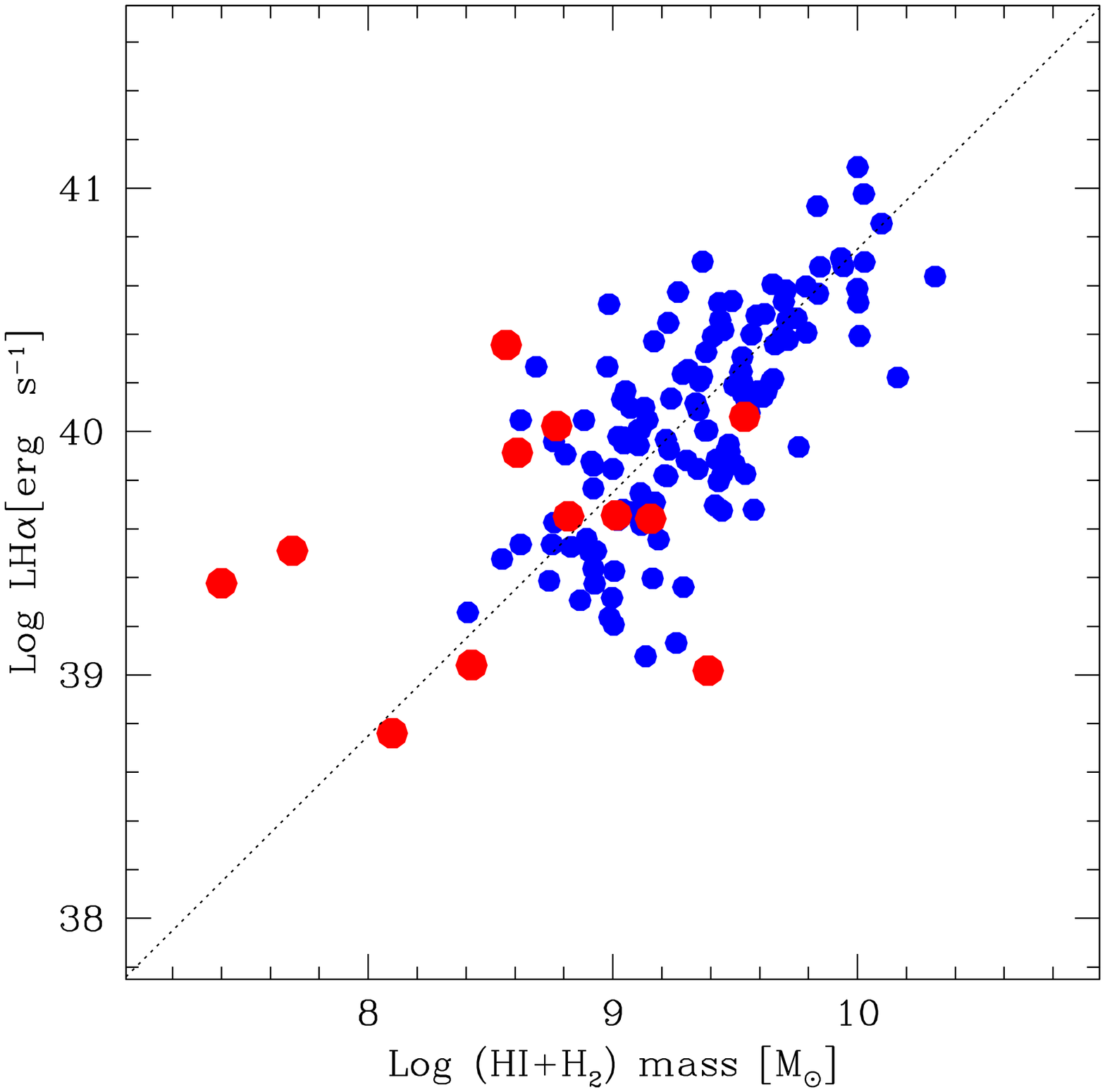}
\caption{Relation between the gas (HI+$\rm H_2$) mass and the H$\alpha$ luminosity separately for a set of 131 LTGs from HRS (blue) and 12 ETGs from this work (red). 
The dotted line is to guide the eye on the direct proportionality relation.}
\label{gasHalum}  
\end{figure}

The relation between the frequency of H$\alpha$ detections and the gaseous content (HI or $\rm H_2$; Figure  \ref{gasHalum})
shows an obvious correlation:
of the 55 H$\alpha$ detected ETGs, 33 (60\%) are found to retain gas. On the other hand, only 7 (8\%) of the 92 undetected ETGs have gas.
However, when considering the 12 detected objects that have positive HI and $\rm H_2$ detections (we note that 11 of them are S0 and only 1 is an E), 
we find a barely significant correlation between the total (HI+$\rm H_2$) 
gas mass and the H$\alpha$ luminosity (see the red symbols in Figure \ref{gasHalum}).
When we overplot the relation between the similar quantities for a sample of 131 LTGs from the $Herschel$ Reference Sample by Boselli et al. (2010)
that have been detected in H$\alpha$ (Boselli et al. 2015) and in HI and CO (Boselli et al. 2014b), we find that the ETG population is consistent with the
low-mass tail of LTGs. We note that of the 12 detected ETGs with gas in  Figure \ref{gasHalum}, 6 have HII region-like nuclear spectra,
indicating that star formation is occurring in their nuclei. The other 6 are AGNs of some kind. Furthermore, 8 are galaxies with a relatively strong 
H$\alpha$ luminosity
and an EW and gaseous content. The 4 objects with the lowest luminosity and gas content are members of the Virgo cluster (within 7 degrees of the projected angular 
separation from M87). \\

\begin{figure}
\centering
\includegraphics[angle=0, scale=0.39]{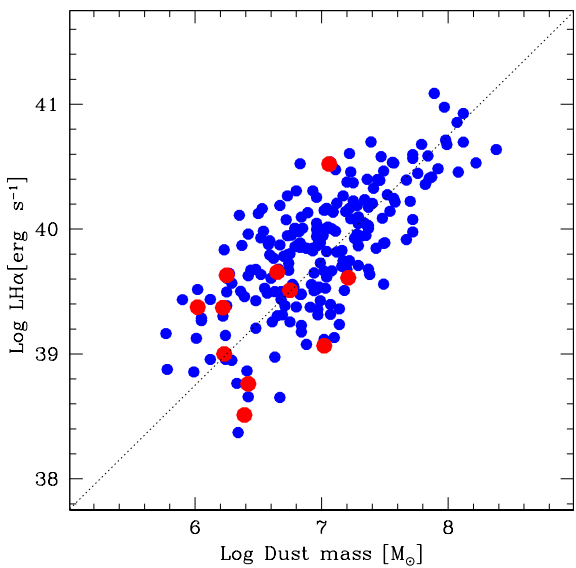}
\caption{Relation between the dust mass and the H$\alpha$ luminosity separately for a set of 214 LTGs from HRS (blue) and 11 ETGs from this work (red). 
The dotted line is to guide the eye on the direct proportionality relation.}
\label{dustHalum}  
\end{figure}

To assess the comparison of the ETGs with the LTGs from HRS in the H$\alpha$ luminosity/dust mass plane, we cross-correlated the HRS catalog 
(Ciesla et al. 2014) with
the  $\rm ATLAS^{3D}$ sample of 147 ETGs  observed by us in H$\alpha$, finding 11 matches. In Figure \ref{dustHalum} we plot (with similar 
symbols as in Figure \ref{gasHalum}) the 
relation between H$\alpha$ luminosity and dust mass for 214 LTGs detected at 22 micron by WISE and at 250 micron by SPIRE on board $Herschel$, 
providing an estimate of the dust mass. 
It is again evident that the 11 ETGs do not deviate significantly from the sequence occupied by LTGs.

The majority of $\rm ATLAS^{3D}$ targets have been observed with HST, providing high-resolution optical images.
Many of them appear to harbor a distinct disc structure at their interior. 
Interestingly, 30 of 55 (55\%) of the detected ETGs show evidence of structures (disc or dust filaments) in the HST images. Conversely,
only 4 (4\%) of the  92 undetected ETGs have similar structures. Eight of 12 gas-rich ETGs show dusty discs in the HST images.
The galaxy morphological mix in the observed sample is 68\% S0s versus 32\% Es. In the detected galaxies this mix does not change
at all: 68\%  S0s  and 32\% Es.

When we divide the galaxies into H$\alpha$ detected and undetected, these ratios change as follows:
of the detected galaxies, 32 of 55 (58\%) are AGN of various types, 11 of 55 (20\%) are HII regions, and 12 of 55 (22\%) are PAS.
Of the undetected galaxies,  13 of 92 (14\%) are AGN of various types,  none of the 92 (0\%) are HII regions, and 74 of 92 (80\%) are PAS.
To summarize, the relative majority of detected galaxies are AGNs, while the relative majority of the undetected galaxies are PAS.

\section{Conclusions and summary}

We have analyzed H$\alpha$ imaging observations of 147 early-type galaxies selected from the
(260) targets in the  $\rm ATLAS^{3D}$ catalog, assumed to be representative of the ETG population in the local Universe.
For the totality of our sample, we also gathered nuclear spectroscopy, either from the literature, or by 
new observations (36 objects), allowing their nuclear classification.  
Fifty-five ETGs (37\%) were detected as H$\alpha$  emitters above our threshold limit  (H$\alpha$ $EW$ $\leq$ -1 $\AA$), and 21 were above H$\alpha$ $EW$ $\leq$ -5 $\AA$.
Seventy-six percent of the strong emitters were found to be associated with low-mass ($M_*\sim 10^{10} M_\odot$) S0 galaxies, showing a conspicuous gas (HI+$H_2$) content, 
extended stellar disks,
and star formation even in their nuclei. All but two of them are fast rotators. The remaining 33 weak detections were 
found to be associated with more massive ($M_*\sim 10^{11} M_\odot$) gas-poor targets
that often harbor an AGN in their nucleus. Two-thirds of them are fast rotators and 64\% are associated with S0 galaxies.
The majority of the remaining (92) undetected systems are gas
poor and diskless
and show a passive spectrum even in their nucleus. Eighty-eight
percent of them are associated with fast rotators and 66\%  with S0 galaxies.

These pieces of evidence, considered in the light of the cinematic measurements provided by $\rm ATLAS^{3D}$ (Cappellari, 2016), 
reinforce the evolutionary picture where the majority of the
current low-mass ETGs are in fact the outcome of a secular evolution of disky gas-rich systems
governed by rotation and star formation, both on the disc and on the nuclear scale.
In contrast, the most massive ($M_*> 10^{10.5} M_\odot$) ETGs are genuine products of dry merging, which dissipated their angular momentum, 
and they provide the evolutionary track of giant galaxies at the center of rich clusters of galaxies.

\begin{acknowledgements}
This research has made use of the GOLDmine database (Gavazzi et
al. 2003, 2014b) and of the NASA/IPAC Extragalactic Database (NED) which is
operated by the Jet Propulsion Laboratory, California Institute of Technology, under contract with the National Aeronautics and Space Administration.  
Funding for the Sloan Digital Sky Survey (SDSS) and SDSS-II h
as been pro-vided  by  the  Alfred  P.  Sloan  Foundation,  the  Participating  Iinstitutions,  
the National  Science  Foundation,  the  U.S.  Department  of  Energy,  
the  National  Aeronautics and Space Administration, the Japanese Monbukagakusho, 
and the Max Planck Society, and the Higher Education  Funding Councill  
for England. The  SDSS  Web  site  is http://www.sdss.org/.
The  SDSS  is  managed  by  the Astrophysical  Research  Consortium  (ARC)  for  the  Participating  
Institutions.The  Participating  Institutions  are  the  American  Museum  of  Natural  History, 
Astrophysical Institute Potsdam, University of Basel, University of Cambridge, 
Case Western Reserve University, The University of Chicago, 
Drexel University, Fermilab,  the  Institute  for  Advanced  Study,  the  Japan  Participation  Group, 
The  Johns  Hopkins  University,  the  Joint  Institute  for  Nuclear  Astrophysics, 
the  Kavli   Institute   for  Particle   Astrophysics   and   Cosmology,  
the  Korean Scientist  Group,  the  Chinese  Academy  of  Sciences  (LAMOST),
Los  Alamos National  Laboratory,  the  Max-Planck-Institute   for  Astronomy  
(MPIA),  the Max-Planck-Institute  for  Astrophysics  (MPA),  New  Mexico  State  University, 
Ohio  State  University,  University  of  Pittsburgh,  University  of  Portsmouth, 
Princeton  University,  the United States  Naval Observatory,  and the University of Washington.
M. Fossati acknowledges the support of the Deutsche Forschungsgemeinschaft via Project ID 387/1-1.
M. Fumagalli acknowledges support by the Science and Technology Facilities Council [grant number ST/P000541/1].
\end{acknowledgements}

\section{APPENDIX A}
  
\begin{table}
\centering
\begin{tabular}{|l| c c|c c|}
\hline
night & \multicolumn{2}{c|} {2015} & \multicolumn{2}{c|} {2016}\\
         &   obs & obs.cond. & obs. &  obs.cond.\\
\cline{2-5}
1 & N &  - & Y & P \\
2 & N &  - & N & - \\
3 & N &  - & Y & P \\
4 & N &  - & N & - \\
5 & Y & nP & Y & P \\
6 & Y & nP & Y & P \\
7 & Y &  P & Y & P \\
8 & Y &P/nP& N & - \\
\hline \hline 
&\multicolumn{4}{c|} { Log Zero-point} \\
\cline{2-5}
 & \multicolumn{2}{c|}{} & \multicolumn{2}{c|} {}\\
 & \multicolumn{2}{c|}{\large -15.29 $\pm$ 0.03 } & \multicolumn{2}{c|} {\large-14.97 $\pm$ 0.05}\\
 & \multicolumn{2}{c|}{} & \multicolumn{2}{c|} {}\\
\hline
\end{tabular}
\caption{Separately for the 2015 and 2016 runs, we list the usable nights and the photometric quality (P=photometric;  nP= not photometric). 
The average ZPs with uncertainty are given in the last line. The 0.3 dex difference between the two ZPs derives from a different
setup of the CCD readout electronics.}
\label{logbook}
\end{table}

  The logbook of the observations is provided in Table \ref{logbook}. 

  The 147 galaxies selected from $\rm ATLAS^{3D}$ for H$\alpha$ observations (either from this work or from the literature) are listed in Table \ref{generaltab}, 
  organized as follows:\\
  Column (1): galaxy name; \\
  Column (2) and (3): J2000 celestial coordinates;\\
  Column (4): recessional velocity in km/s; \\
  Column (5): assumed distance in Mpc; \\
  Column (6): morphological classification in $\rm ATLAS^{3D}$ (Cappellari et al. 2011); \\
  Column (7): SDSS magnitude  in  $g$ from Consolandi et al. (2016); \\
  Column (8): SDSS magnitude  in  $r$ from Consolandi et al. (2016); \\
  Column (9):  log of stellar mass $M_star$ computed by us from the $r$ -band absolute magnitude and the $g-r$ color index using the prescription of 
  Zibetti et al. (2009). \\
  Column (10): kinematic classification in $\rm ATLAS^{3D}$ (Emsellem et al. 2011) as fast (F) or slow (S) rotators; \\
  Column (11): log of the molecular hydrogen mass  (H$_{2}$) in $\rm M_{\odot}$ with sign, as given by Young et al. (2011); 
  when a CO spectrum is available from the NED,   but an estimate of the H$_{2}$ mass is not given, a "H2" is reported. \\
  Column (12): log of the atomic hydrogen mass (HI) in $\rm M_{\odot}$ with sign, as given by Serra et al. (2012); 
  when a HI spectrum is available from NED (mostly from ALFALFA),   but an estimate of the HI mass is not given, a "HI" is reported.\\

  The observational parameters of the target galaxies are given in Table \ref{obspar} as follows: \\
  Column (1): galaxy name; \\
  Column (2): reference to the H$\alpha$ observation (see also Column 16 in Table \ref{photomtab});\\
  Column (3): air mass during the ON-band exposure in 2015;\\
  Column (4): air mass during the OFF-band exposure in 2015;\\
  Column (5): air mass during the ON-band exposure in 2016;\\
  Column (6): air mass during the OFF-band exposure in 2016;\\
  Column (7): exposure time (in seconds) of the individual ON-band exposure in 2015; three equal exposures were combined.\\
  Column (8): exposure time (in seconds) of the individual OFF-band exposure in 2015; three equal exposures were combined.\\
  Column (9): exposure time (in seconds) of the individual ON-band exposure in 2016; three equal exposures were combined.\\
  Column (10): exposure time (in seconds) of the individual OFF-band exposure in 2016; three equal exposures were combined.\\

  The results of our $\rm H\alpha$ imaging campaign are listed in Table \ref{photomtab}, organized as follows:\\
  Column (1): galaxy name; \\
  Columns (2, 3): (EW) H$\alpha$+[NII] (in \AA) measured in a circular aperture of 3$^{\prime\prime}$ positioned on the galaxy nucleus, with associated uncertainty;\\
  Columns (4, 5): log of the H$\alpha$+[NII] flux (erg cm$^{-2}$ s$^{-1}$) measured in a circular aperture of 3$^{\prime\prime}$ 
  positioned on the galaxy nucleus, with associated uncertainty;\\
  Columns (6, 7): total (EW) H$\alpha$+[NII] (in \AA) measured in a circular aperture containing the whole galaxy, with associated uncertainty;\\
  Columns (8, 9): total log of the H$\alpha$+[NII] flux (erg cm$^{-2}$ s$^{-1}$)  measured in a circular aperture containing the whole galaxy, 
  with associated uncertainty;\\
  Column (10): log of the total star formation rate derived from  the H$\alpha$ flux as prescribed by Kennicutt (1998), but adapted for a Chabrier IMF.

Column (11): availability of a nuclear spectrum: LOI = taken at the Loiano 1.5m telescope (this work); SDSS = taken from the SDSS database; 
  NED =  taken from NED; HO= in Ho et al. 1995;
  NEDMH= from NED taken at Mount Hopkins 1.5m  telescope.\\
  Column (12): nuclear spectral classification; HII, SEY, AGN, PSB, RET, PAS, according to the criteria of Gavazzi et al. (2012; 2013);\\ 
  Column (13): 1=strong H$\alpha$ detection; 2=weak detection, 3=undetected.\\
  Column (14): H$\alpha$ morphological classification; d=disc, D=diffuse, c=centrally peaked, F=filamentary;\\
  Column (15): HST imaging availability: -: not available; 0: available, featureless; 1: available, showing a prominent structure (disc, dust ring, or filaments); \\
  Column (16): reference to H$\alpha$ imaging. 1: this work; 2: Koopmann \& Kenney (2006); 3: Boselli at al. (2015); 4: Young et al. (1996); 5: Macchetto et al. (1996); 6: Koopmann et al.
  (2001); 7:  Boselli \& Gavazzi (2002);  8: Kennicutt \& Kent (1983); 9: Trinchieri \& Di Serego Alighieri (1991); 10: Gavazzi et al. (2000), 11: Kenney et al. (2008).
 
Filler targets are listed in  Table \ref{genfiller}, 
 their observational parameters  are given in Table \ref{obsfiller}, and 
  the photometric parameters are listed in Table \ref{photomfiller}.
  \newpage 
  \begin{onecolumn}
  \begin{landscape}
  \vskip -2cm
    \small
\begin{longtable}{l l c c c c c c c c r r}
    \hline
    \endfoot
    \hline 
       Galaxy  &           RA    &      Dec     &   cz   &  D     &   Type    &    $g$    &      $r$  & $\rm logM_{star}$   & Rot  & log $\rm M(H_2)$ & log $\rm M(HI)$   \\ 
               &         J2000.0 &  J2000.0     & km/s   & Mpc    &          &     mag   &  mag     & $M_{\odot}$         &      &  $M_{\odot}$  & $M_{\odot}$\\ 
    \hline
    \endhead
      UGC03960  &   07h40m22.75s & +23d16m30.0s & 2255  &  33.2  &      E     &  14.06    &   13.31   &     -    &      S/F  & $<$7.81 &      7.79    \\   
      NGC2481   &   07h57m13.75s & +23d46m04.0s & 2157  &  32.0  &      S0    &  13.29    &   12.36   &   10.57  &       F   & $<$7.79 &      $<$7.42         \\   
      NGC2577   &   08h22m43.45s & +22d33m11.1s & 2062  &  30.8  &      S0    &  12.77    &   11.90   &   10.63  &       F   & $<$7.71 &      $<$7.35         \\   
      NGC2592   &   08h27m08.05s & +25d58m13.1s & 1979  &  25.0  &      E     &  12.97    &   12.12   &   10.32  &       F   & $<$7.54 &      $<$7.18         \\   
      NGC2594   &   08h27m17.16s & +25d52m43.7s & 2362  &  35.1  &      S0    &  14.12    &   13.27   &   10.16  &       F   & $<$7.83 &      8.91    \\   
      NGC2679   &   08h51m32.94s & +30d51m55.3s & 2027  &  31.1  &      S0    &  13.55    &   12.80   &   10.08  &       F   & $<$7.87 &      $<$7.35         \\   
      NGC2764   &   09h08m17.47s & +21d26m36.0s & 2706  &  39.6  &      S0    &  13.20    &   12.54   &   10.24  &       F   &    9.19 &      9.28    \\   
      NGC2778   &   09h12m24.37s & +35d01m39.1s & 2025  &  22.3  &      E     &  13.05    &   12.25   &   10.09  &       F   & $<$7.48 &      $<$7.06         \\
\end{longtable}
    
    \end{landscape}
      \end{onecolumn}
    \small
    \begin{landscape}
    \begin{longtable}{l c c c c c c c c r}
    \hline
         Galaxy &       RA     &      Dec     &   cz  &  D     &  Type  & $g$        &  $r$  & $\rm logM_{star}$       \\  
              &       J2000.0  &  J2000.0     & km/s  & Mpc    &     & mag   &  mag  & $M_{\odot}$             \\  
           \hline                                                                                             
     NGC2480  &   07h57m10.44s & +23d46m47.3s &  2345 &  34.6  &   ScB    & 14.36   &  13.83   &   9.39  \\
     NGC2779  &   09h12m28.29s & +35d03m13.5s &  2105 &  31.8  &    -   & 14.52   &    13.80   &   9.64  \\
     NGC5636  &   14h29m39.02s & +03d15m58.7s &  1745 &  26.9  &   SABr   & 13.84   &  13.05   &   8.18  \\
     J125830  &   12h58m30.84s & +26d59m41.8s &  7453 &  103.0 &    -    & 17.80   &    17.65   &   8.91  \\
     J125838  &   12h58m38.69s & +27d00m47.5s &  7015 &  97.0  &    -   & 17.44   &    16.98   &   8.71  \\
     J125853  &   12h58m53.79s & +27d23m19.1s &  7654 &  105.0 &    -    & 17.95   &    17.52   &   8.60  \\
\end{longtable}
    \end{landscape}
    \newpage 
    \begin{landscape}
    \small
    \begin{longtable}{l c c c c c c c c c}
    \hline
    \endfoot
    \endlastfoot
    \hline 
     Galaxy &   ref H$\alpha$  & $\rm Am_{ON}$ & $\rm Am_{OFF}$ &  $\rm Am_{ON}$ & $\rm Am_{OFF}$       &  $\rm Texp_{ON}$ & $\rm Texp_{OFF}$ &  $\rm Texp_{ON}$ & $\rm Texp_{OFF}$  \\
            &             & 2015      &  2015      &  2016      & 2016    & 2015       &  2015          &  2016        & 2016          \\               
    \hline    
    \endhead  
  UGC03960  &  1   & 1.01  & 1.00  &   -    &   -     &  300  & 60    &  -    &  -     \\
  NGC2481   &  1   & 1.00  & 1.00  &   -    &   -     &  300  & 60    &  -    &  -     \\
  NGC2577   &  1   & 1.02  & 1.03  &   -    &   -     &  300  & 60    &  -    &  -     \\
  NGC2592   &  1   & 1.03  & 1.03  &   -    &   -     &  300  & 60    &  -    &  -     \\
  NGC2594   &  1   & 1.01  & 1.01  &   -    &   -     &  300  & 60    &  -    &  -     \\
  NGC2679   &  1   & 1.00  & 1.00  &   -    &   -     &  300  & 60    &  -    &  -     \\
  NGC2764   &  1   & 1.03  & 1.03  &   -    &   -     &  300  & 60    &  -    &  -     \\
  NGC2778   &  1   & 1.00  & 1.01  &   -    &   -     &  300  & 60    &  -    &  -     \\
\end{longtable}
        \end{landscape}
        \begin{landscape}
    \newpage
    \small
    \begin{longtable}{l c c c c c c c c c}
    \hline
    \endfoot
    \endlastfoot
    \hline 
     Galaxy &    Ref      & $\rm Am_{ON}$ & $\rm Am_{OFF}$ &  $\rm Am_{ON}$ & $\rm Am_{OFF}$ &  $\rm Texp_{ON}$ & $\rm Texp_{OFF}$ &  $\rm Texp_{ON}$ & $\rm Texp_{OFF}$  \\
            &             & 2015      &  2015      &  2016      & 2016    & 2015       &  2015          &  2016        & 2016          \\               
    \hline    
    \endhead  
  NGC2480   &  1   & 1.00   & 1.00   &   -    &   -     &  300  & 60   &  -    &  -     \\
  NGC2779   &  1   & 1.00   & 1.01   &   -    &   -     &  300  & 60   &  -    &  -     \\
  NGC5636   &  1   & 1.14   & 1.14   &   -    &   -     &  300  & 60   &  -    &  -     \\
  J125830   &  1   &  -     &  -     &  1.03 &  1.02   &   -   &  -    & 600   & 120 \\
  J125838   &  1   &  -     &  -     &  1.10 &  1.09   &   -   &  -    & 600   & 120 \\
  J125853   &  1   &  -     &  -     &  1.19 &  1.17   &   -   &  -    & 600   & 120 \\
  J125916   &  1   &  -     &  -     &  1.13 &  1.12   &   -   &  -    & 600   & 120 \\
  J130045   &  1   &  -     &  -     &  1.24 &  1.22   &   -   &  -    & 600   & 120 \\
\end{longtable}
    \end{landscape}
    \newpage
    \begin{landscape}
    \small
    \begin{longtable}{l r c c c r c c c c c c c c c  c }
    \hline
    \endfoot
    \endlastfoot
    \hline 
     Galaxy &   $EW_{3"}$ & $\pm$ & $LogF_{3"}$                &  $\pm$                  & $EW_{Tot}$    & $\pm$    &   $LogF_{Tot}$             &  $\pm$                     &    log SFR              &  Spec  & Spec Ty & det &  morf H$\alpha$  &  HST &  ref H$\alpha$  \\                    
            &   \AA       & \AA   & $\rm erg ~cm^{-2} s^{-1}$   &  $\rm erg~cm^{-2} s^{-1}$  & \AA             &  \AA     &   $\rm erg ~cm^{-2} s^{-1}$ &  $\rm erg ~cm^{-2} s^{-1}$  &     $M_{\odot} yr^{-1}$  &     &             & &                 &      &                 \\                       
    \hline    
    \endhead  
      UGC03960  &    -2.47    & 3.07      &   -14.60    &   0.35    &  -3.89   &  3.76      &   -13.38  &  0.29        &        -2.15             &  SDSS   & PAS & 2 &  c      &   0   &  1     \\  
      NGC2481   &    -0.43    & 2.98      & $<$-14.79   &   0.89    &   3.66   &  2.99      &      -    &     -        &          -               &  LOI    & RET & 3 &  -      &   0   &  1     \\  
      NGC2577   &    -6.29    & 3.02      &   -13.72    &   0.16    &   1.71   &  3.07      &      -    &     -        &          -               &  LOI    & PAS & 3 &  -      &   -   &  1     \\  
      NGC2592   &     2.32    & 2.93      &      -      &      -    &   4.82   &  3.27      &      -    &     -        &          -               &  LOI    & PAS & 3 &  -      &   0   &  1     \\  
      NGC2594   &    -1.14    & 3.01      &   -14.43    &   0.56    &   2.93   &  2.99      &      -    &     -        &          -               &  LOI    & PAS & 3 &  -      &   -   &  1     \\  
      NGC2679   &    -0.28    & 2.98      & $<$-15.26   &   1.05    &   1.41   &  4.16      &      -    &     -        &          -               &  SDSS   & PAS & 3 &  -      &   -   &  1     \\  
      NGC2764   &   -81.01    & 4.69      &   -13.37    &   0.01    & -17.25   &  3.74      &   -12.61  &  0.08        &        -1.23             &  LOI    & HII & 1 &  d      &   -   &  1     \\  
      NGC2778   &    -6.53    & 3.02      &   -13.70    &   0.16    &  -1.13   &  3.15      &   -13.48  &  0.57        &        -2.60             &  LOI    & PAS & 2 &  c      &   0   &  1     \\
\end{longtable}
    \end{landscape}
   \begin{landscape}

   \small
   \begin{longtable}{l c c c c c c c c c }
   \hline
   \endfoot
   \endlastfoot
   \hline 
    Galaxy &   $EW_{3"}$ & $\pm$ & $LogF_{3"}$            &  $\pm$              & $EW_{Tot}$    & $\pm$  &  $LogF_{Tot}$         &  $\pm$                                 & log SFR            \\
           &   \AA       & \AA   & $\rm erg ~cm^{-2} s^{-1}$   &  $\rm erg ~cm^{-2} s^{-1}$  & \AA        &  \AA   &  $\rm erg ~cm^{-2} s^{-1}$ &  $\rm erg ~cm^{-2} s^{-1}$ & $M_{\odot} yr^{-1}$\\
   \hline    
   \endhead  
   NGC2480 & -180.5   & 8.90  & -13.38   & 0.01  &  -28.15 &   4.74 & -12.81  &    0.06     &        -1.55  \\
   NGC2779 & -4.14    & 3.22  & -14.77   & 0.25  &  -13.95 &   3.14 & -13.2   &    0.09     &        -2.01  \\
   NGC5636 & -42.72   & 3.87  & -13.82   & 0.03  &  -2.41  &   3.82 & -13.48  &    0.41     &        -1.72  \\
   J125830 & -215.4   & 8.89  & -14.45   & 0.03  &  -84.28 &   4.55 & -13.93  &    0.06     &        -1.83  \\
   J125838 & -41.59   & 4.07  & -14.73   & 0.08  &  -32.7  &   3.40 & -13.99  &    0.08     &        -2.30  \\
   J125853 & -27.99   & 4.13  & -14.88   & 0.15  &  -22.1  &   3.77 & -14.53  &    0.17     &        -1.20  \\
   J125916 & -42.96   & 4.02  & -13.98   & 0.07  &  -39.71 &   3.87 & -13.38  &    0.07     &        -2.72  \\
   J130045 & -0.59    & 8.71  & -17.38   & 1.63  &  -35.39 &   6.18 & -14.9   &    0.17     &        -1.68  \\
\end{longtable}
   \end{landscape}

     \begin{onecolumn}
     \begin{figure*}
     \centering
     \includegraphics[scale=0.25]{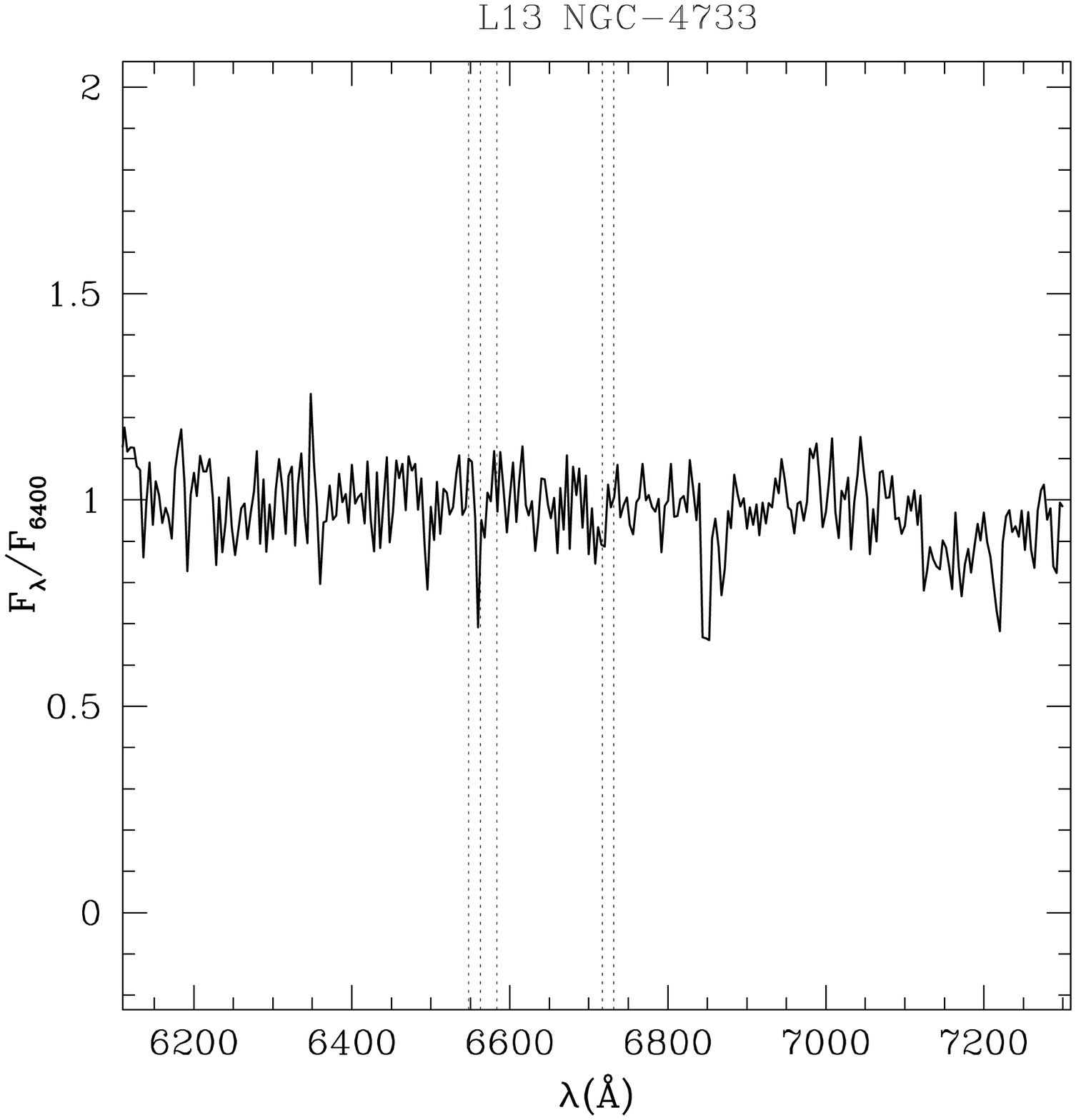}\includegraphics[scale=0.25]{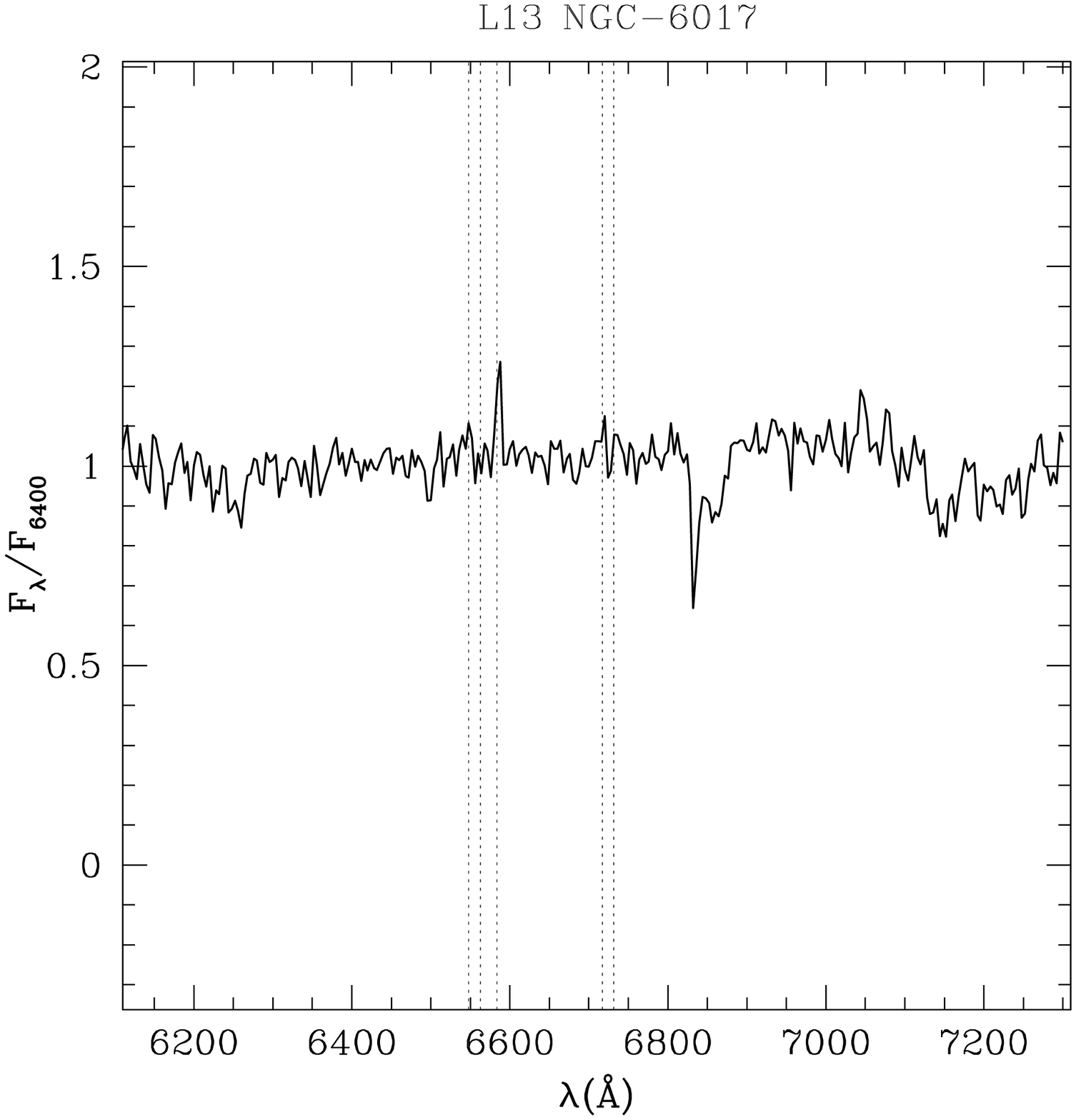}\includegraphics[scale=0.25]{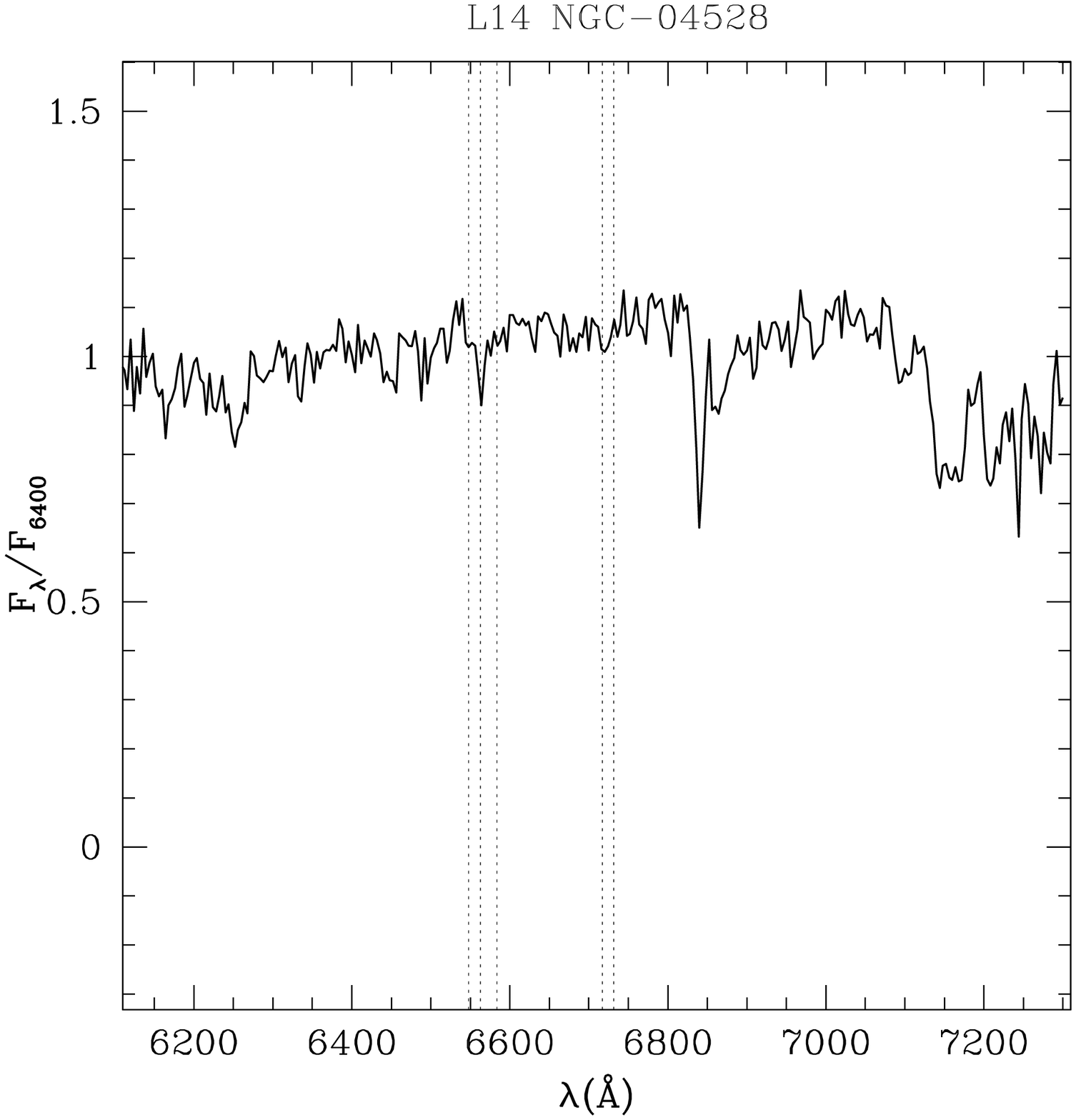}\\
     \includegraphics[scale=0.25]{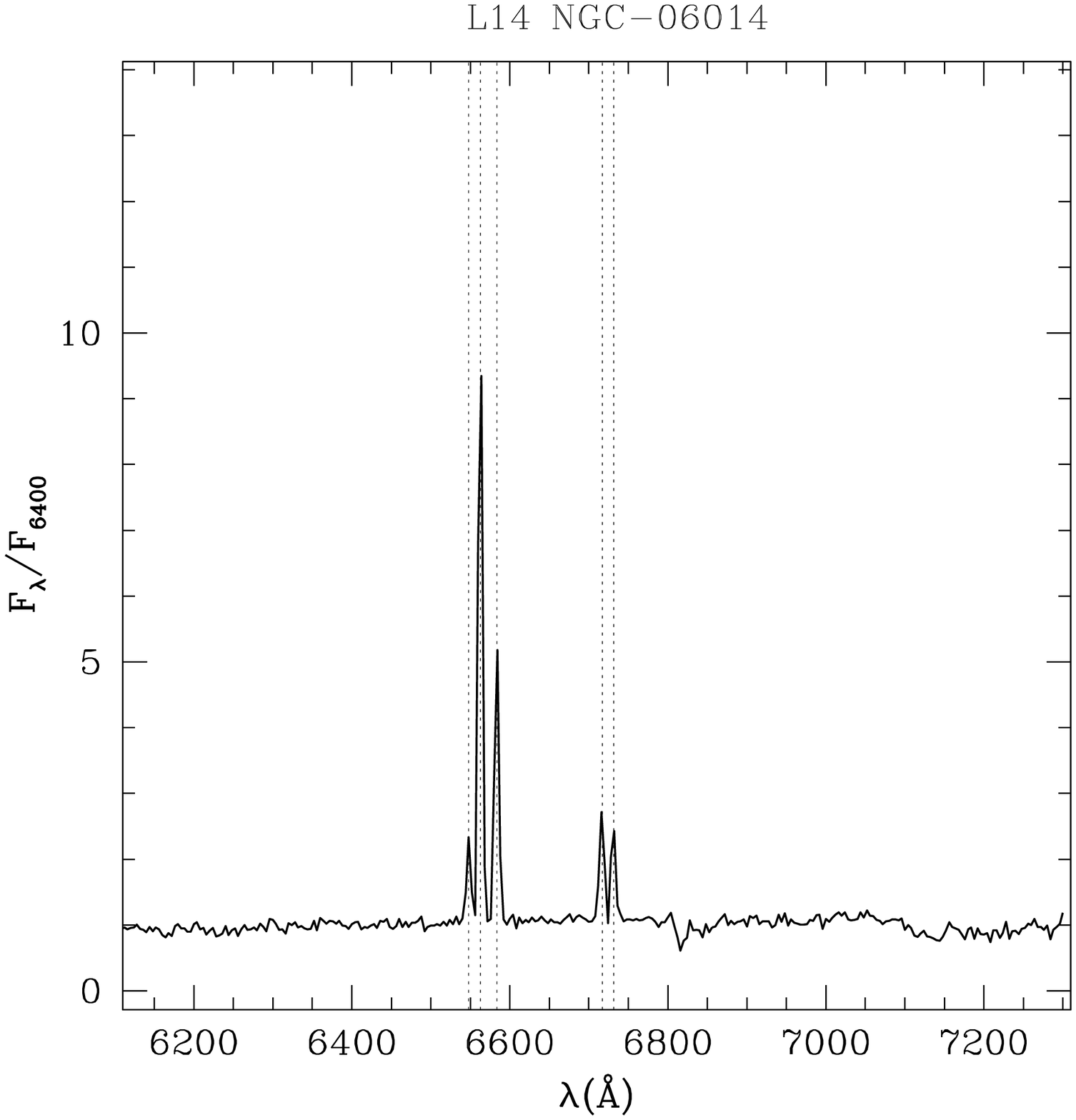}\includegraphics[scale=0.25]{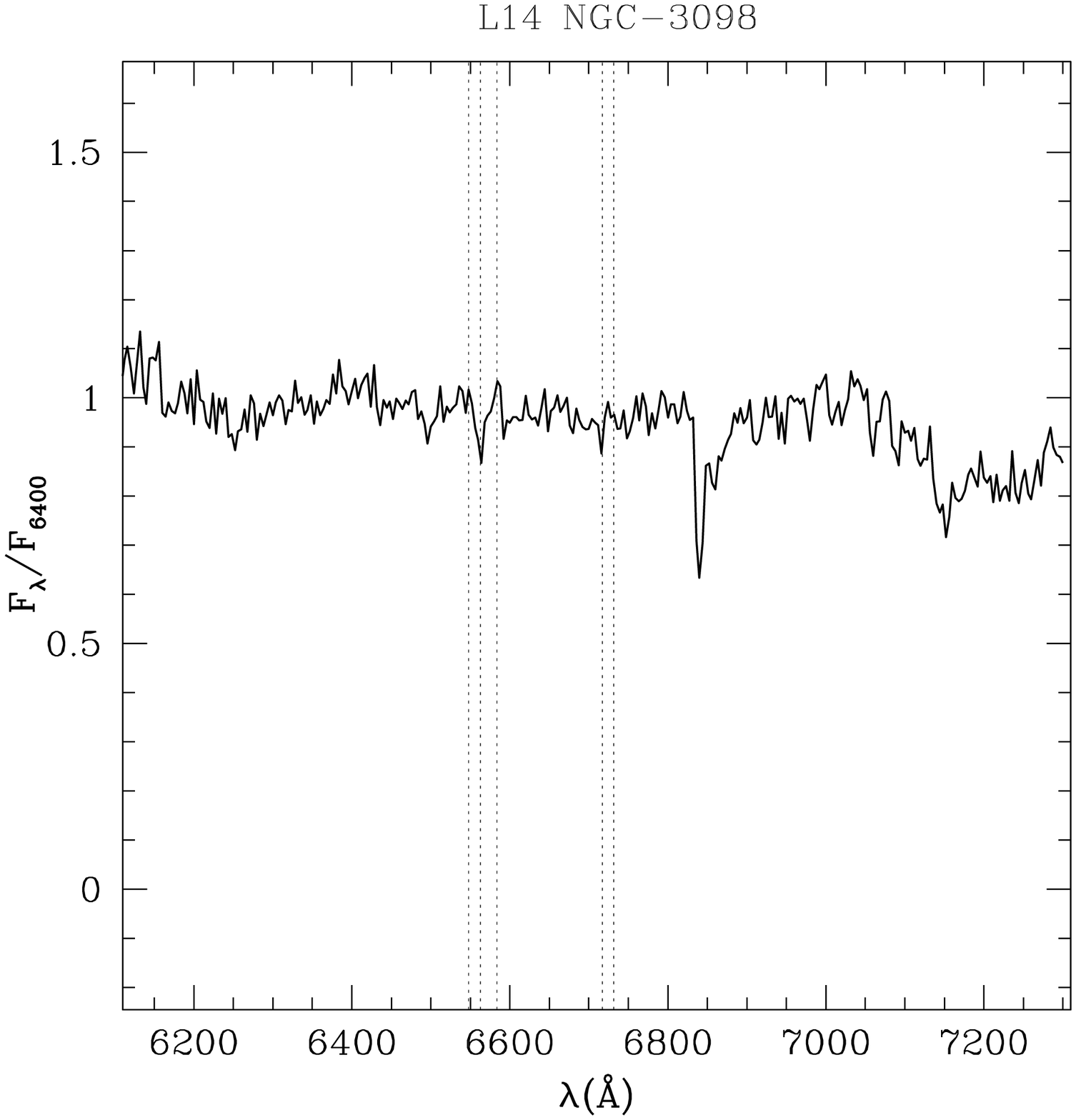}\includegraphics[scale=0.25]{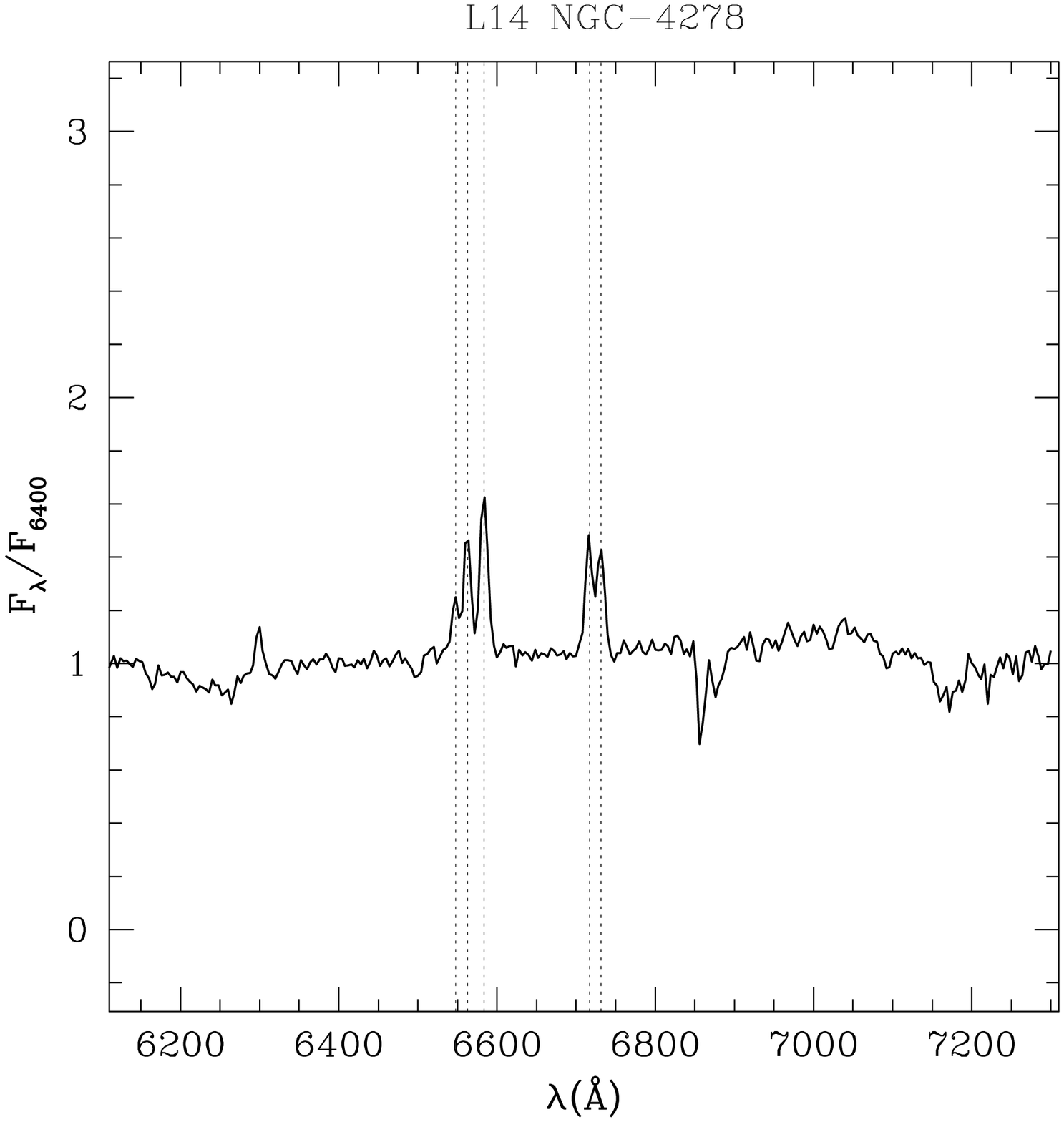}\\
     \includegraphics[scale=0.25]{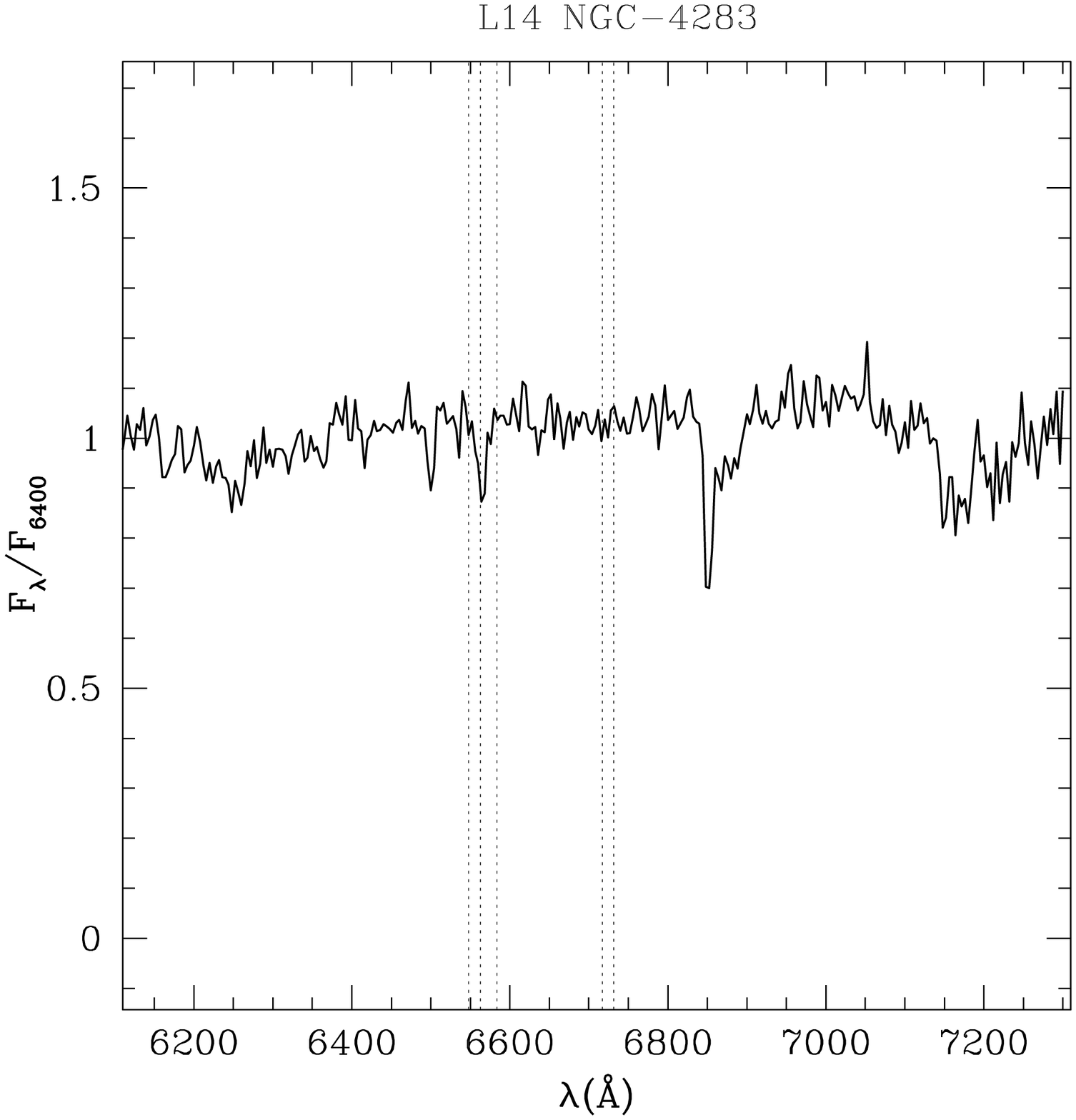}\includegraphics[scale=0.25]{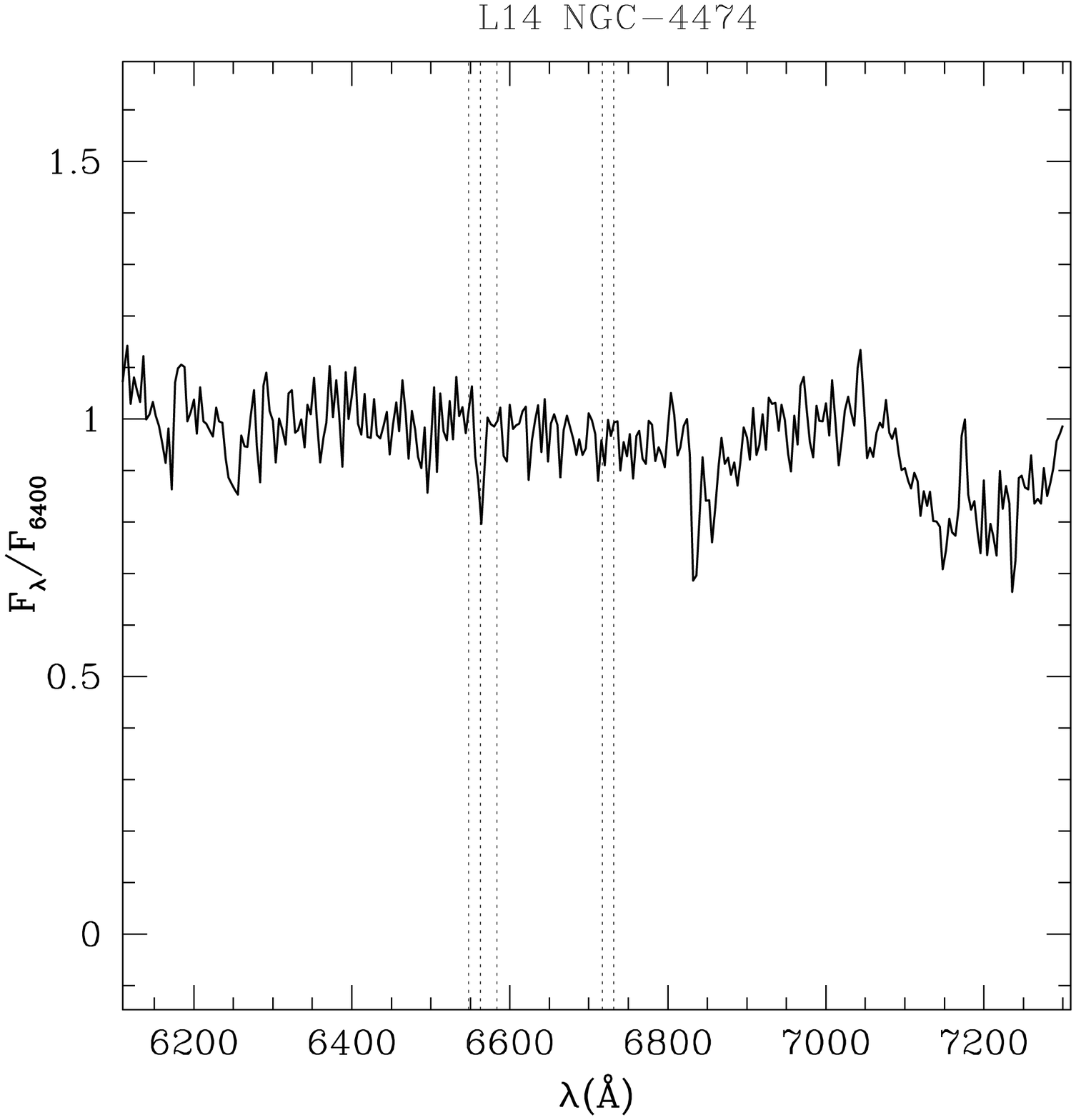}\includegraphics[scale=0.25]{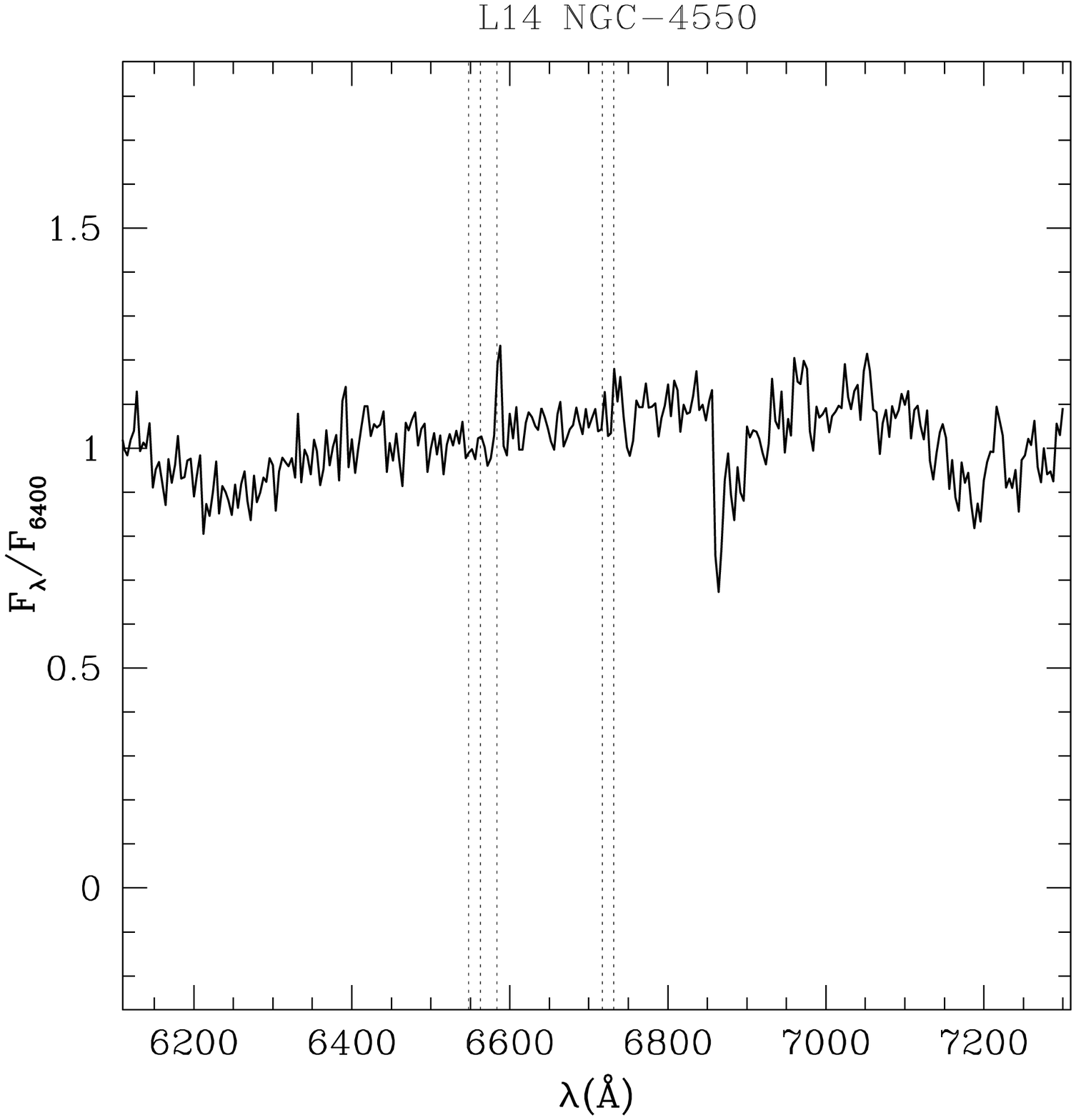}\\
     \includegraphics[scale=0.25]{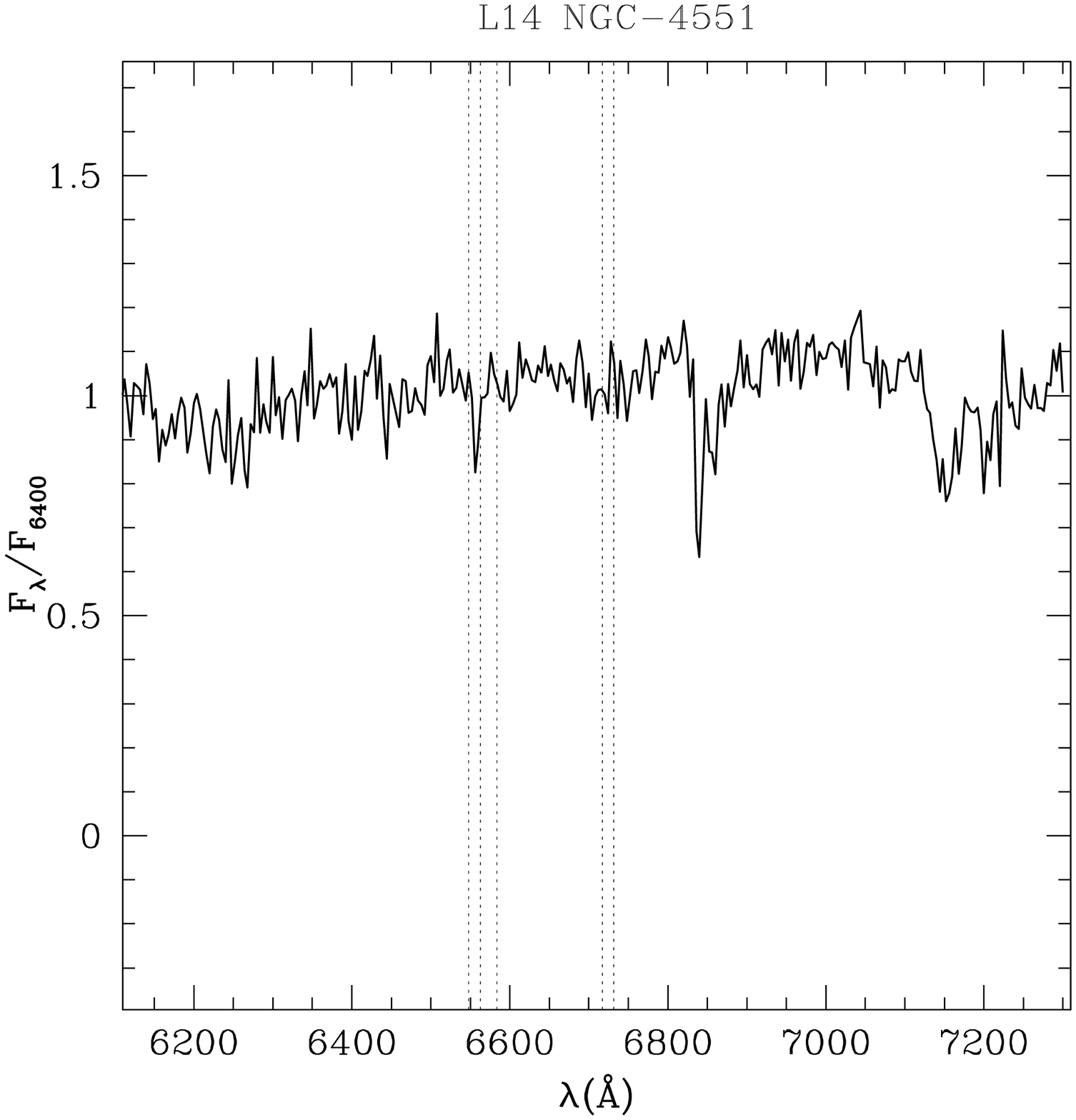}\includegraphics[scale=0.25]{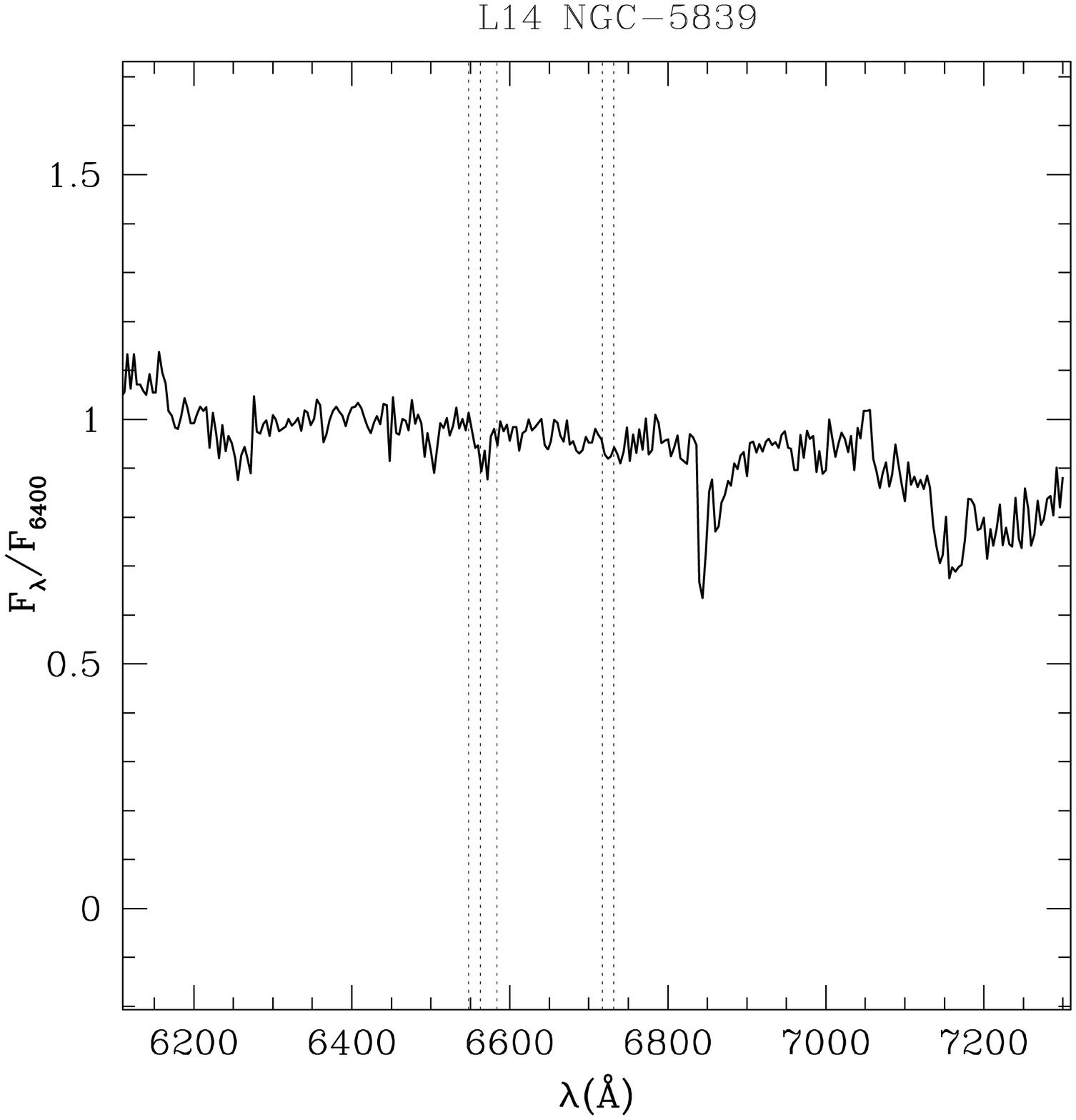}\includegraphics[scale=0.25]{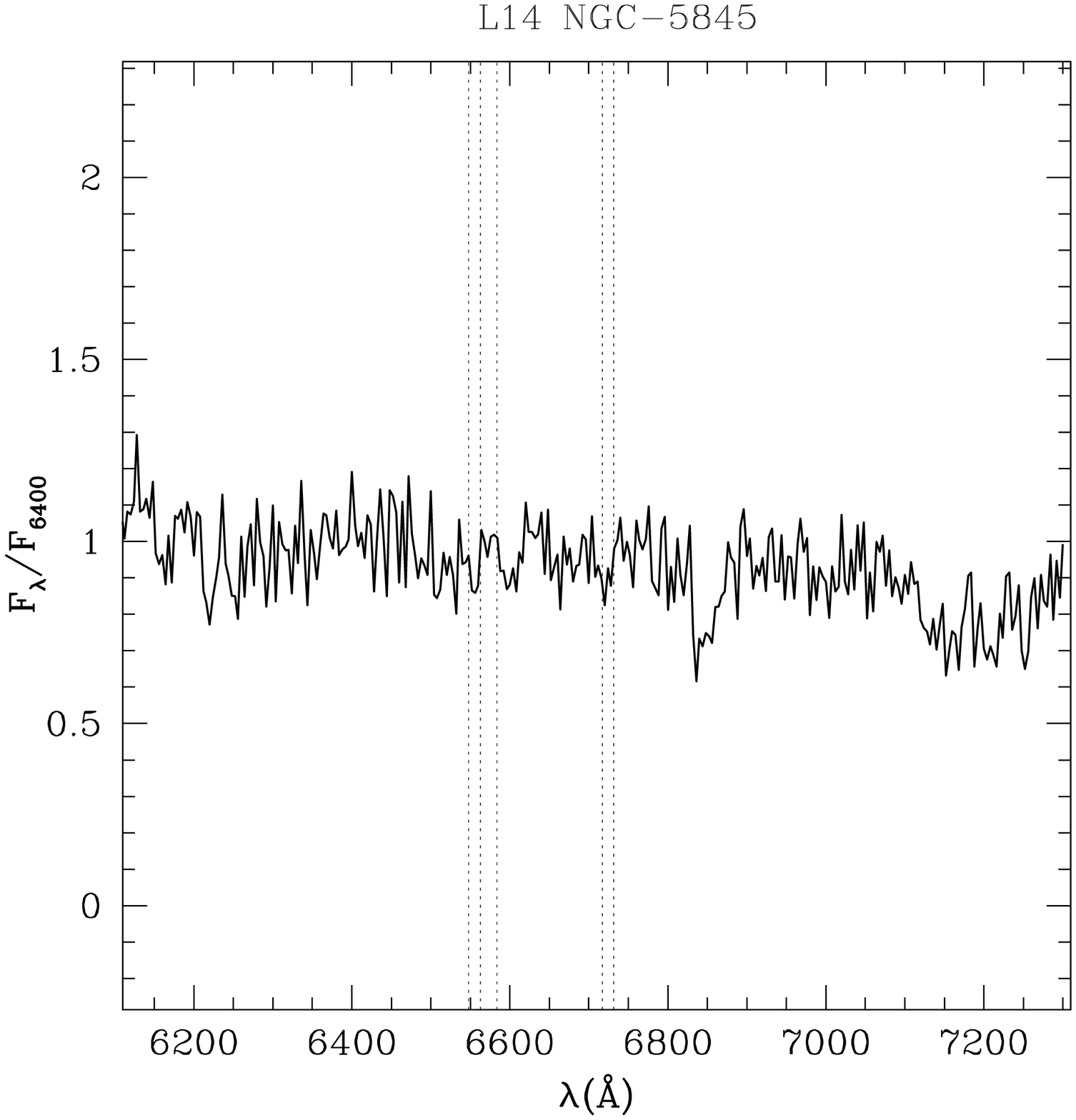}\\
     \label{spectra}  
     \end{figure*}
     \begin{figure*}
     \centering
     \includegraphics[scale=0.25]{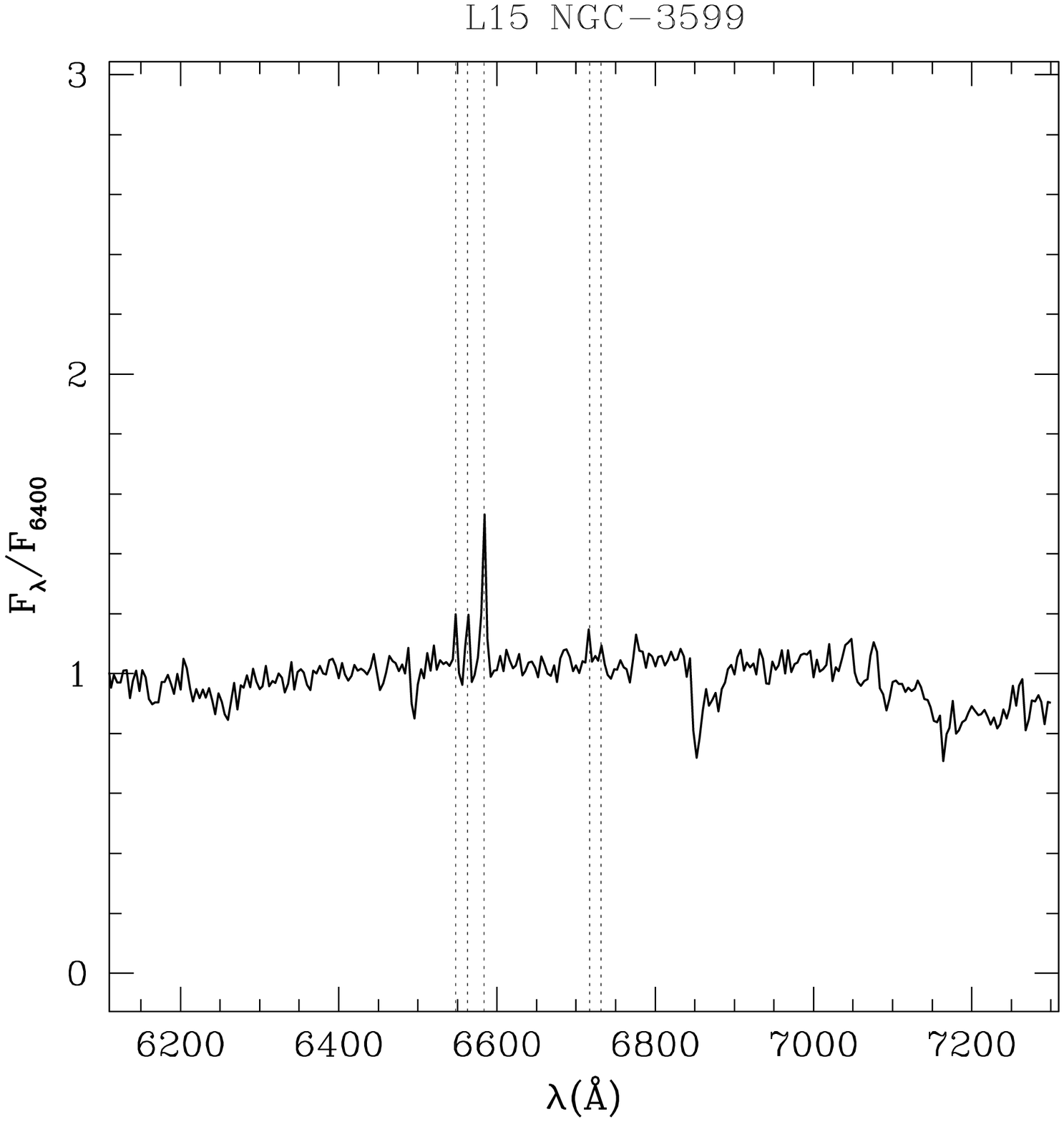}\includegraphics[scale=0.25]{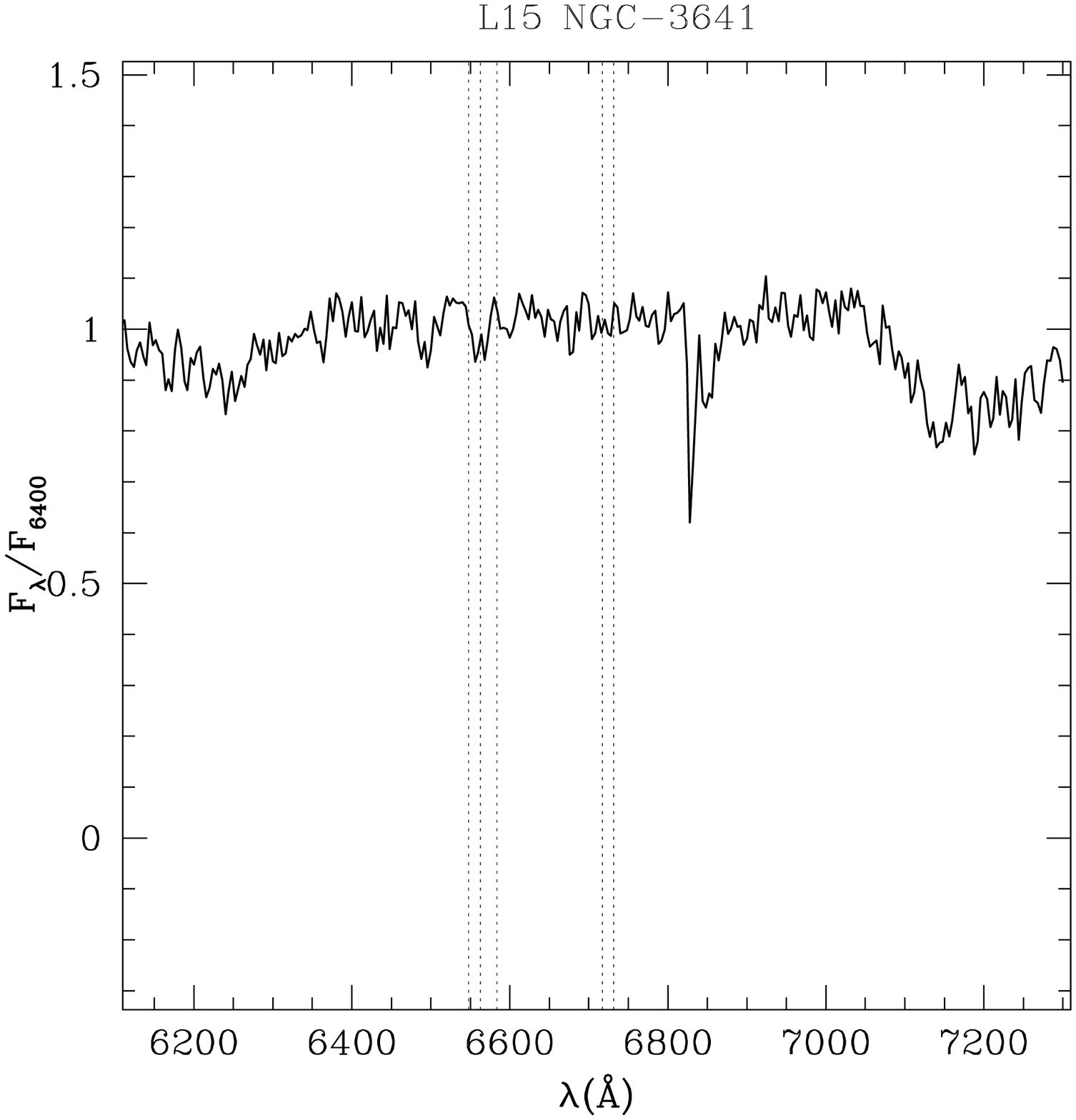}\includegraphics[scale=0.25]{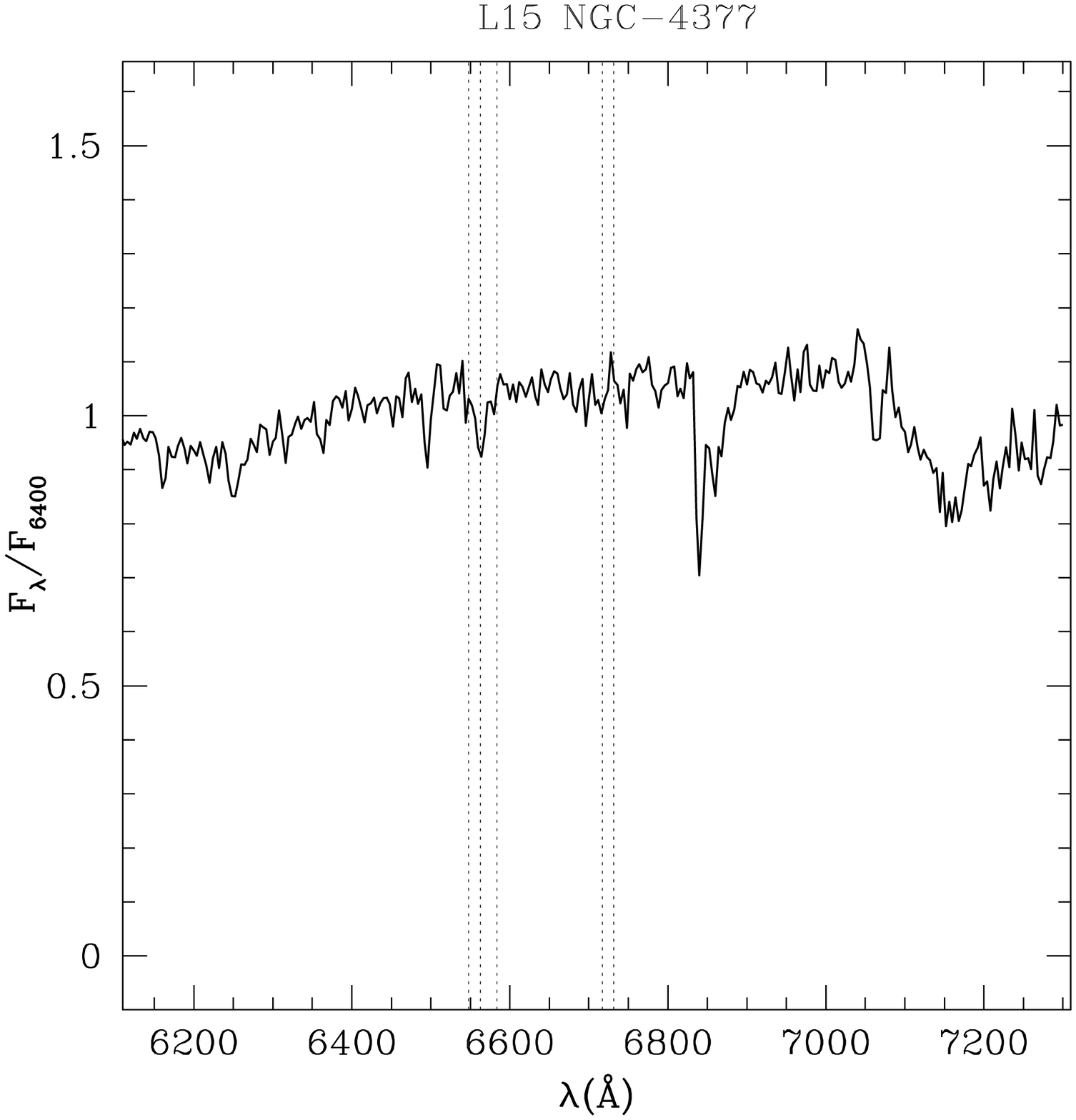}\\
     \includegraphics[scale=0.25]{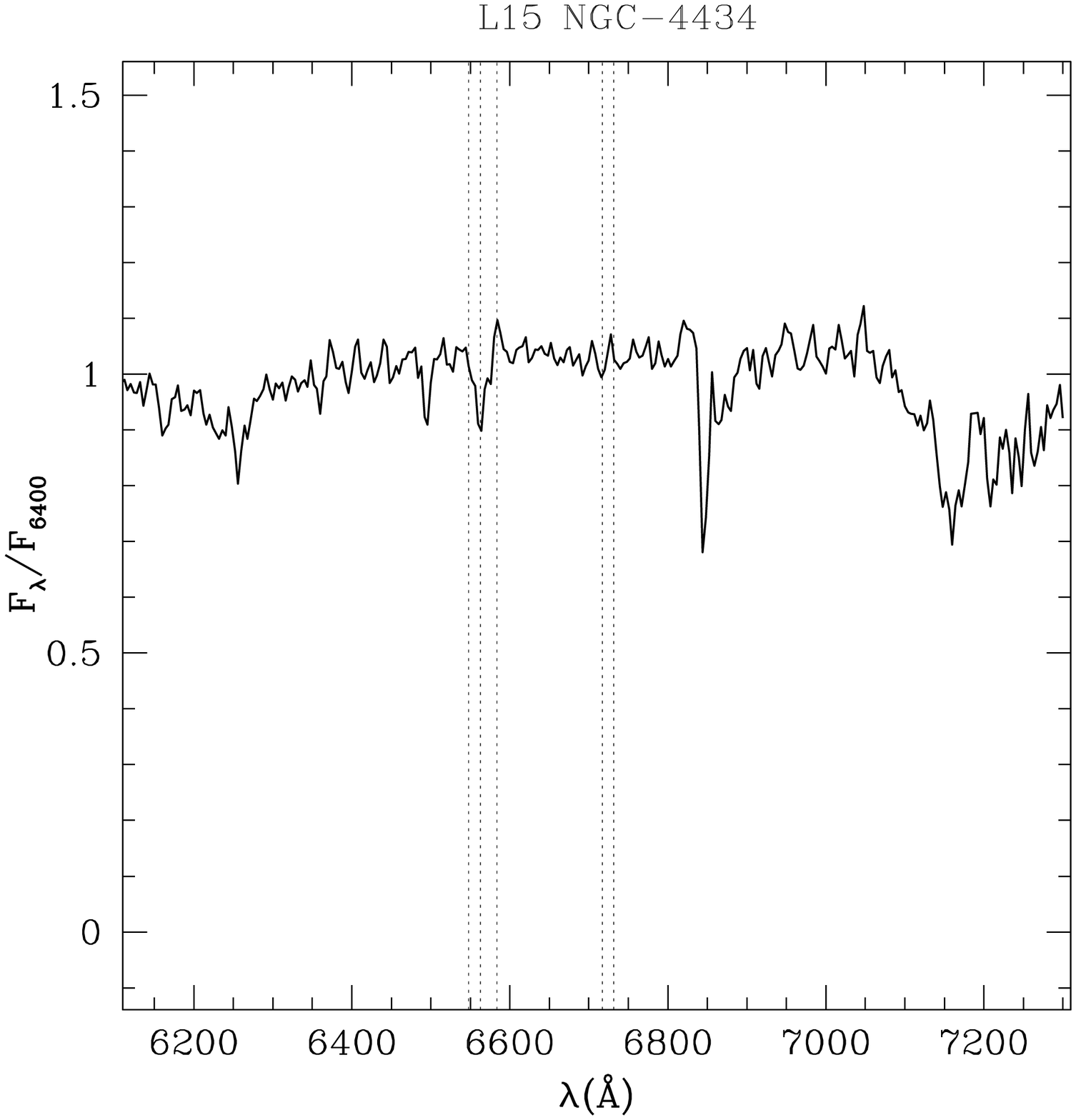}\includegraphics[scale=0.25]{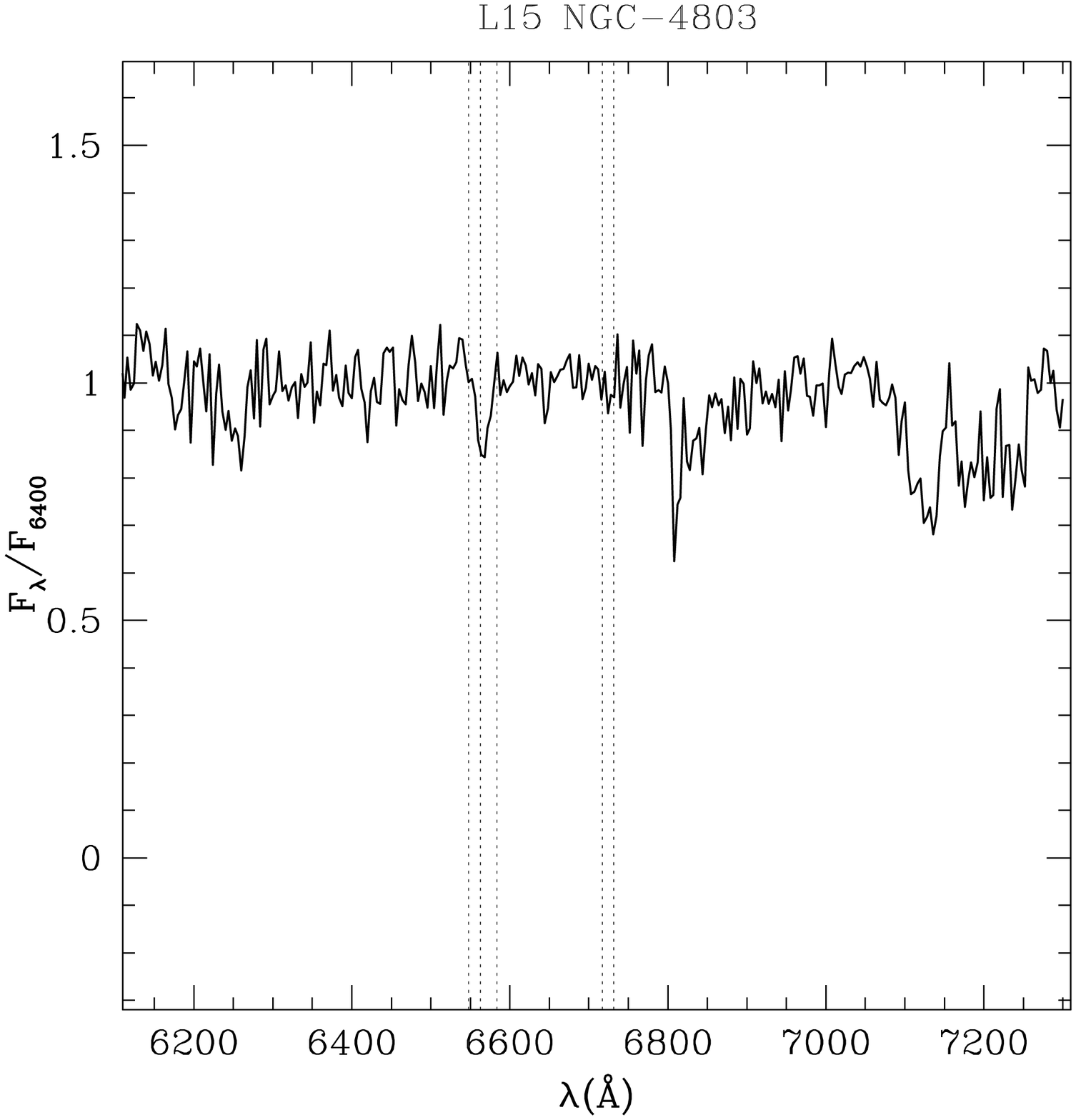}\includegraphics[scale=0.25]{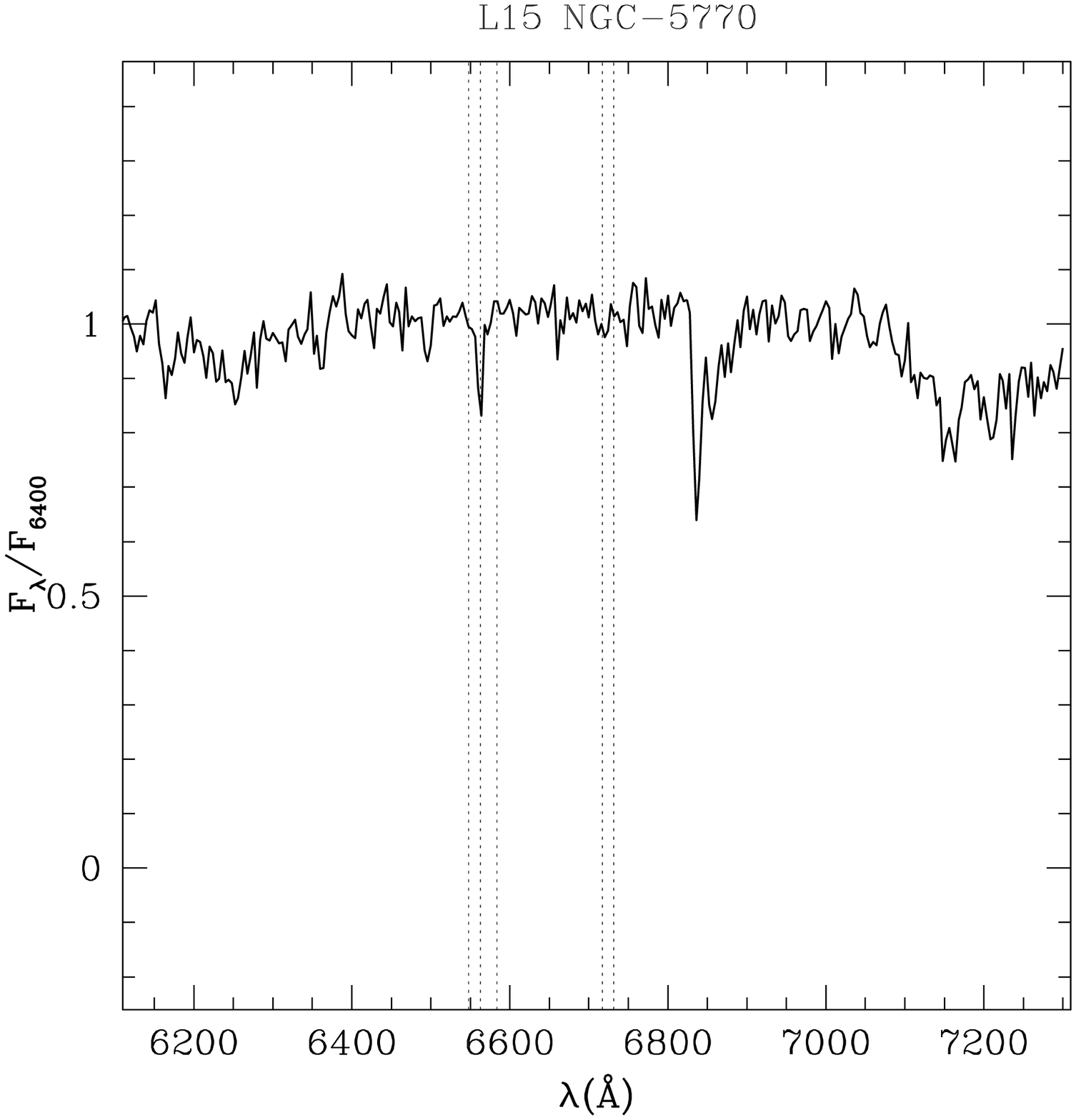}\\
     \includegraphics[scale=0.25]{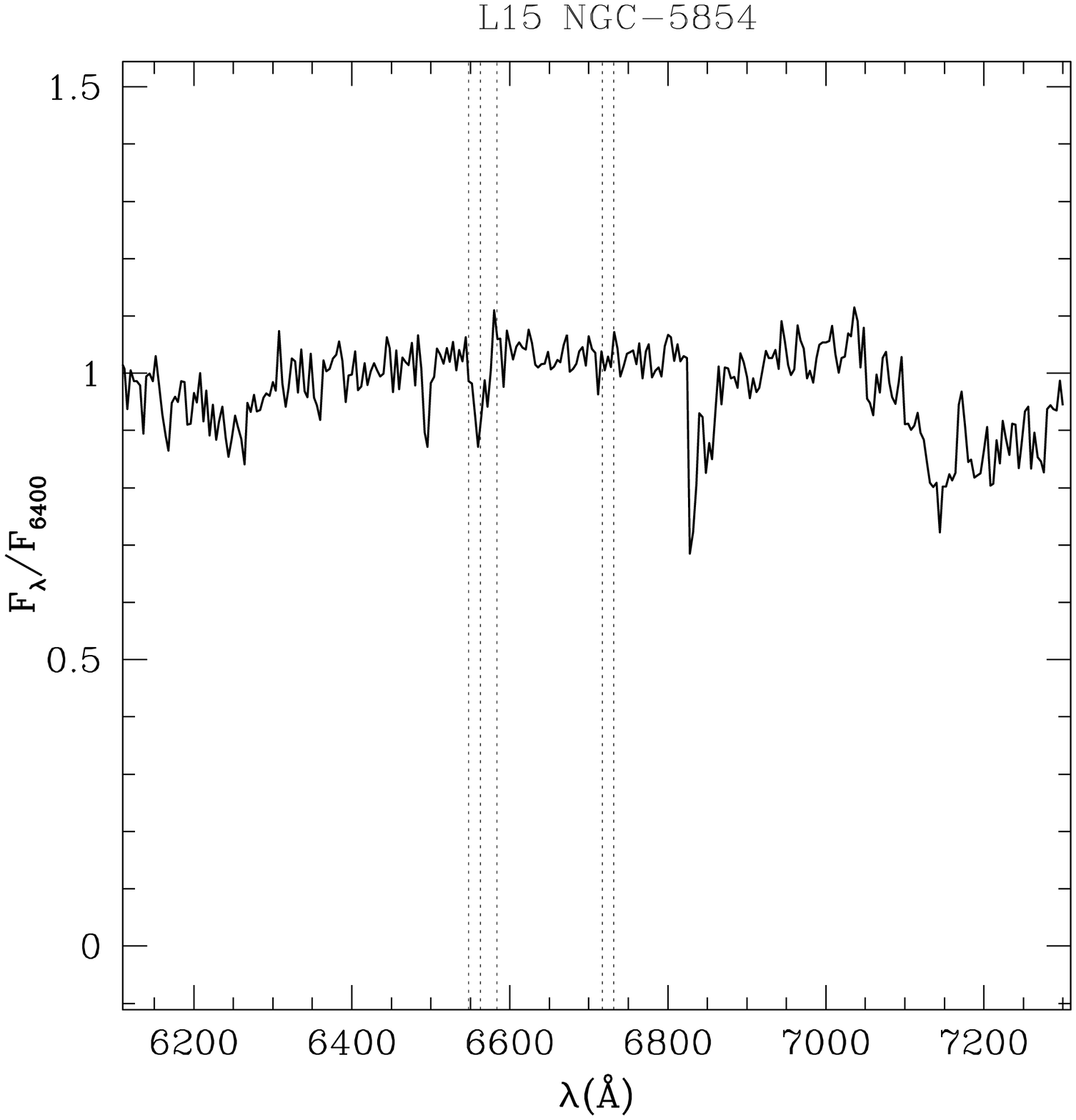}\includegraphics[scale=0.25]{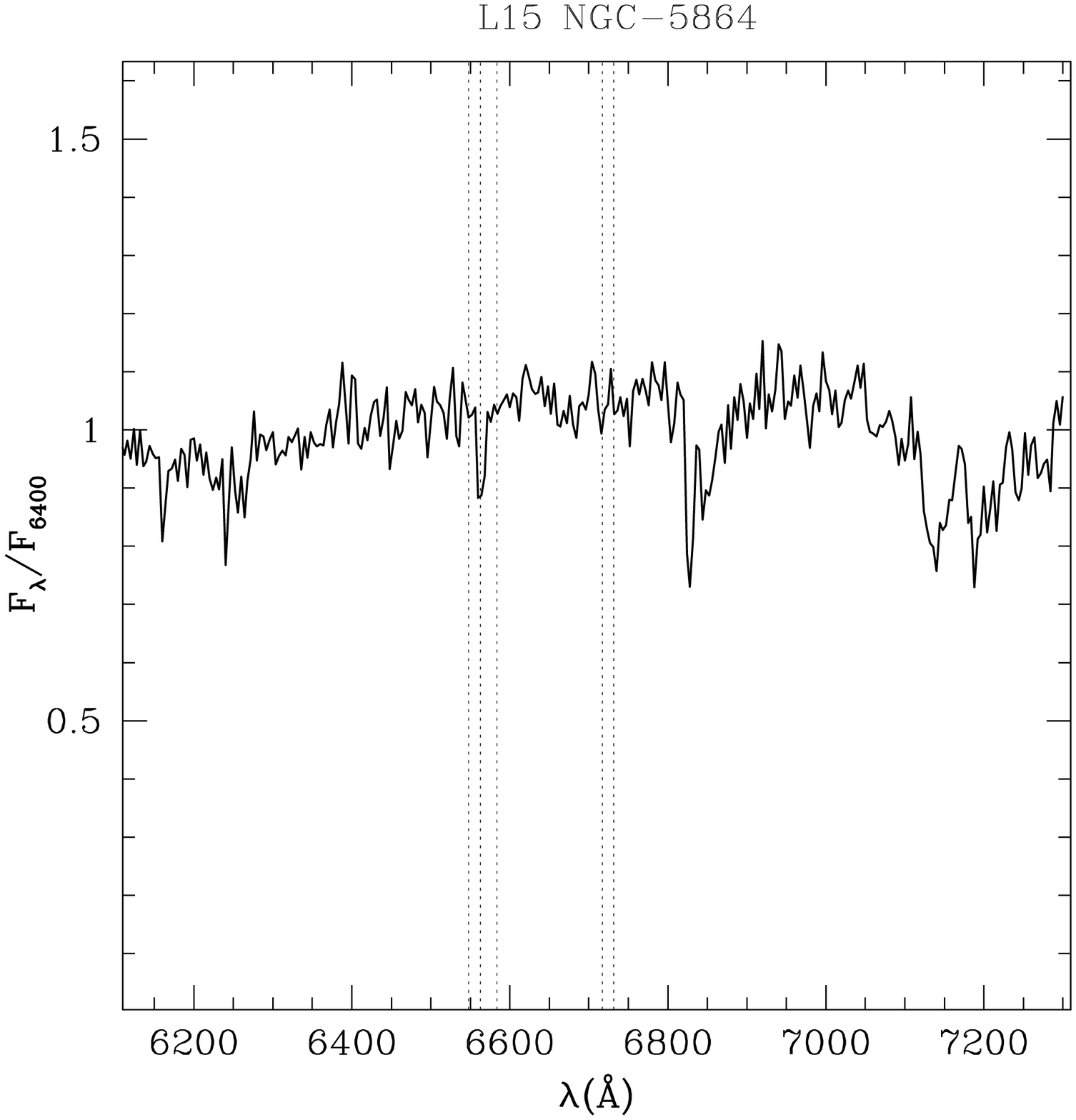}\includegraphics[scale=0.25]{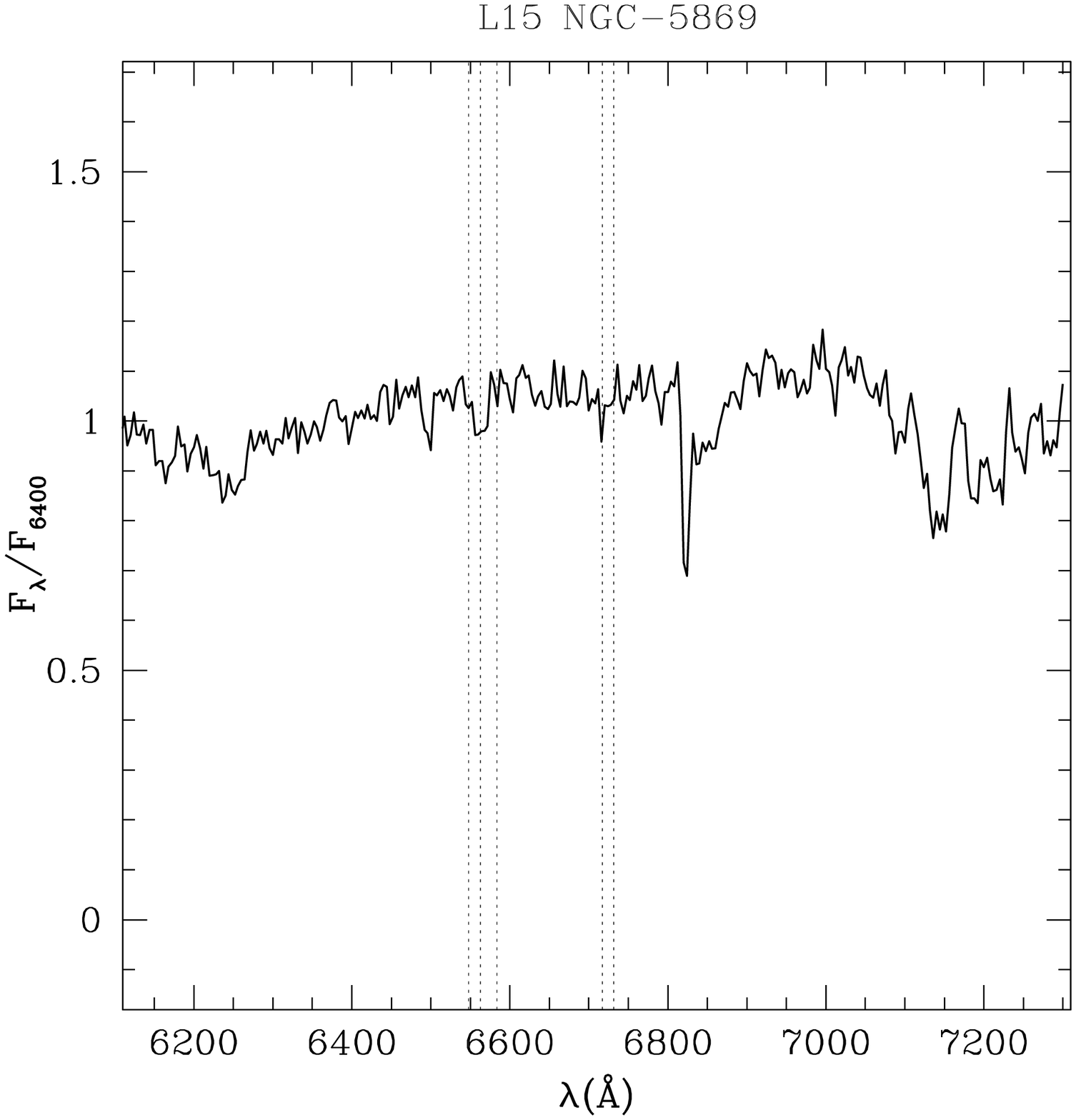} \\
     \includegraphics[scale=0.25]{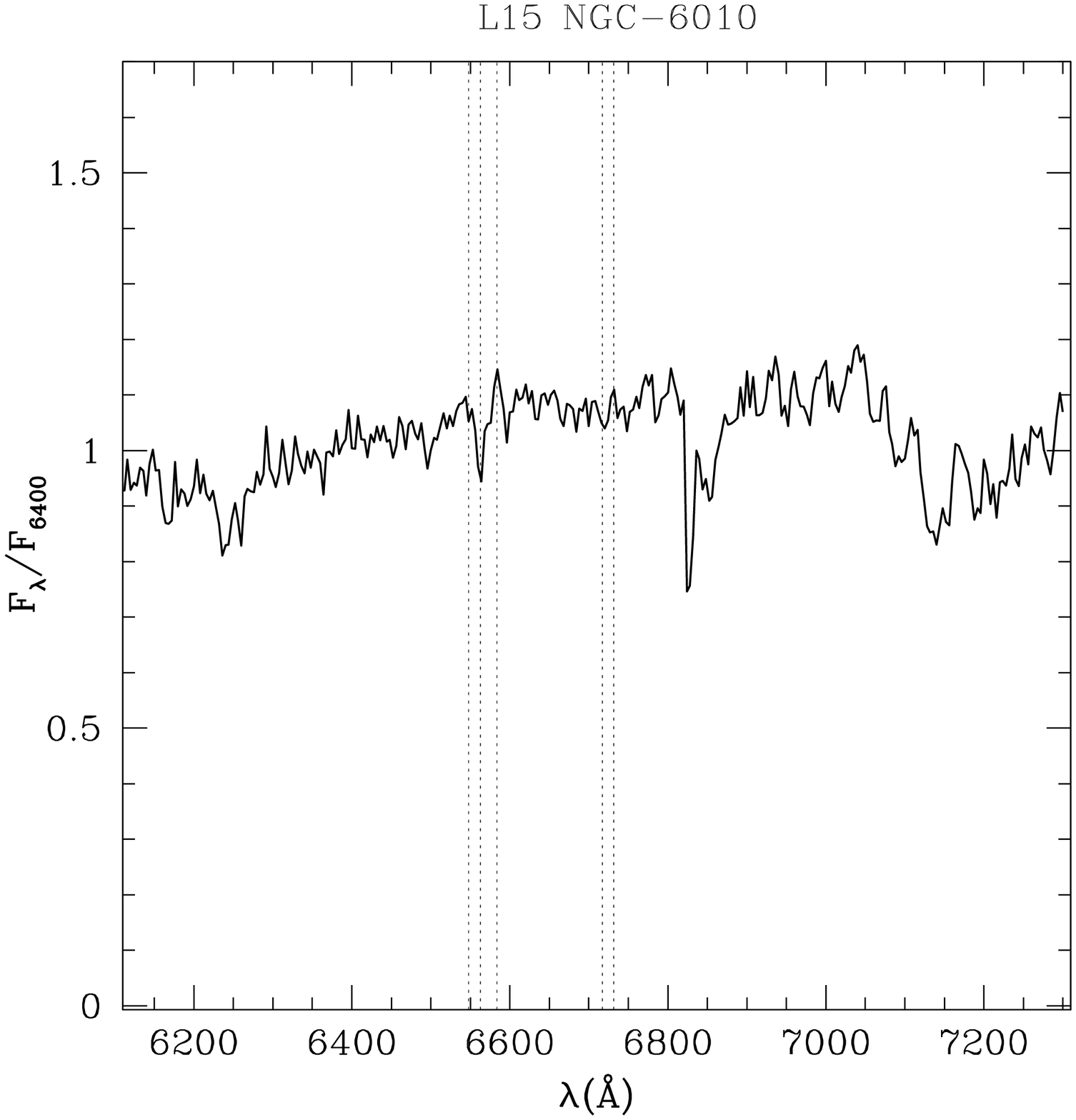}\includegraphics[scale=0.25]{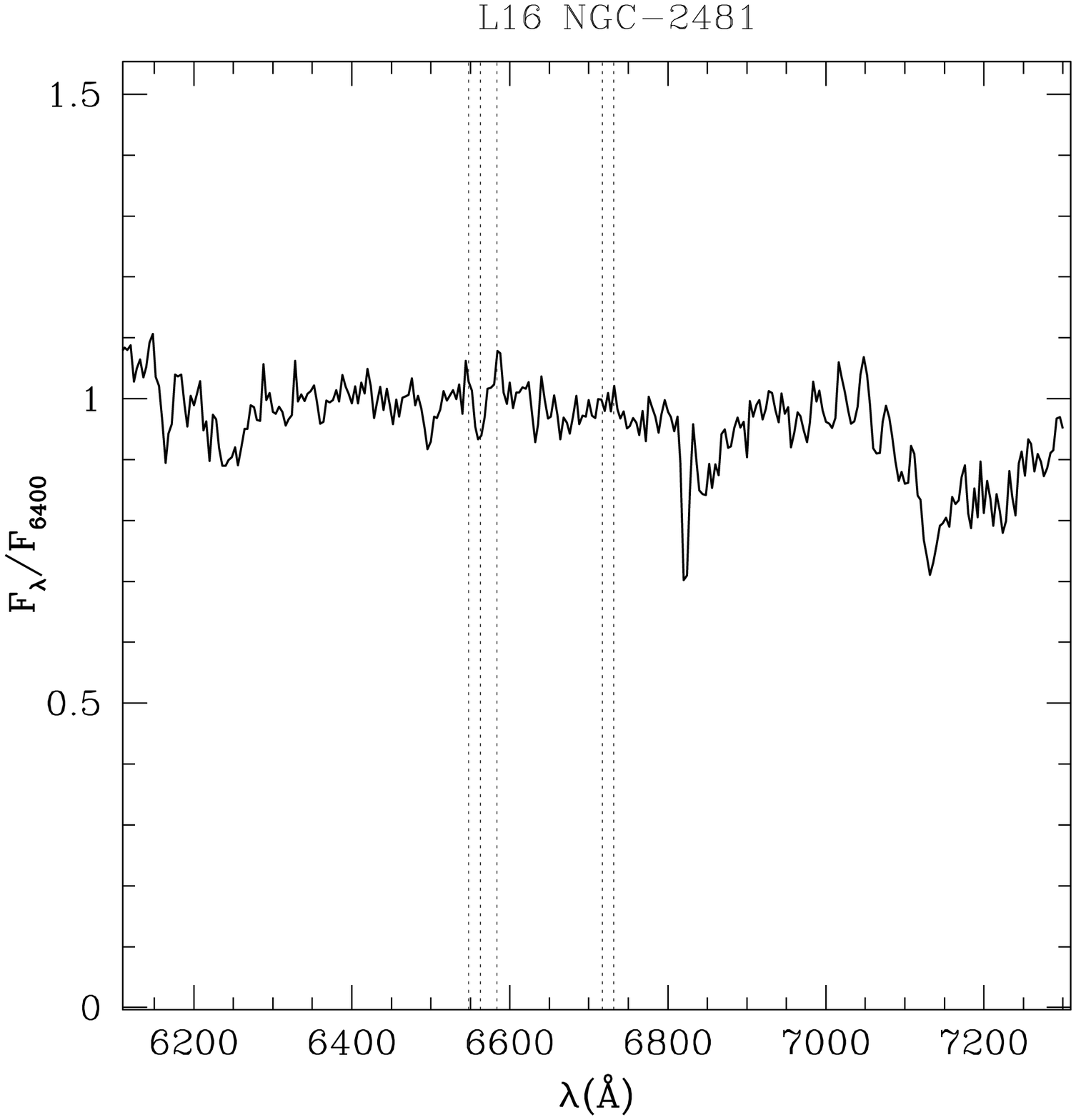}\includegraphics[scale=0.25]{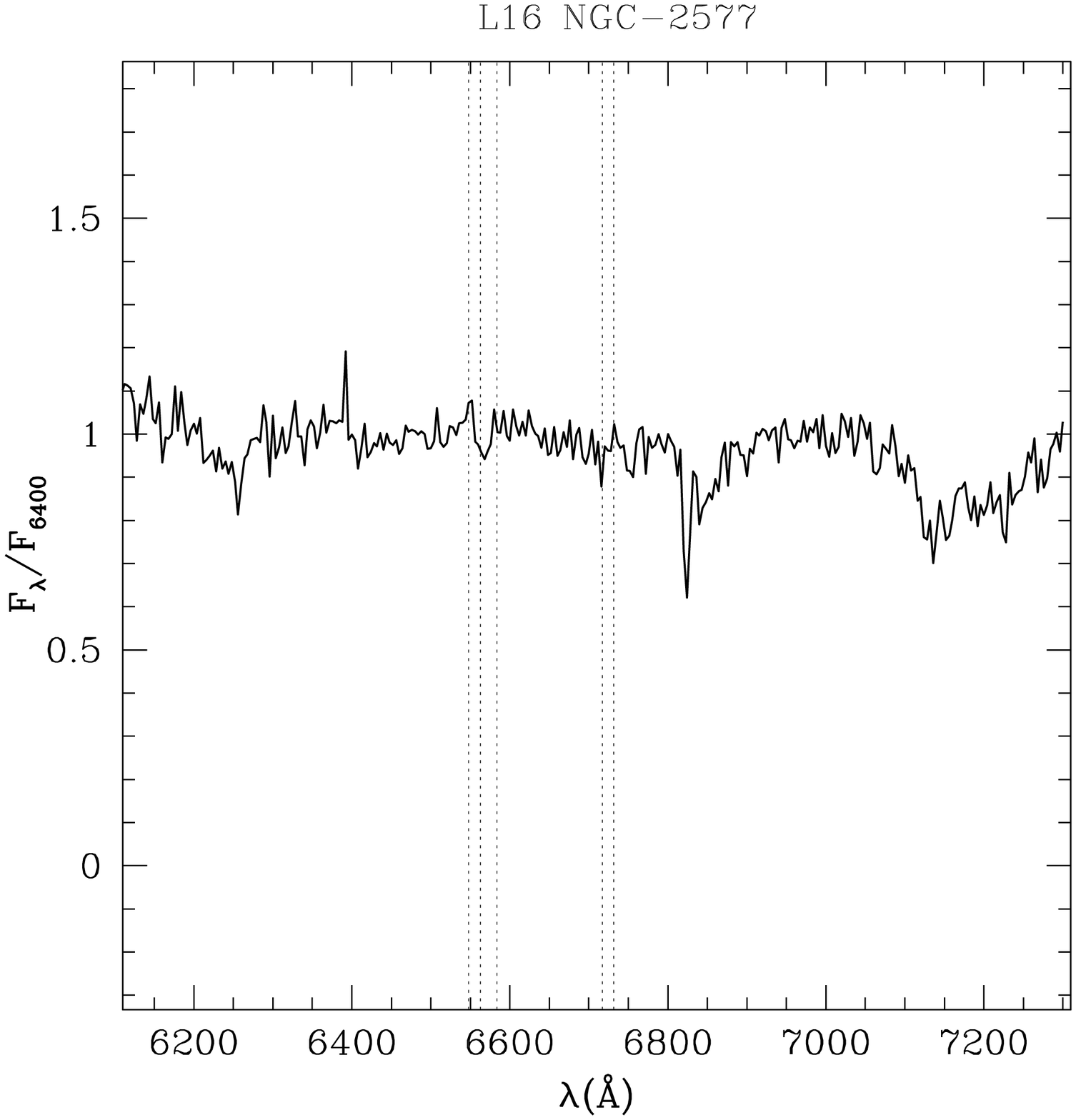}\\
     \label{spectra2}  
     \end{figure*}
     \begin{figure*}
     \centering
     \includegraphics[scale=0.25]{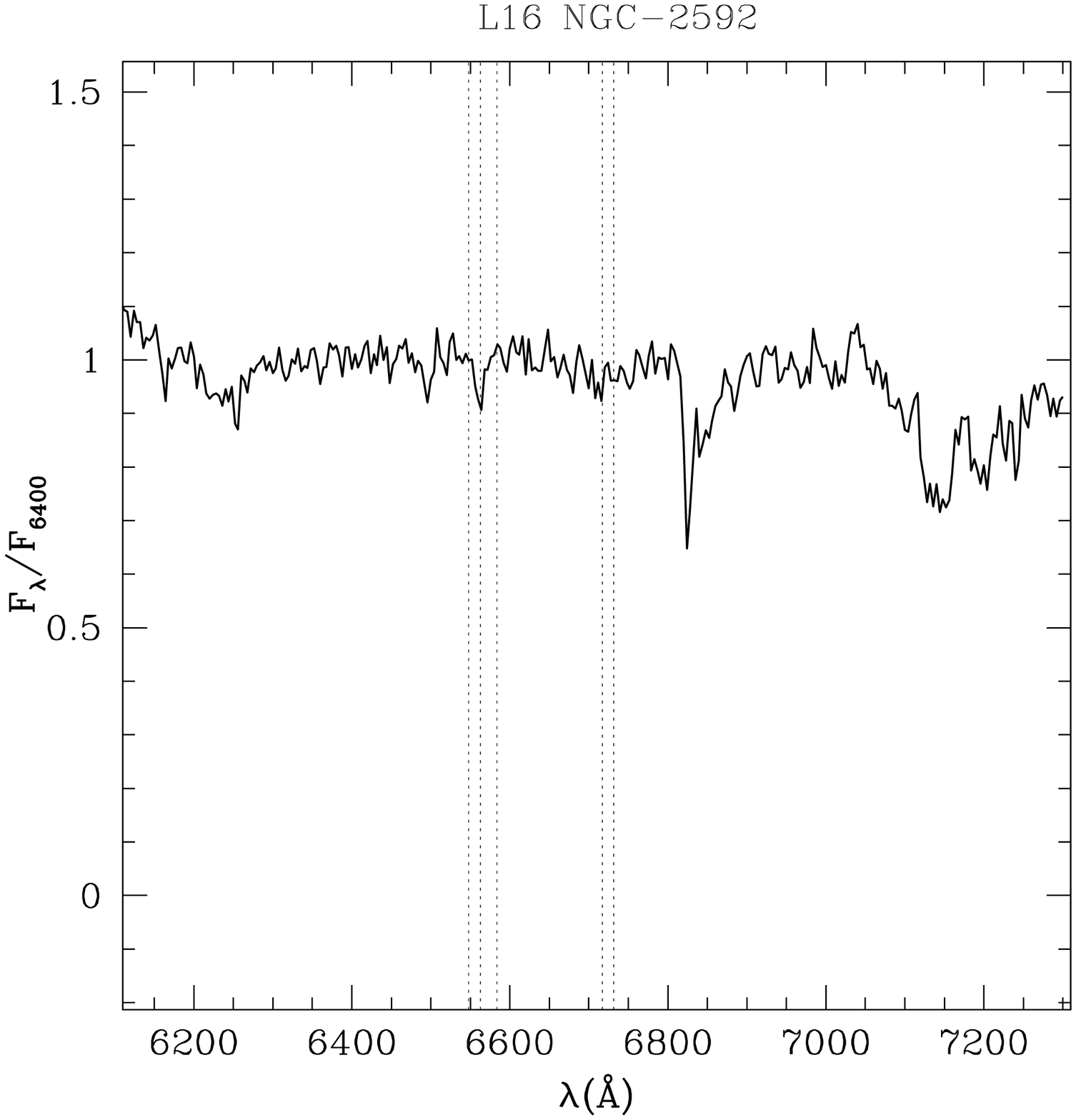}\includegraphics[scale=0.25]{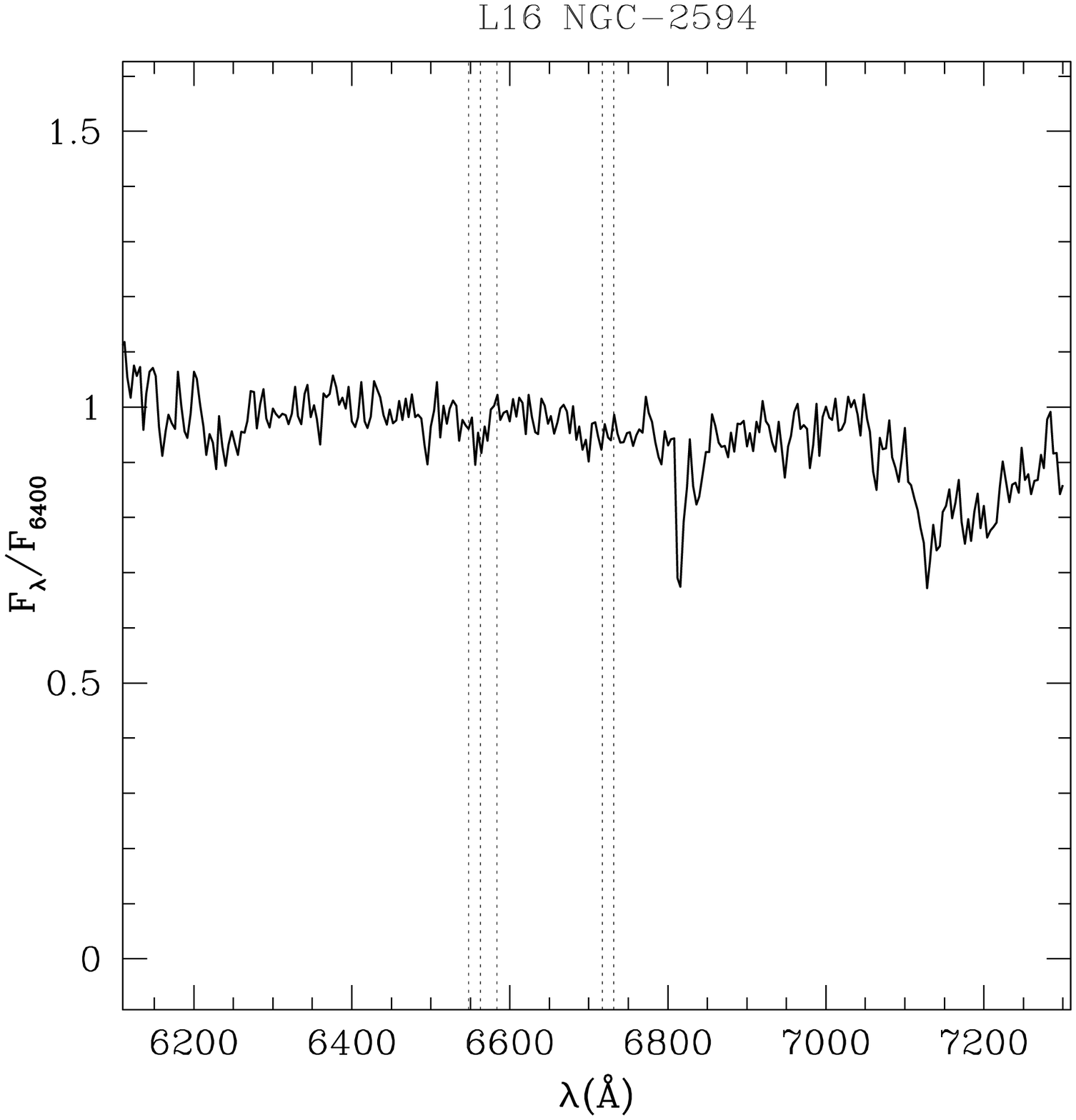}\includegraphics[scale=0.25]{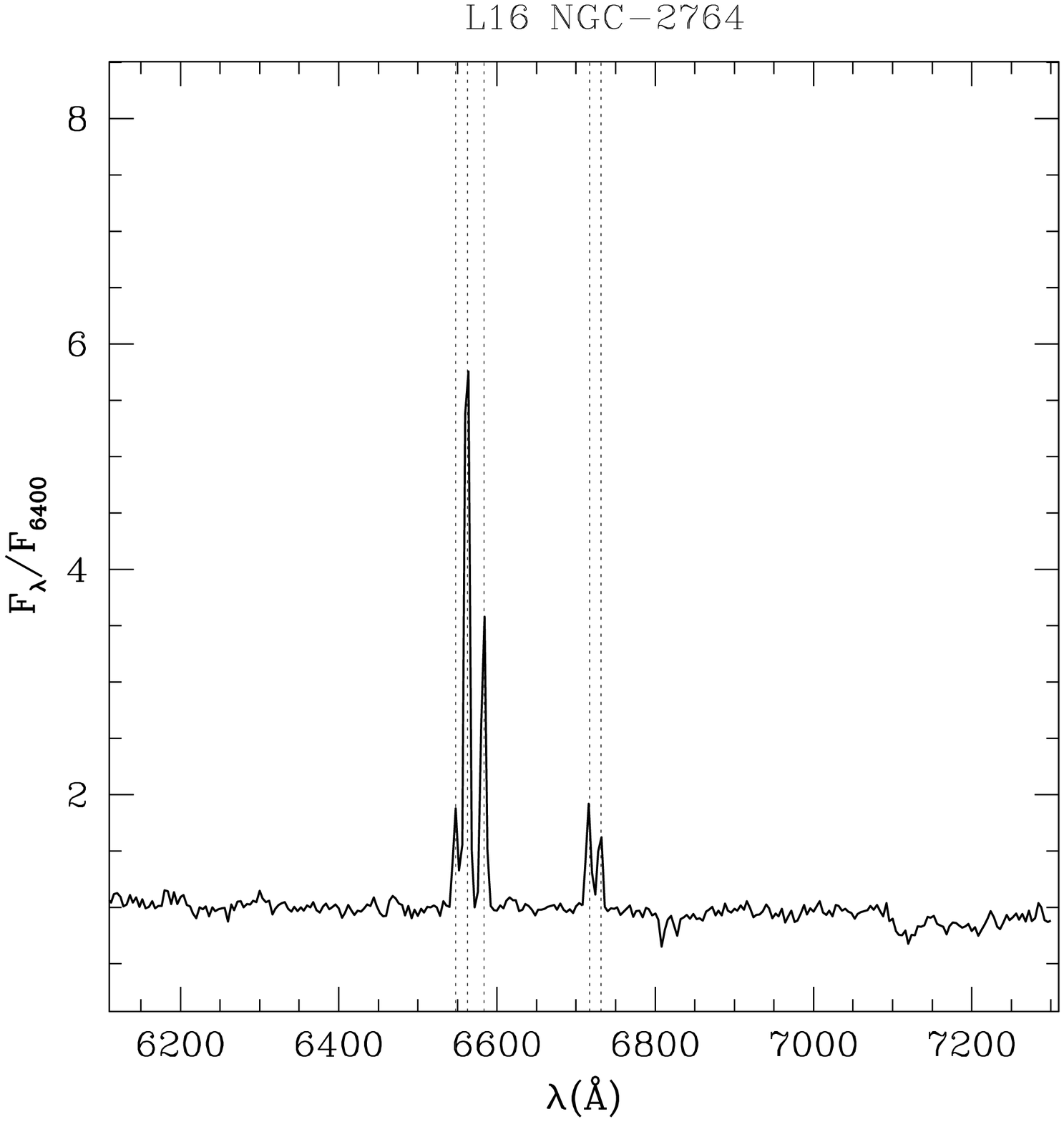}\\
     \includegraphics[scale=0.25]{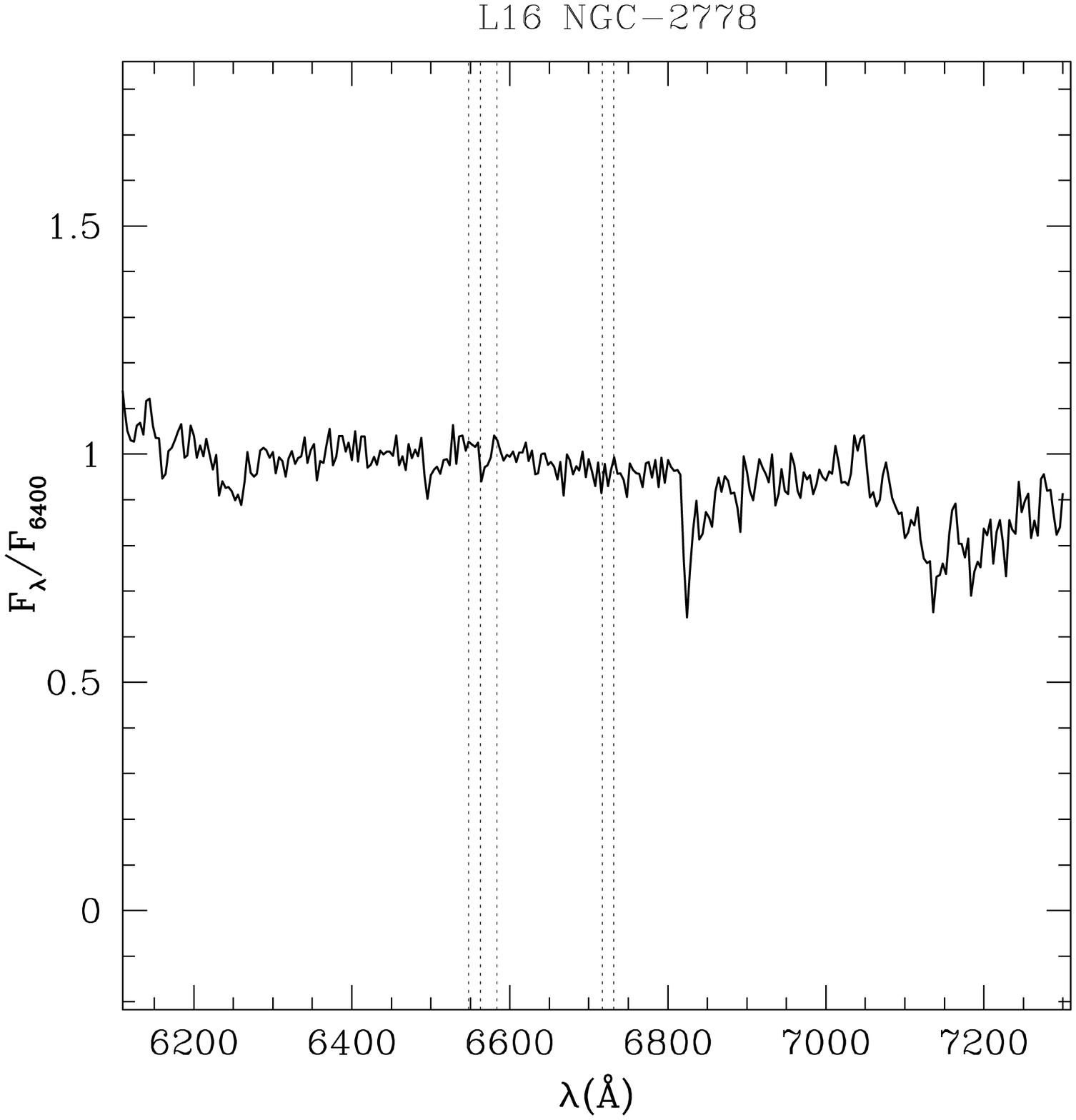}\includegraphics[scale=0.25]{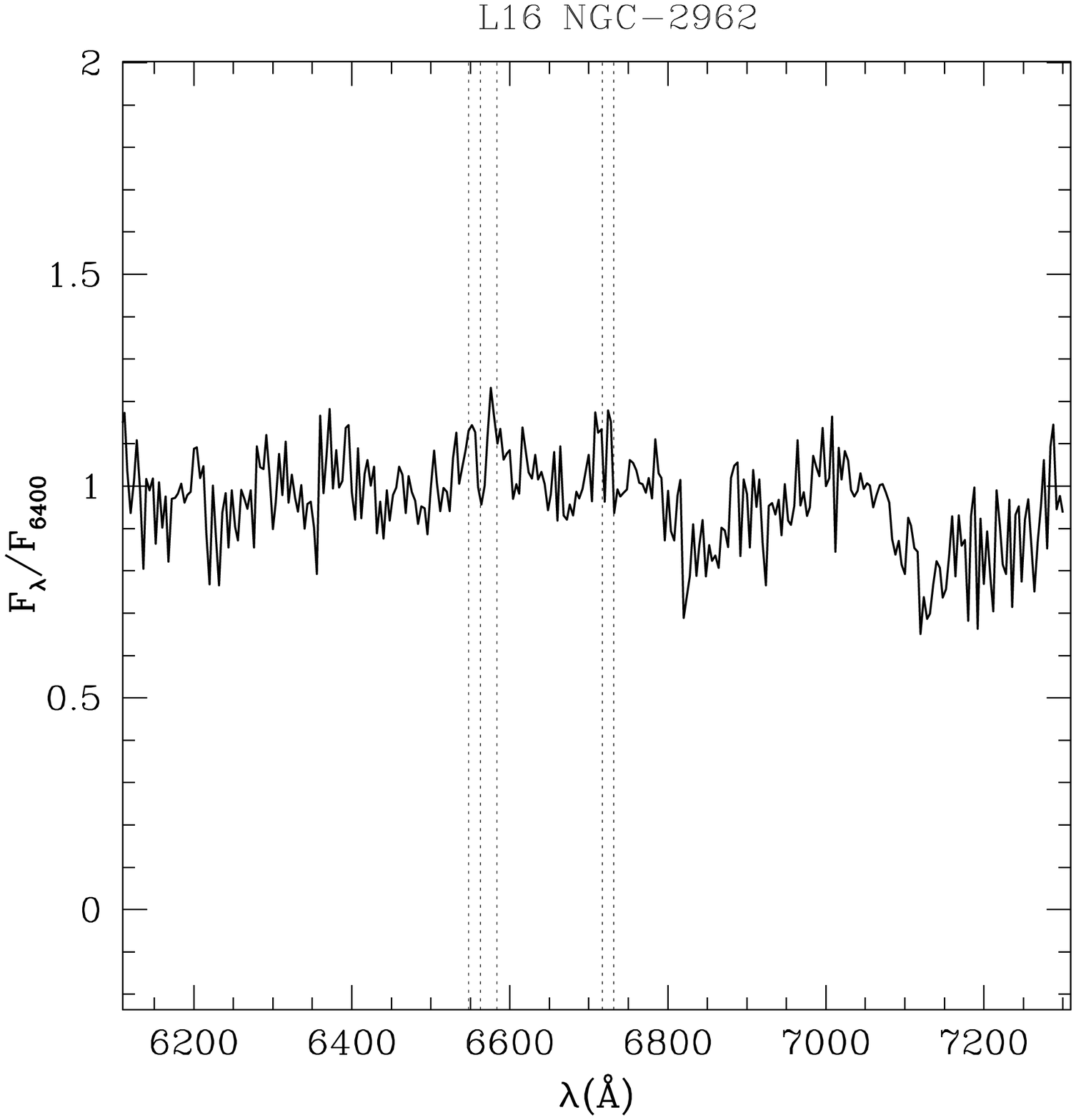}\includegraphics[scale=0.25]{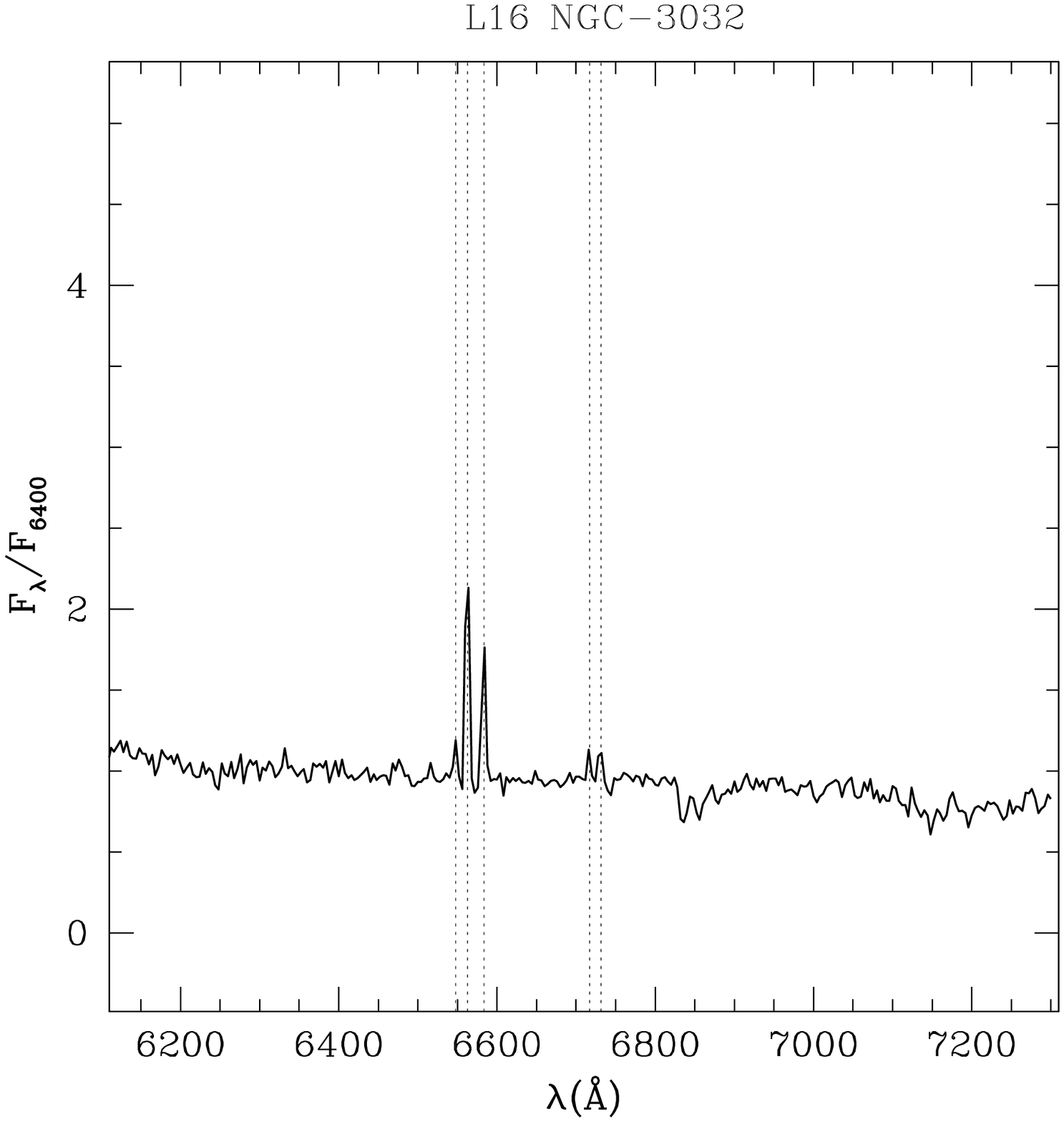}\\
     \includegraphics[scale=0.25]{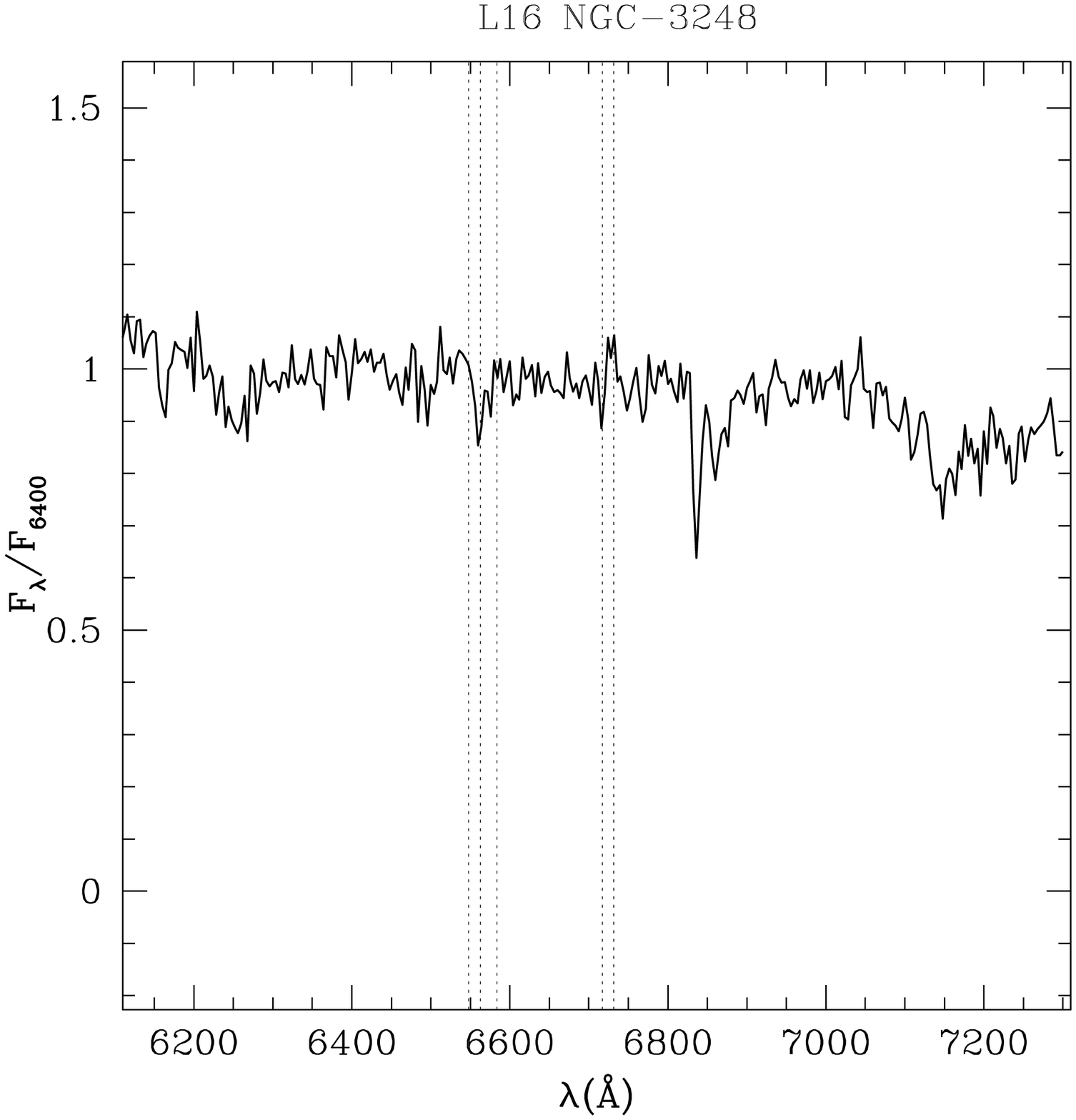}\includegraphics[scale=0.25]{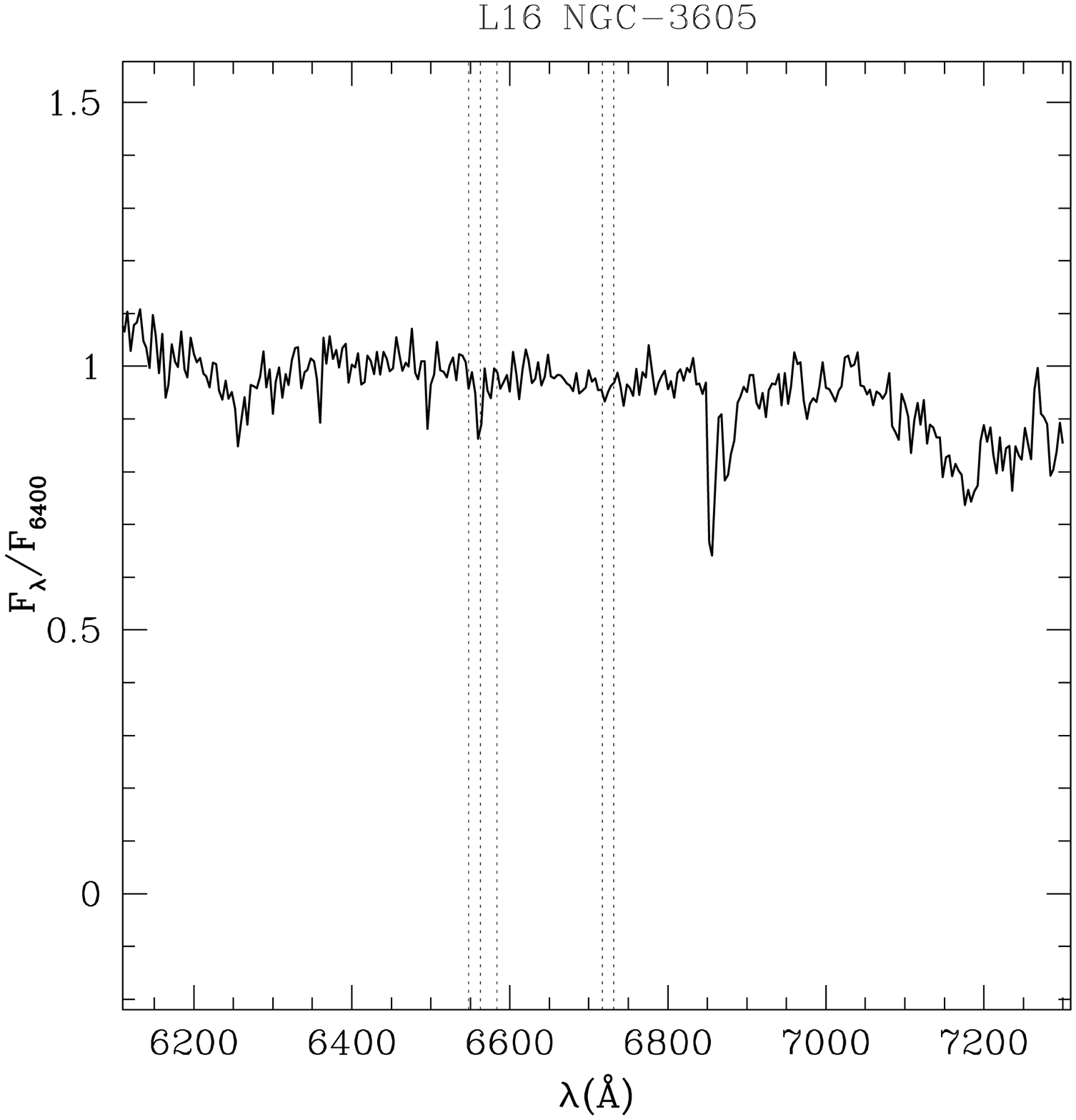}\includegraphics[scale=0.25]{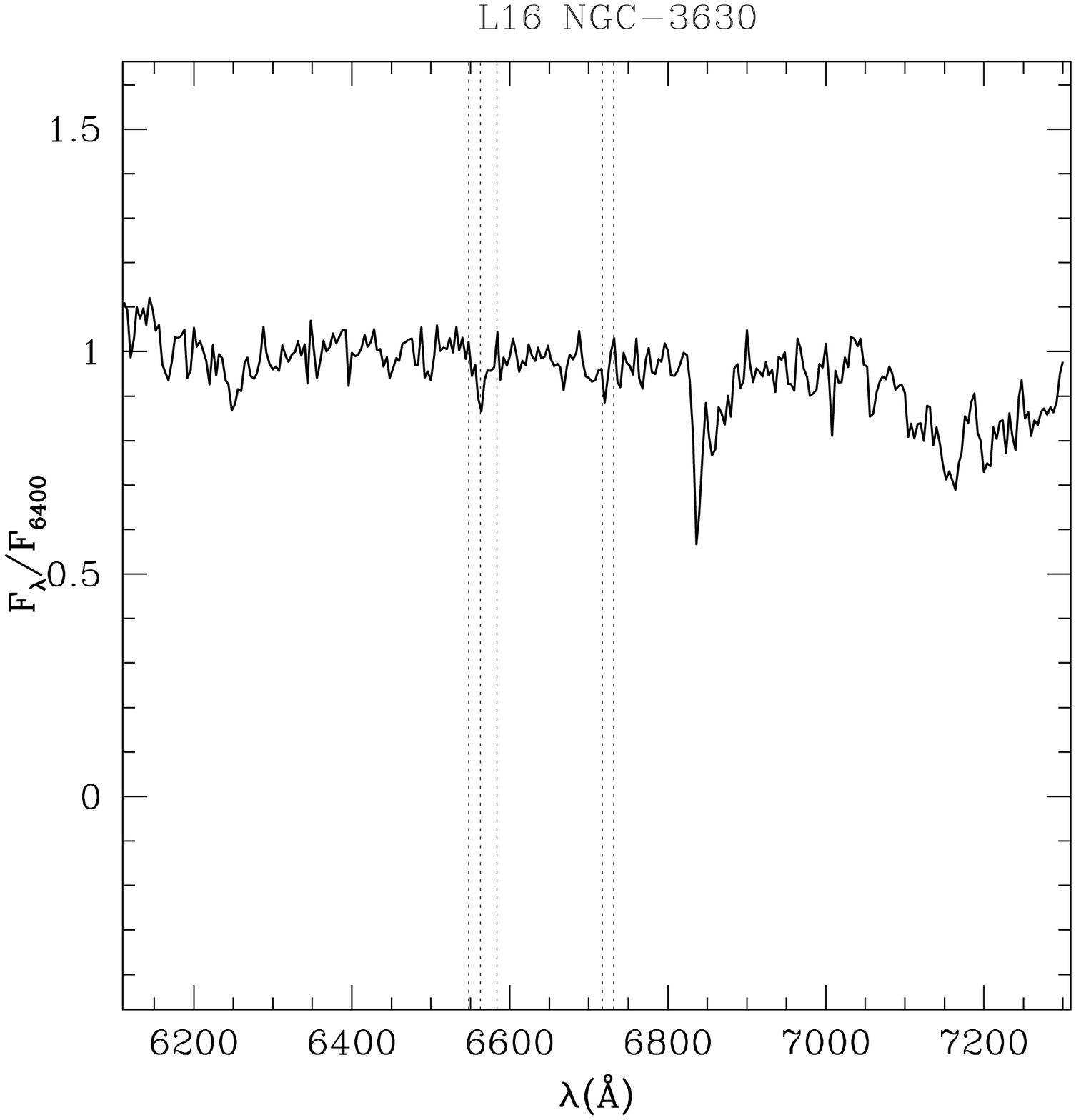}\\
     \includegraphics[scale=0.25]{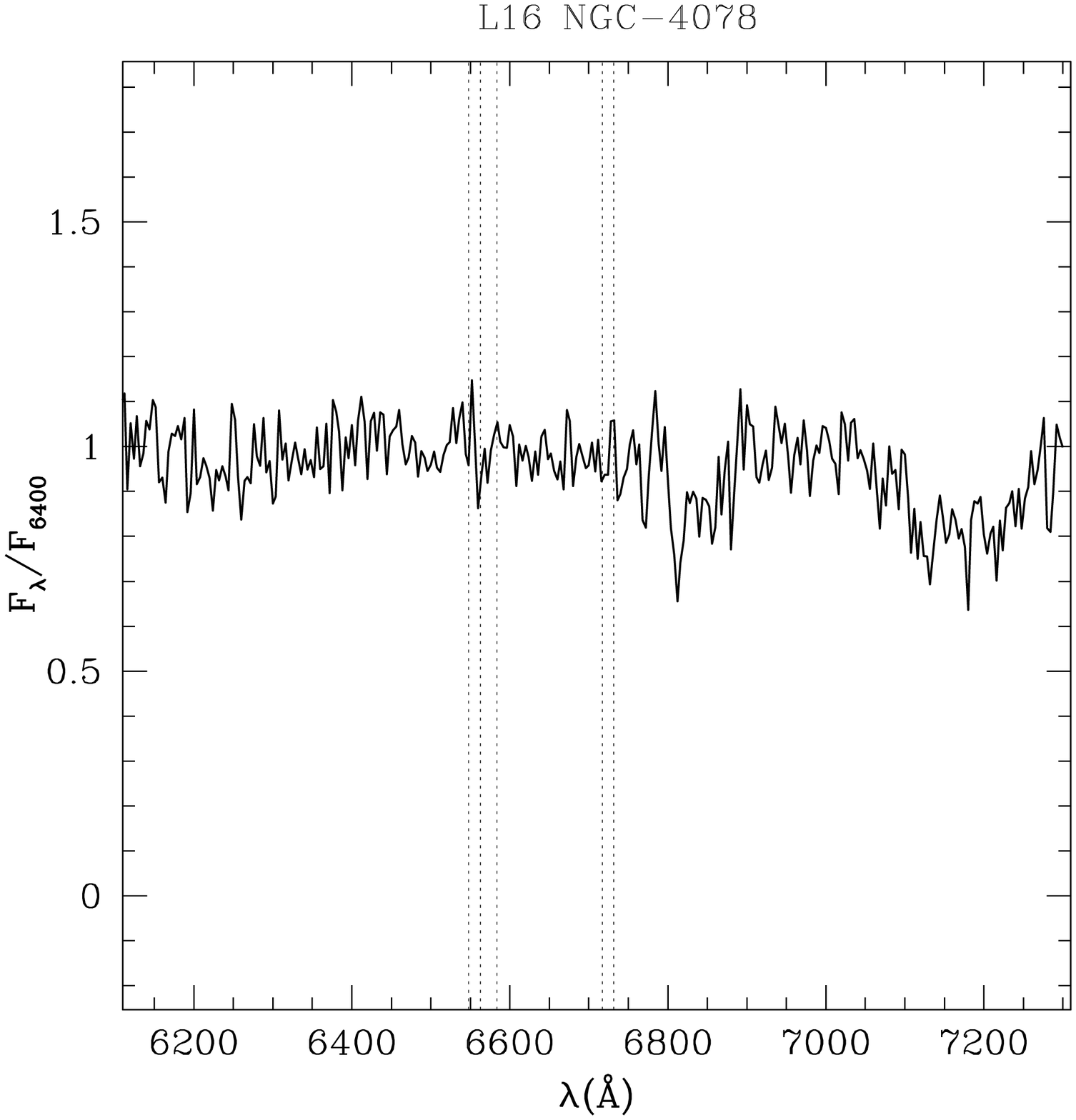}\includegraphics[scale=0.25]{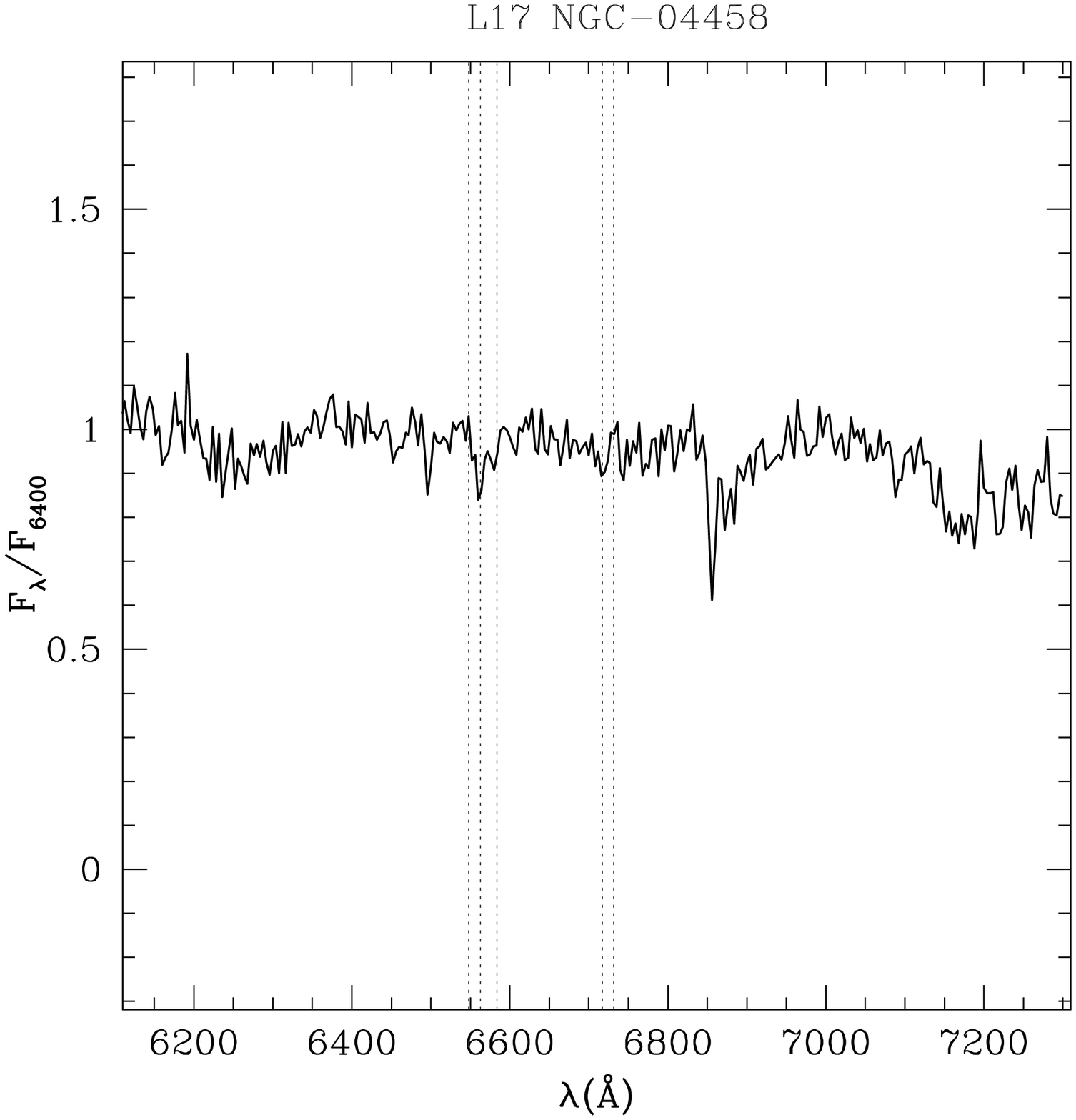}\includegraphics[scale=0.25]{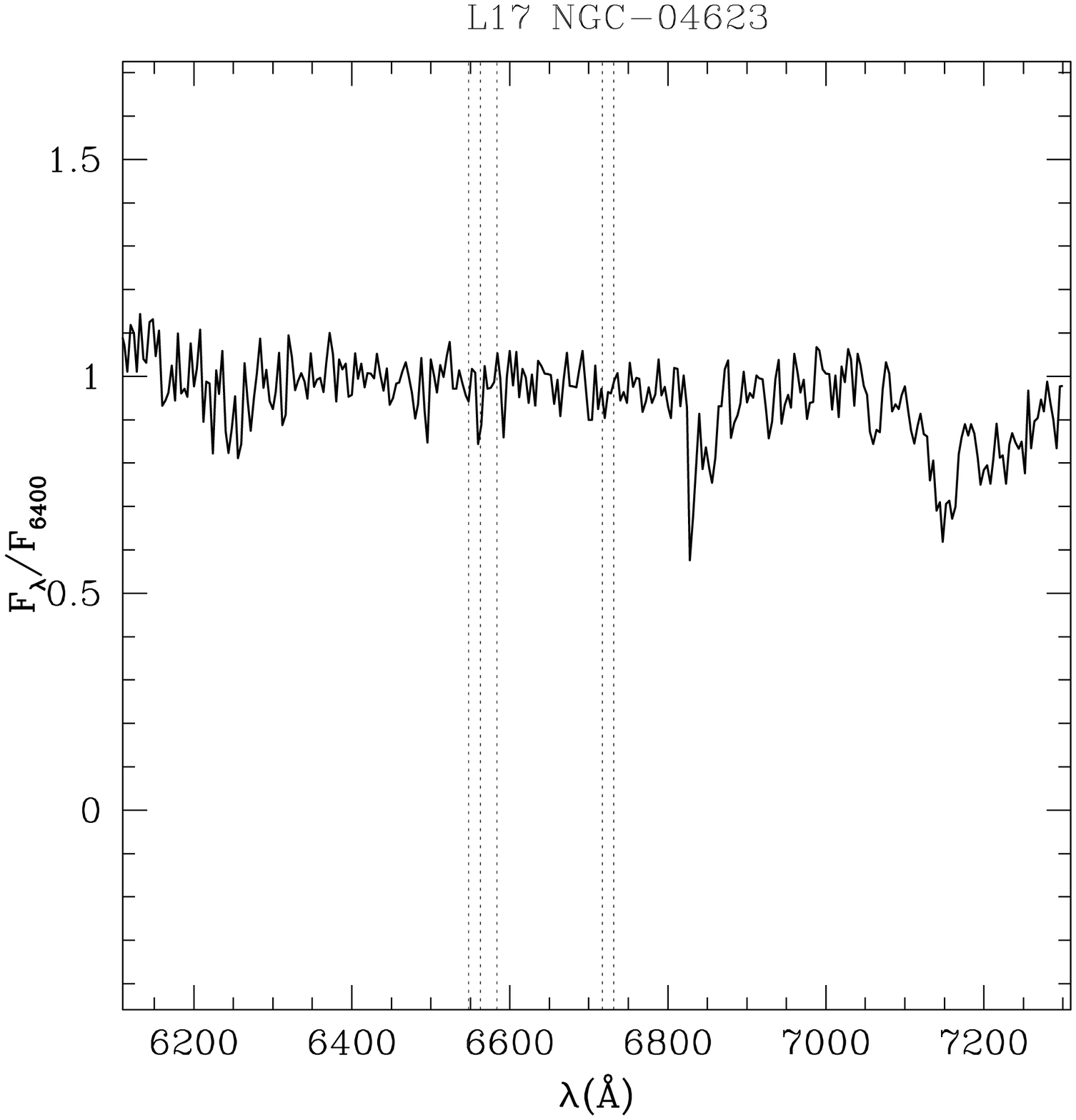}\\
     \caption{Spectra taken at Loiano with the red grism covering from approximately 6200 to 7200 \AA. The spectra have been Doppler shifted to $\lambda_0$ and
     normalized to the flux in the interval 6400-6500 $\AA$. The vertical broken lines mark the rest-frame position of [NII]$\lambda$6549; H$\alpha\lambda$6563;[NII]$\lambda$6584;
     [SII]$\lambda$6717, and [SII]$\lambda$6731.
     }.
     \label{spectra3}  
     \end{figure*}

     \begin{figure*}
     \centering
     \includegraphics[scale=0.30]{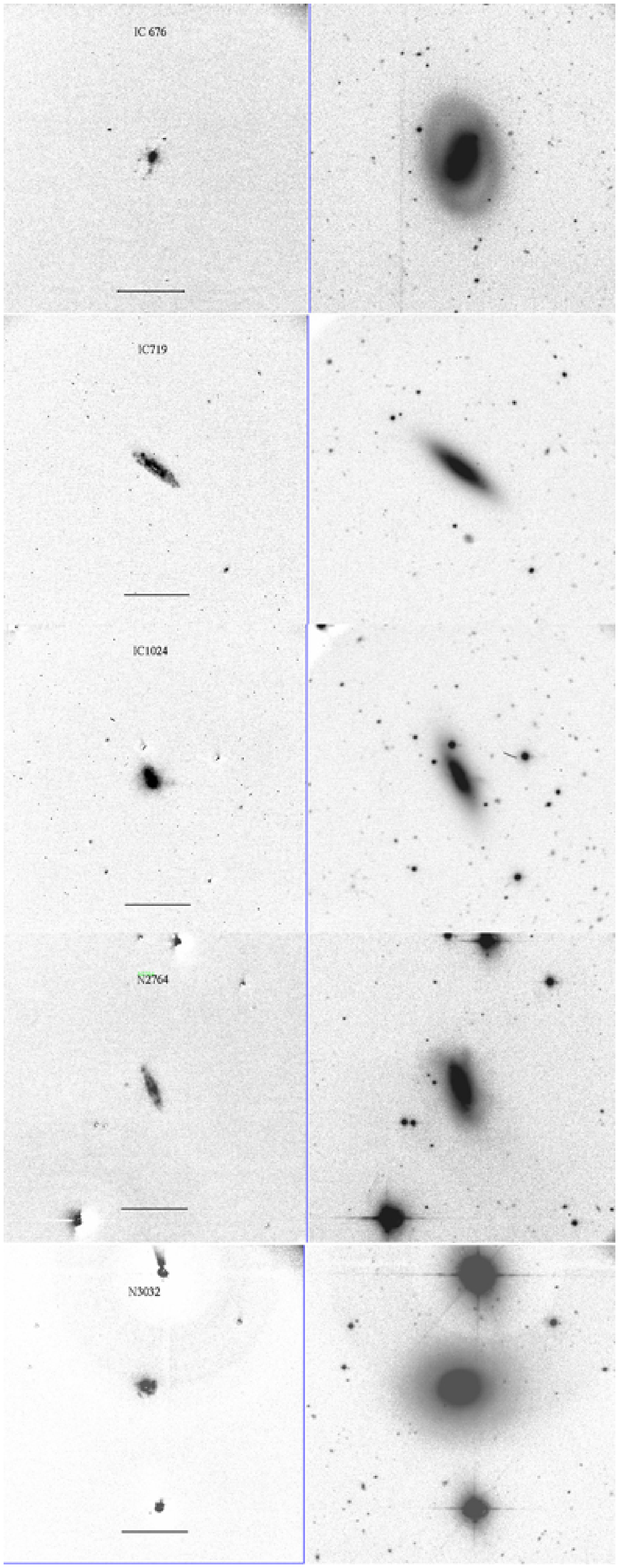} \hskip 2mm \includegraphics[scale=0.30]{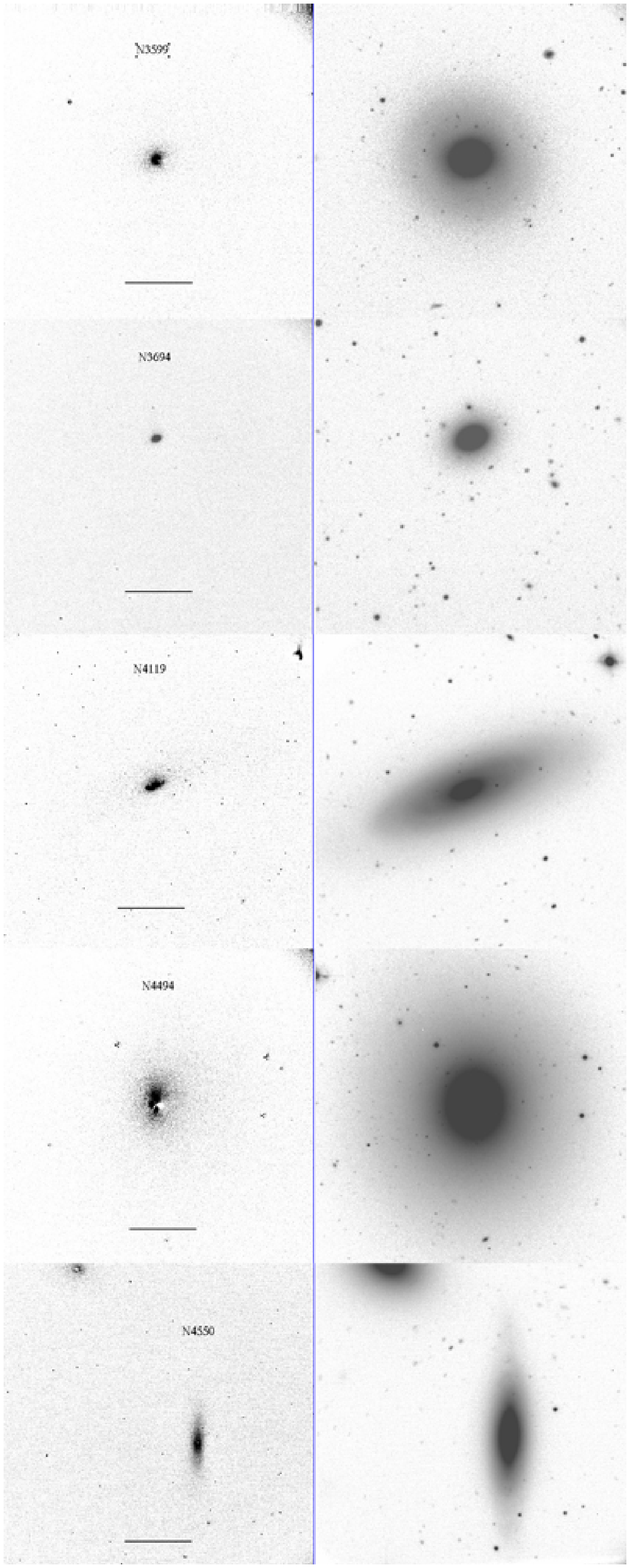}\\
     \label{images5}  
     \end{figure*}
     \begin{figure*}
     \centering
     \includegraphics[scale=0.30]{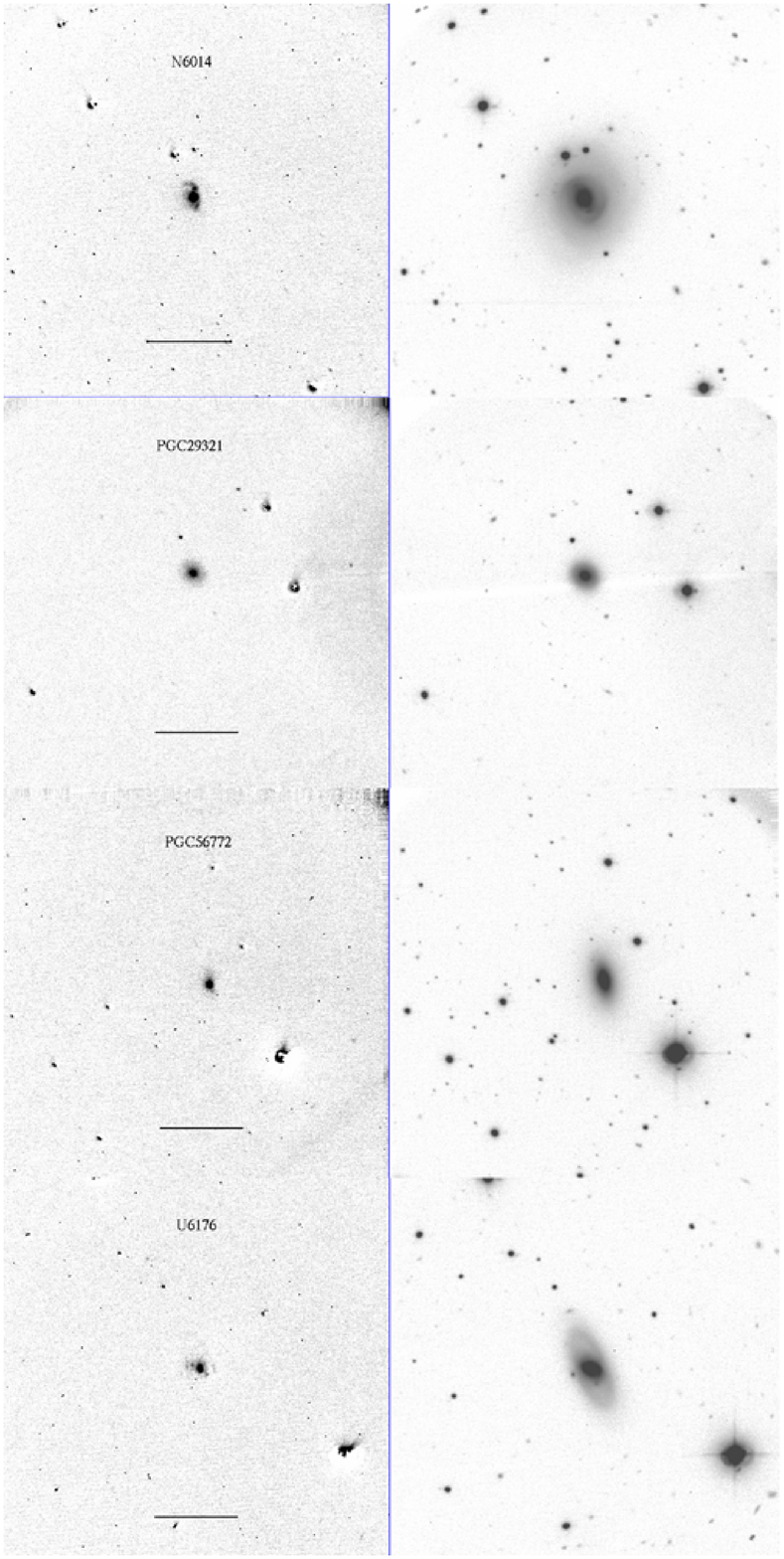} 
     \label{images6}  
     \caption{NET (left) and OFF (right) images of 14 galaxies with strong H$\alpha$ detections (in the present observation campaign). 
     North is up and east is to the left. A 1 arcmin bar is given in all images.}
     \end{figure*}

    \end{onecolumn}

\begin{thebibliography}{}
\bibitem[Abazajian et al.(2009)]{2009ApJS..182..543A} Abazajian, K.~N., Adelman-McCarthy, J.~K., Ag{\"u}eros, M.~A., et al.\ 2009, \apjs, 182, 543-558 
\bibitem[Alam et al.(2015)]{2015ApJS..219...12A} Alam, S., Albareti, F.~D., Allende Prieto, C., et al.\ 2015, \apjs, 219, 12 
\bibitem[Albareti et al.(2016)]{22016} Albareti, F. D., et al. 2016, preprint
\bibitem[Baldry et al.(2004)]{2004ApJ...600..681B} Baldry, I.~K., Glazebrook, K., Brinkmann, J., et al.\ 2004, \apj, 600, 681 
\bibitem[Balogh et al.(2004)]{2004ApJ...615L.101B} Balogh, M.~L., Baldry, I.~K., Nichol, R., et al.\ 2004, \apjl, 615, L101 
\bibitem[Belfiore et al.(2016)]{2016MNRAS.461.3111B} Belfiore, F., Maiolino, R., Maraston, C., et al.\ 2016, \mnras, 461, 3111 
\bibitem[Bell et al.(2003)]{2003ApJS..149..289B} Bell, E.~F., McIntosh, D.~H., Katz, N., \& Weinberg, M.~D.\ 2003, \apjs, 149, 289 
\bibitem[Bois et al.(2011)]{2011MNRAS.416.1654B} Bois, M., Emsellem, E., Bournaud, F., et al.\ 2011, \mnras, 416, 1654 
\bibitem[Boselli \& Gavazzi(2002)]{2002A&A...386..124B} Boselli, A., \& Gavazzi, G.\ 2002, \aap, 386, 124 
\bibitem[Boselli \& Gavazzi(2006)]{2006PASP..118..517B} Boselli, A., \& Gavazzi, G.\ 2006, \pasp, 118, 517 
\bibitem[Boselli et al.(2010)]{2010PASP..122..261B} Boselli, A., Eales, S., Cortese, L., et al.\ 2010, \pasp, 122, 261 
\bibitem[Boselli et al.(2013)]{2013A&A...550A.114B} Boselli, A., Hughes, T.~M., Cortese, L., Gavazzi, G., \& Buat, V.\ 2013, \aap, 550, A114 
\bibitem[Boselli \& Gavazzi(2014)]{2014A&ARv..22...74B} Boselli, A., \& Gavazzi, G.\ 2014, \aapr, 22, 74 
\bibitem[Boselli et al.(2015)]{2015A&A...579A.102B} Boselli, A., Fossati, M., Gavazzi, G., et al.\ 2015, \aap, 579, A102 
\bibitem[Boselli et al.(2014)]{2014A&A...570A..69B} Boselli, A., Voyer, E., Boissier, S., et al.\ 2014, \aap, 570, A69 
\bibitem[Boselli et al.(2014b)]{2014A&A...564A..65B} Boselli, A., Cortese, L., \& Boquien, M.\ 2014, \aap, 564, A65 
\bibitem[Boselli et al.(2015)]{2015A&A...579A.102B} Boselli, A., Fossati, M., Gavazzi, G., et al.\ 2015, \aap, 579, A102 
\bibitem[Bryant et al.(2015)]{2015MNRAS.447.2857B} Bryant, J.~J., Owers, M.~S., Robotham, A.~S.~G., et al.\ 2015, \mnras, 447, 2857 
\bibitem[Buson et al.(1993)]{1993A&A...280..409B} Buson, L.~M., Sadler, E.~M., Zeilinger, W.~W., et al.\ 1993, \aap, 280, 409 
\bibitem[Cappellari et al.(2011)]{2011MNRAS.413..813C} Cappellari, M., Emsellem, E., Krajnovi{\'c}, D., et al.\ 2011, \mnras, 413, 813 
\bibitem[Cappellari et al.(2011b)]{2011MNRAS.416.1680C} Cappellari, M., Emsellem, E., Krajnovi{\'c}, D., et al.\ 2011, \mnras, 416, 1680 
\bibitem[Cappellari(2016)]{2016ARA&A..54..597C} Cappellari, M.\ 2016, \araa, 54, 597 
\bibitem[Ciesla et al.(2014)]{2014A&A...565A.128C} Ciesla, L., Boquien, M., Boselli, A., et al.\ 2014, \aap, 565, A128 
\bibitem[Consolandi et al.(2016)]{2016A&A...591A..38C} Consolandi, G., Gavazzi, G., Fumagalli, M., Dotti, M., \& Fossati, M.\ 2016, \aap, 591, A38 
\bibitem[di Serego Alighieri et al.(2007)]{2007A&A...474..851D} di Serego Alighieri, S., Gavazzi, G., Giovanardi, C., et al.\ 2007, \aap, 474, 851 
\bibitem[Emsellem et al.(2011)]{2011MNRAS.414..888E} Emsellem, E., Cappellari, M., Krajnovi{\'c}, D., et al.\ 2011, \mnras, 414, 888 
\bibitem[Gavazzi et al.(2000)]{2000A&A...361....1G} Gavazzi, G., Boselli, A., V{\'{\i}}lchez, J.~M., Iglesias-Paramo, J., \& Bonfanti, C.\ 2000, \aap, 361, 1 
\bibitem[Gavazzi et al.(2002)]{2002A&A...386..114G} Gavazzi, G., Boselli, A., Pedotti, P., Gallazzi, A., \& Carrasco, L.\ 2002, \aap, 386, 114 
\bibitem[Gavazzi et al.(2003)]{2003A&A...400..451G} Gavazzi, G., Boselli, A., Donati, A., Franzetti, P., \& Scodeggio, M.\ 2003, \aap, 400, 451 
\bibitem[Gavazzi et al.(2011)]{2011A&A...534A..31G} Gavazzi, G., Savorgnan, G., \& Fumagalli, M.\ 2011, \aap, 534, A31 
\bibitem[Gavazzi et al.(2010)]{2010A&A...517A..73G} Gavazzi, G., Fumagalli, M., Cucciati, O., \& Boselli, A.\ 2010, \aap, 517, A73 
\bibitem[Gavazzi et al.(2012)]{2012A&A...545A..16G} Gavazzi, G., Fumagalli, M., Galardo, V., et al.\ 2012, \aap, 545, A16 
\bibitem[Gavazzi et al.(2013)]{2013A&A...558A..68G} Gavazzi, G., Consolandi, G., Dotti, M., et al.\ 2013, \aap, 558, A68 
\bibitem[Gavazzi et al.(2015)]{2015A&A...580A.116G} Gavazzi, G., Consolandi, G., Dotti, M., et al.\ 2015, \aap, 580, A116 
\bibitem[Grossi et al.(2009)]{2009A&A...498..407G} Grossi, M., di Serego Alighieri, S., Giovanardi, C., et al.\ 2009, \aap, 498, 407 
\bibitem[Gualandi \& Merighi (2001)]{} Gualandi R., \& Merighi, R. Thecnical report 2001, Bologna Astronomical Observatory 
\bibitem[Ho et al.(1995)]{1995ApJS...98..477H} Ho, L.~C., Filippenko, A.~V., \& Sargent, W.~L.\ 1995, \apjs, 98, 477 
\bibitem[Kenney et al.(2008)]{2008ApJ...687L..69K} Kenney, J.~D.~P., Tal, T., Crowl, H.~H., Feldmeier, J., \& Jacoby, G.~H.\ 2008, \apjl, 687, L69 
\bibitem[Kauffmann et al.(1993)]{1993MNRAS.264..201K} Kauffmann, G., White, S.~D.~M., \& Guiderdoni, B.\ 1993, \mnras, 264, 201 
\bibitem[Kauffmann et al.(2003)]{2003MNRAS.341...54K} Kauffmann, G., Heckman, T.~M., White, S.~D.~M., et al.\ 2003, \mnras, 341, 54 
\bibitem[Kauffmann et al.(2003b)]{2003MNRAS.346.1055K} Kauffmann, G., Heckman, T.~M., Tremonti, C., et al.\ 2003b, \mnras, 346, 1055 
\bibitem[Kennicutt(1998)]{1998ARA&A..36..189K} Kennicutt, R.~C., Jr.\ 1998, \araa, 36, 189 
\bibitem[Kennicutt \& Kent(1983)]{1983AJ.....88.1094K} Kennicutt, R.~C., Jr., \& Kent, S.~M.\ 1983, \aj, 88, 1094 
\bibitem[Kennicutt(1992)]{1992ApJ...388..310K} Kennicutt, R.~C., Jr.\ 1992, \apj, 388, 310 
\bibitem[Koopmann et al.(2001)]{2001ApJS..135..125K} Koopmann, R.~A., Kenney, J.~D.~P., \& Young, J.\ 2001, \apjs, 135, 125 
\bibitem[Koopmann et al.(2006)]{2006AJ....131..716K} Koopmann, R.~A., Haynes, M.~P., \& Catinella, B.\ 2006, \aj, 131, 716 
\bibitem[Koopmann \& Kenney(2006)]{2006ApJS..162...97K} Koopmann, R.~A., \& Kenney, J.~D.~P.\ 2006, \apjs, 162, 97 
\bibitem[Macchetto et al.(1996)]{1996A&AS..120..463M} Macchetto, F., Pastoriza, M., Caon, N., et al.\ 1996, \aaps, 120, 463 
\bibitem[Massey et al.(1988)]{1988ApJ...328..315M} Massey, P., Strobel, K., Barnes, J.~V., \& Anderson, E.\ 1988, \apj, 328, 315 
\bibitem[Michielsen et al.(2004)]{2004MNRAS.353.1293M} Michielsen, D., de Rijcke, S., Zeilinger, W.~W., et al.\ 2004, \mnras, 353, 1293 
\bibitem[Moore et al.(1999)]{1999MNRAS.304..465M} Moore, B., Lake, G., Quinn, T., \& Stadel, J.\ 1999, \mnras, 304, 465 
\bibitem[Paturel et al.(2003)]{2003A&A...412...45P} Paturel, G., Petit, C., Prugniel, P., et al.\ 2003, \aap, 412, 45 
\bibitem[Penoyre et al.(2017)]{2017arXiv170300545P} Penoyre, Z., Moster, B.~P., Sijacki, D., \& Genel, S.\ 2017, arXiv:1703.00545 
\bibitem[S{\'a}nchez et al.(2012)]{2012A&A...538A...8S} S{\'a}nchez, S.~F., Kennicutt, R.~C., Gil de Paz, A., et al.\ 2012, \aap, 538, A8 
\bibitem[Serra et al.(2012)]{2012MNRAS.422.1835S} Serra, P., Oosterloo, T., Morganti, R., et al.\ 2012, \mnras, 422, 1835 
\bibitem[Spector et al.(2012)]{2012MNRAS.419.2156S} Spector, O., Finkelman, I., \& Brosch, N.\ 2012, \mnras, 419, 2156 
\bibitem[Theios et al.(2016)]{2016ApJ...822...45T} Theios, R.~L., Malkan, M.~A., \& Ross, N.~R.\ 2016, \apj, 822, 45 
\bibitem[Trinchieri \& di Serego Alighieri(1991)]{1991AJ....101.1647T} Trinchieri, G., \& di Serego Alighieri, S.\ 1991, \aj, 101, 1647 
\bibitem[Veale et al.(2017)]{2017MNRAS.464..356V} Veale, M., Ma, C.-P., Thomas, J., et al.\ 2017, \mnras, 464, 356 
\bibitem[Y{\i}ld{\i}z et al.(2017)]{2017MNRAS.464..329Y} Y{\i}ld{\i}z, M.~K., Serra, P., Peletier, R.~F., Oosterloo, T.~A., \& Duc, P.-A.\ 2017, \mnras, 464, 329 
\bibitem[York et al.(2000)]{2000AJ....120.1579Y} York, D.~G., Adelman, J., Anderson, J.~E., Jr., et al.\ 2000, \aj, 120, 1579 
\bibitem[Young et al.(1996)]{1996AJ....112.1903Y} Young, J.~S., Allen, L., Kenney, J.~D.~P., Lesser, A., \& Rownd, B.\ 1996, \aj, 112, 1903 
\bibitem[Young et al.(2011)]{2011MNRAS.414..940Y} Young, L.~M., Bureau, M., Davis, T.~A., et al.\ 2011, \mnras, 414, 940 
\bibitem[Zibetti et al.(2009)]{2009MNRAS.400.1181Z} Zibetti, S., Charlot, S., \& Rix, H.-W.\ 2009, \mnras, 400, 1181 
\end{thebibliography}
\end{document}